\documentclass[prl,twocolumn,superscriptaddress,floatfix]{revtex4-2}
\usepackage[utf8]{inputenc}
\usepackage{graphicx}
\usepackage{float}
\usepackage{amsmath,amssymb,amstext,dsfont,tikz,graphicx,mathtools,bm}
\usepackage[notrig]{physics}
\renewcommand{\vec}[1]{\vb{#1}}
\usepackage{color}
\usepackage[hidelinks]{hyperref}

\usepackage{soul,xcolor}

\begin{document}

\setstcolor{red}

\title{Measurement-Induced Power-Law Negativity in an Open Monitored Quantum Circuit}

\author{Zack Weinstein}
\affiliation{Department of Physics, University of California, Berkeley, CA 94720, USA}

\author{Yimu Bao}
\affiliation{Department of Physics, University of California, Berkeley, CA 94720, USA}

\author{Ehud Altman}
\affiliation{Department of Physics, University of California, Berkeley, CA 94720, USA}
\affiliation{Materials Sciences Division, Lawrence Berkeley National Laboratory, Berkeley, CA 94720, USA}

\date{\today}

\begin{abstract}
	Generic many-body systems coupled to an environment lose their quantum entanglement due to decoherence and evolve to a mixed state with only classical correlations.
	Here, we show that measurements can stabilize quantum entanglement within open quantum systems.
	Specifically, in random unitary circuits with dephasing at the boundary, we find both numerically and analytically that projective measurements performed at a small nonvanishing rate result in a steady state with an $L^{1/3}$ power-law scaling entanglement negativity within the system.
	Using an analytical mapping to a statistical mechanics model of directed polymers in a random environment, we show that the power-law negativity scaling can be understood as Kardar-Parisi-Zhang fluctuations due to the random measurement locations.
	%
	%
	Further increasing the measurement rate leads to a phase transition into an area-law negativity phase, which is of the same universality as the entanglement transition in monitored random circuits without decoherence.
	%
\end{abstract}
\maketitle

The dynamics of quantum entanglement is being investigated extensively as a potential resource for quantum information processing~\cite{kim2013ballistic,kaufman2016quantum,nahum_quantum_2017,li_quantum_2018,skinner2019measurement,arute2019quantum,choi2021emergent,noel2021observation,ippoliti_entanglement_2020}.
%
Recent theoretical developments have shown that large-scale quantum entanglement can be established in monitored quantum systems undergoing unitary evolution interspersed by measurements~\cite{li_quantum_2018,skinner2019measurement,li_measurement-driven_2019,choi_quantum_2020,gullans2020dynamical,ippoliti_postselection-free_2020,ippoliti_fractal_2021,lu_spacetime_2021}.
For moderate measurement rates below a threshold, the entanglement generated by the unitary evolution can overcome the disentangling effect of measurements, leading to volume-law scaling of the entanglement entropy in individual quantum state trajectories at late times.
%
%
Increasing the measurement rate beyond a critical value drives a measurement-induced phase transition (MIPT) to a steady state with area-law scaling of the entanglement entropy~\cite{li_quantum_2018,skinner2019measurement}.

%
%

Studies of monitored systems thus far have largely focused on dynamics involving only unitary gates and projective measurements, which preserves the purity of the quantum state. Insofar as such monitored circuits can be understood as models for entanglement dynamics in generic many-body systems, they are missing an important ingredient. Real systems always exhibit unintended decoherent interactions with their environment, leading inevitably to mixed-state dynamics. Such effects typically destroy internal entanglement, as the system degrees of freedom become entangled with the infinite bath instead of with each other.
%
%
%
%
For example, in monitored random circuits, a nonvanishing rate of decoherence throughout the bulk inevitably results in a short-range entangled steady state at late times~\cite{bao_theory_2020}.
It is therefore natural to ask if a monitored system with weaker decoherence can sustain large-scale entanglement in the steady state.

\begin{figure}[t!]
    \includegraphics[width=\linewidth]{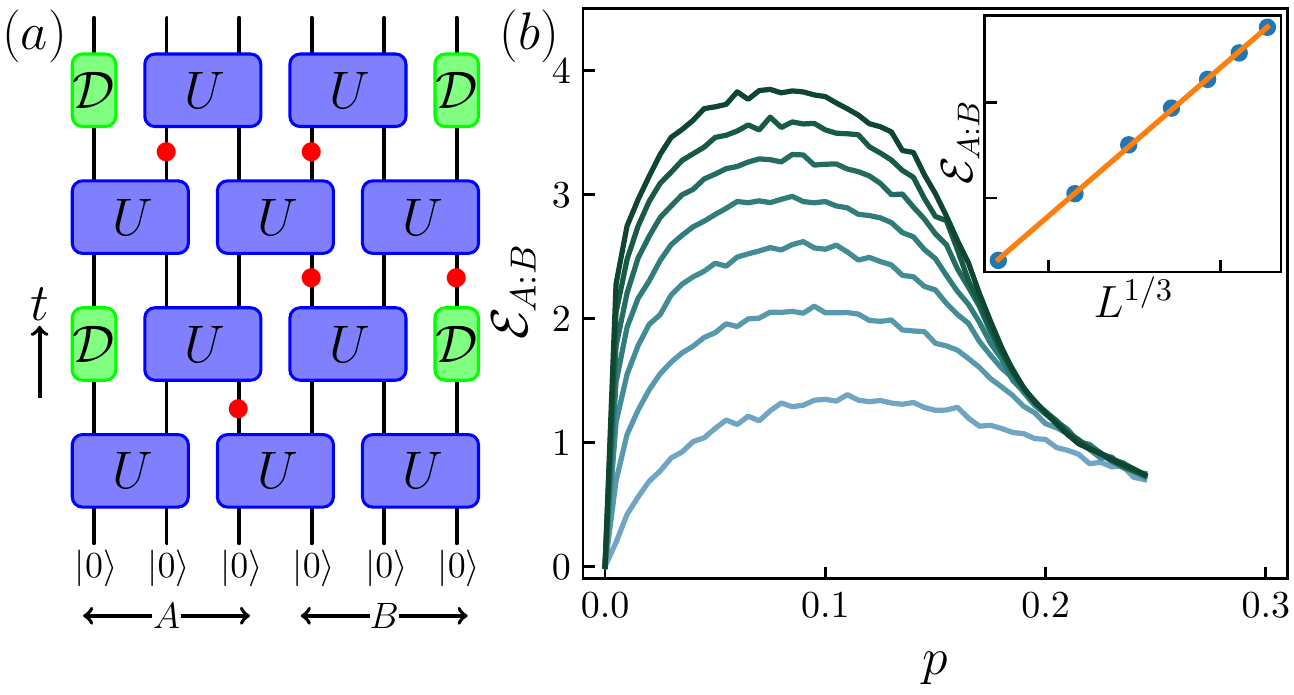}
	\caption{(a) Circuit diagram for the model studied. Qudits are evolved under random unitary gates (blue) and projective measurements (red dots) occurring randomly at a rate $p$, along with dephasing channels (green) applied on the first and last qudit between each layer of unitary gates.
	(b) Late-time logarithmic negativity between subsystems $A$ and $B$, taken to be the left and right halves of the qubit chain, as a function of measurement rate $p$. Different curves indicate various system sizes $L$, ranging from 40 (light blue) to 280 (dark green). Inset: Logarithmic negativity as a function of $L^{1/3}$ (blue dots) at $p = 0.1$, along with the fitting curve $y = c_1 L^{1/3} + c_2$ (orange line) with $c_1 \approx 0.779$ and $c_2 \approx -1.307$. The numerical results are averaged over 200 random circuit realizations.}
	\label{fig:circuit}
\end{figure}

In this Letter, we address this question in models of one-dimensional quantum circuits consisting of random unitary gates and measurements, coupled to an infinite bath at the boundary implemented as a dephasing quantum channel [see Fig.~\ref{fig:circuit}(a)]. Using the logarithmic entanglement negativity as a measure of mixed-state entanglement \cite{vidal_computable_2002,plenio_logarithmic_2005,calabrese_entanglement_2012,calabrese_entanglement_2013,lu_singularity_2019,lu_detecting_2020,wu_entanglement_2020,lu_entanglement_2020,sang_entanglement_2020,shapourian_entanglement_2020}, we employ both numerical simulations of Clifford circuits and an analytical mapping to a statistical mechanics model to assess the scaling of internal entanglement in the circuit qubits.

While the entanglement negativity vanishes in the absence of measurements as expected, we find numerically that the half-system negativity exhibits an $L^{1/3}$ power-law scaling with system size for nonzero measurement rates below the MIPT critical point [see Fig.~\ref{fig:circuit}(b)]. This power law persists until a critical measurement rate $p_c$, above which the negativity exhibits an area law. 

To develop a theoretical understanding of the observed power-law negativity, we build on previous works mapping the dynamics of entanglement entropy in random circuits to effective statistical mechanics models~\cite{hayden_holographic_2016,nahum_quantum_2017,zhou_emergent_2019,bao_theory_2020,jian_measurement-induced_2020}. Using a similar replica formalism to previous works, we show that the negativity can be calculated using the same effective model of ferromagnetic spins with different boundary conditions~\cite{dong_holographic_2021}. The vanishing of the volume law contribution to the negativity in the presence of dephasing channels is immediately seen to be a consequence of symmetry-breaking boundary conditions imposed by dephasing.

The exponent 1/3 has previously been observed in \textit{subleading} contributions to the bipartite entanglement entropy in the pure state dynamics of circuit models, both in $t^{1/3}$ subleading growth of entanglement entropy over time in pure unitary circuits \cite{nahum_quantum_2017, zhou_emergent_2019}, and in $\ell^{1/3}$ subleading scaling of late-time entanglement entropy with subsystem size in monitored circuits within the volume-law phase \cite{li_entanglement_2021}. This exponent was explained as the Kardar-Parisi-Zhang (KPZ) fluctuations of the domain walls, interpreted as directed polymers in a random environment \cite{kardar_roughening_1985,huse_huse_1985,kardar_dynamic_1986,kardar_statistical_2007}. 
Here, we analytically derive a mapping relating the negativity to a collection of directed polymers within a limit of large qudit dimension $d \rightarrow \infty$ and verify its prediction of $L^{1/3}$ negativity scaling. 
Building on previous works \cite{zhou_emergent_2019,li_entanglement_2021,agrawal_entanglement_2021}, we explicitly demonstrate the role of measurements in generating a random attractive potential on the polymers, which naturally lead to KPZ fluctuations in the negativity for nonzero measurement rates.


\textit{Model.---} 
We consider a chain of $L$ $d$-qudits with open boundary conditions, initialized in the product state $\ket{0}^{\otimes L}$, and evolved under a brick-wall random unitary circuit [see Fig.~\ref{fig:circuit}(a)], where each gate is independently drawn from the Haar ensemble. 
In between layers of unitary gates, each $i$th qudit is measured in the computational basis $\qty{ \ket{a} }_{a = 0}^{d-1}$ with probability $p$, which collapses the system onto the state $\rho \mapsto P^a_i \rho P^a_i/\tr (P^a_i \rho)$ with probability $\tr (P^a_i \rho)$ given by the Born rule, where $P^a_i = \dyad{a}_i$ projects the $i$th qudit onto the state $\ket{a}$. 
To model the coupling to an infinite bath, the boundary qudits $i=1$ and $i=L$ are subjected to local dephasing described by $\mathcal{D}_i[\rho] = \sum_{a=0}^{d-1}P^a_i \rho P^a_i$~\cite{nielsen_quantum_2010,lidar_lecture_2020}. This coupling can also be understood as a measurement in which we average the density matrix over all possible measurement outcomes.

The addition of dephasing channels results in open-system dynamics and inevitably drives the system into a mixed state, for which the von Neumann entropy is no longer a meaningful measure of entanglement \cite{bennett_mixed-state_1996,horodecki_mixed-state_1998}. 
To quantify quantum entanglement within the system at late times, we employ the logarithmic negativity~\cite{plenio_logarithmic_2005,calabrese_entanglement_2012,calabrese_entanglement_2013,lu_singularity_2019,lu_detecting_2020,lu_entanglement_2020,wu_entanglement_2020,shapourian_entanglement_2020,sang_entanglement_2020,dong_holographic_2021}, a measure of mixed-state bipartite entanglement and a rigorous upper bound to the distillable entanglement of a mixed state~\cite{peres_separability_1996,horodecki_separability_1996,horodecki_mixed-state_1998,vidal_irreversibility_2001,vidal_computable_2002,horodecki_five_2020}:
\begin{equation}
\label{eq:logEN}
	\mathcal{E}_{A:B}[\rho] = \log \norm{\rho^{T_B}}_1 ,
\end{equation}
where $\rho^{T_B}$ is the partial transpose of $\rho$ in subsystem $B$, and $\norm{\cdot}_1$ denotes the trace norm. Throughout this Letter, we take $A$ and $B$ to respectively consist of the left and right halves of the qudit chain.
Note that Ref. \cite{sang_entanglement_2020} previously used the logarithmic negativity to characterize the conformal field theory underlying the MIPT without decoherence.


\textit{Numerical Results.---}
To efficiently simulate the circuit, we employ random Clifford unitary gates acting on $d = 2$ qubits using the stabilizer formalism~\cite{gottesman_heisenberg_1998,aaronson_improved_2004,hamma2005ground,hamma2005bipartite,nahum_quantum_2017,li_measurement-driven_2019,SOM}.
While the Clifford gates are not generic, they form a unitary 3-design~\cite{webb2015clifford} and are expected to give the same qualitative behavior as the Haar random circuit.
The late-time negativity as a function of measurement rate $p$ for system sizes up to $L = 280$ is shown in Fig.~\ref{fig:circuit}(b).

In the case without measurements (i.e. $p=0$), the late-time negativity is uniformly zero independent of system size. 
This is to be expected both from general physical considerations and from Page's theorem \cite{page_average_1993,bhosale_entanglement_2012,lu_entanglement_2020}: if the dephasing channels are understood as an effective coupling to an infinitely large bath, then the system becomes maximally entangled with the bath at late times and no bipartite entanglement within the system remains. 

Remarkably, the negativity sharply increases as $p$ increases from zero and exhibits nontrivial scaling with the system size. 
At moderate measurement rates, for example $p=0.1$, the scaling of the negativity is consistent with a power law of the form $\mathcal{E}_{A:B} = c_1 L^{1/3} + c_2$ for two fitting parameters $c_{1,2}$ as shown in the inset.

At sufficiently high measurement rates, the negativity begins to decrease as a function of measurement strength. 
This culminates in a measurement-induced transition at $p_c$ in which the power-law coefficient $c_1$ vanishes. 
Since our circuit model differs from previous pure-state circuits only in its boundary conditions, we expect the bulk critical behavior to be identical to that of the ordinary MIPT without dephasing. In the Supplemental Material \cite{SOM} we perform a finite-size scaling analysis and find $p_c \simeq 0.16$ consistent with previous works \cite{li_quantum_2018,li_measurement-driven_2019,zabalo2020critical}, but we cannot reliably extract a correlation length exponent $\nu$ due to the numerical smallness of the negativity.

\textit{Effective statistical mechanics model.---} 
Our numerical results can be understood analytically by relating the averaged logarithmic negativity to the free energy of directed polymers in a random environment.
Here, we consider the Haar random circuit acting on $d$-qudits with $d\rightarrow \infty$ allowing for greater analytical control~\cite{nahum_quantum_2017,zhou_emergent_2019,bao_theory_2020,jian_measurement-induced_2020,SOM}.
Within the effective model, the $L^{1/3}$ negativity scaling can be understood as KPZ fluctuations of the directed polymers.
Complete details of the statistical mechanics model can be found in the Supplemental Material \cite{SOM}.


The $n$th R\'enyi negativity \cite{calabrese_entanglement_2012,calabrese_entanglement_2013,lu_singularity_2019,lu_detecting_2020} (properly defined for $n \geq 4$) for a \textit{fixed} set of measurement locations $\vec{X}$ in spacetime, averaged over Haar unitary gates $\mathcal{U} = \qty{U_{ij,t}}$ and measurement outcomes $\vec{m}$, is given by
\begin{equation}
\label{eq:avgRenyiEN}
	\overline{\mathcal{E}^{(n)}_{A:B}(\vec{X})} = \mathbb{E}_{\mathcal{U}} \sum_{\vec{m}} p_{\vec{m}} \frac{1}{2-n} \log \qty{ \frac{\tr[(\rho_{\vec{m}}^{T_B})^n]}{\tr \rho_{\vec{m}}^n} } ,
\end{equation}
where $\rho_{\vec{m}}$ is the unnormalized density matrix obtained along the measurement trajectory $\vec{m}$, and $p_{\vec{m}} = \tr \rho_{\vec{m}}$ is the probability for achieving the measurement outcomes $\vec{m}$ conditioned on the locations of the measurements $\vec{X}$ and the unitary realization $\mathcal{U}$.
The logarithmic negativity [Eq.~(\ref{eq:logEN})] is obtained from Eq. (\ref{eq:avgRenyiEN}) using the peculiar limit $n \rightarrow 1$ along \textit{even} $n$ \cite{calabrese_entanglement_2012,calabrese_entanglement_2013}. 

To facilitate the mapping, we employ the replica trick \cite{nishimori_statistical_2001,kardar_statistical_2007} to write $\overline{\mathcal{E}^{(n)}_{A:B}} = \lim_{k \rightarrow 0} \mathcal{E}^{(n,k)}_{A:B}$, where $\mathcal{E}^{(n,k)}_{A:B}$ can be interpreted as being proportional to the difference of two free energies:
\begin{equation}
	\mathcal{E}^{(n,k)}_{A:B}(\vec{X}) = - \frac{1}{k(n-2)} \log \qty{ \frac{Z^{(n,k)}}{Z^{(n,k)}_0} } ,
\end{equation}
where the two ``partition functions'' $Z^{(n,k)}$ and $Z^{(n,k)}_0$ differ only in their boundary conditions at the final time slice. Note that these partition functions contain the averages over unitary realizations and measurement outcomes, but \textit{not} the locations of measurements; following Ref.~\cite{agrawal_entanglement_2021}, and in contrast to previous works \cite{bao_theory_2020,jian_measurement-induced_2020}, we leave the locations of measurements as quenched disorder.

\begin{figure}[t!]
	\includegraphics[width=\linewidth]{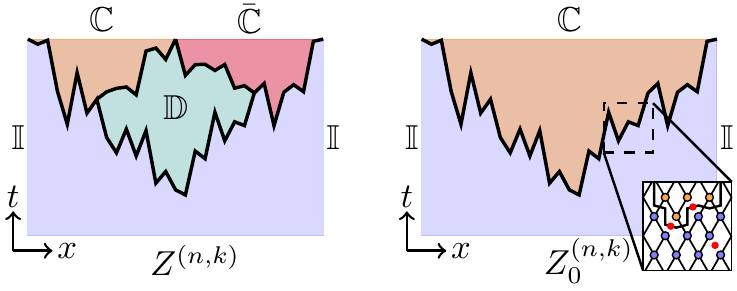}
	\caption{\label{fig:spinmodel}Schematic zero temperature spin configurations of $Z^{(n,k)}$ and $Z^{(n,k)}_0$ for a fixed disorder realization of measurement locations. The final time boundary conditions are shown at the top of each diagram: $Z^{(n,k)}$ contains cyclic permutations $\mathbb{C}$ (orange) at the top of region $A$ and anticyclic permutations $\bar{\mathbb{C}}$ (red) at the top of region $B$, while $Z^{(n,k)}_0$ contains cyclic permutations along the entire top boundary. Dephasing at the left and right boundaries of the chain enforces identity permutations $\mathbb{I}$ (blue) at the left and right boundaries of the effective model. An intermediate domain of spins $\mathbb{D}$ (green) appears in $Z^{(n,k)}$ by a similar mechanism as in Ref. \cite{dong_holographic_2021}. Viewing domain walls as collections of polymers, measurements result in a random attractive potential on the polymers, leading to KPZ fluctuations in the negativity.}
\end{figure}

As in previous works \cite{collins_moments_2003,collins2006integration,nahum_operator_2018,you2018entanglement,zhou_emergent_2019,bao_theory_2020,jian_measurement-induced_2020,agrawal_entanglement_2021}, averaging over Haar random unitary gates results in a sum over pairing configurations between the replicated copies of the density matrix. The bulk effective statistical mechanics model is then a lattice magnet containing permutation-valued spins with ferromagnetic interactions, where a given permutation $\sigma$ from the permutation group $S_{nk+1}$ represents a local tensor contraction between each $\ell$th ket and the $\sigma(\ell)$th bra. The boundary conditions of $Z^{(n,k)}$ and $Z^{(n,k)}_0$ at the final time, which are unique to the calculation of the R\'enyi negativity \cite{dong_holographic_2021}, are shown in Fig.~\ref{fig:spinmodel}: $Z^{(n,k)}$ contains cyclic permutations $\mathbb{C}$ at the top of region $A$ and anticyclic permutations $\bar{\mathbb{C}}$ at the top of region $B$, while $Z^{(n,k)}_0$ contains cyclic permutations along the entire top boundary. $\mathcal{E}^{(n,k)}_{A:B}$ is thus proportional to the free energy cost of imposing a domain wall between $\mathbb{C}$ and $\bar{\mathbb{C}}$ at the interface of $A$ and $B$ at the final time boundary. In the analytically tractable $d \to \infty$ limit, the energetic cost per length of a domain wall away from a measured site is \cite{zhou_emergent_2019,bao_theory_2020,jian_measurement-induced_2020}
\begin{equation}
\label{eq:DWcost}
	\beta E(\sigma_i, \sigma_j) = \abs{\sigma_i^{-1} \sigma_j} \log d \quad (d \rightarrow \infty) ,
\end{equation}
where $\abs{\sigma_i^{-1} \sigma_j}$ is the number of transpositions required to obtain the permutation $\sigma_j$ from $\sigma_i$. The limit $d \rightarrow \infty$ imposes zero temperature, $\beta^{-1} \rightarrow 0$. Reference \cite{zhou_emergent_2019} argued (in the absence of measurements) that a domain wall between permutations $\sigma_i$ and $\sigma_j$ should be viewed as a collection of $\abs{\sigma_i^{-1} \sigma_j}$ ``elementary'' domain walls given by transpositions, which become noninteracting in the $d \rightarrow \infty$ limit according to Eq.~(\ref{eq:DWcost}). Weak interactions between such elementary domain walls can be calculated perturbatively in powers of $1/d$.

The presence of boundary dephasing channels modifies the left and right boundary conditions of the effective model. In contrast to the open boundary conditions for models without decoherence \cite{zhou_emergent_2019}, dephasing imposes identity permutation spins $\mathbb{I}$ at the left and right boundaries, leading to the domain wall structure in Fig.~\ref{fig:spinmodel}; note that an intermediate domain of spins $\mathbb{D}$ (green) can appear in $Z^{(n,k)}$ without additional energy cost provided that $\abs{\sigma^{-1}\mathbb{D}}+\abs{\mathbb{D}^{-1}\tau} = \abs{\sigma^{-1}\tau}$ for $\sigma, \tau = \mathbb{C}, \bar{\mathbb{C}}, \mathbb{I}$~\cite{dong_holographic_2021,SOM}.
Using Eq.~(\ref{eq:DWcost}), this domain wall structure leads to a negativity in the $d \to \infty$ limit of the form
\begin{equation}
\label{eq:EN_DWs}
	\mathcal{E}^{(n,k)}_{A:B}(\vec{X}) = \frac{\log d}{2} \Big\{ \ell_A + \ell_B - \ell_{AB} \Big\} ,
\end{equation}
where $\ell_R$ is the length of the minimal domain wall separating the top boundary of region $R$ from the rest of the system. Note that this quantity is independent of the replica indices $(n,k)$, allowing for the replica limit to be trivially taken. In the absence of measurements these domain walls take straight lines through the system, and the negativity therefore vanishes~\cite{SOM}. This is consistent both with the expectation from Page's theorem and the $p=0$ numerical results of Fig.~\ref{fig:circuit}.


To explain the $L^{1/3}$ scaling of negativity at nonzero measurement rates, we must address the role of measurements in the effective spin model. By keeping the spacetime locations of measurements as unaveraged quenched disorder, we find that measurements effectively eliminate the ferromagnetic bonds between adjacent spins. In the $d \rightarrow \infty$ limit, each domain wall will optimize to pass through as many measurement locations as possible to minimize its energy. Viewing each domain wall as a collection of polymers as in \cite{zhou_emergent_2019}, the elimination of ferromagnetic bonds can be understood as a random attractive potential on the polymers, wherein the energy of a polymer is reduced by $\log d$ for each measured site the polymer passes through. The total energy cost of a single polymer, directed \footnote{In the zero temperature, late time limit, we may assume that the polymer is \textit{directed} in the $x$ direction -- that is, it has no loops or overhangs \cite{kardar_statistical_2007}.} in the $x$ direction with spatial profile $y(x)$, is given by
\begin{equation}
	\beta H[y(x)] = \log d \int \dd{x} \qty{ 1 + \frac{1}{2} (\partial_x y)^2 + V(x,y) } ,
\end{equation}
where $V(x,y)$ is a random potential with mean $p \log d$ and variance $p(1-p)(\log d)^2 \delta(x-x') \delta(y-y')$.
%
%
$\mathcal{E}^{(n,k)}_{A:B}$ is then simply proportional to the sum of the polymer ground state energies in $Z^{(n,k)}$, minus those in $Z^{(n,k)}_0$. Each such energy may then be averaged over measurement locations independently.

The free energy of a directed polymer in a random environment has been well-studied---it is equivalent to the KPZ equation via the Hopf-Cole transformation \cite{huse_pinning_1985,kardar_roughening_1985,huse_huse_1985,kardar_statistical_2007}. Since the polymers here are restricted to the half-plane below the final time slice, a solution for the free energy of each polymer is obtained from the KPZ equation in the half-plane, which can be calculated analytically using Bethe ansatz methods \cite{gueudre_directed_2012,barraquand_half-space_2020,li_entanglement_2021}. The result is an energetic contribution $s_0 \ell + s_1 \ell^{1/3}$ for each polymer of horizontal length $\ell$, where $s_0$ and $s_1$ are nonuniversal positive constants. It can then be seen from Eq.~(\ref{eq:EN_DWs}) that the linear contributions from each polymer cancel as in the $p=0$ case, but the $\ell^{1/3}$ contributions due to KPZ fluctuations in the polymer lengths do not---they yield a positive $L^{1/3}$ growth of the averaged R\'enyi negativity $\overline{\mathcal{E}^{(n)}_{A:B}}$. Although this analytical argument cannot compute the dependence of the power-law coefficient $s_1$ on the measurement rate, the qualitative prediction of $L^{1/3}$ negativity scaling for nonzero measurement rates is consistent with the Clifford numerical results.
%


\textit{Discussion.---} 
We have shown that the active monitoring of a random quantum circuit with decoherence at the boundaries can stabilize large-scale entanglement. This is evinced by the $L^{1/3}$ power-law scaling of late-time entanglement negativity, which is obtained only for nonzero measurement rates below a critical threshold $p_c$. The enhancement of quantum entanglement by measurements in the presence of decoherence stands in contrast with the effect of measurements in random circuits featuring strictly pure-state dynamics \cite{li_quantum_2018,skinner2019measurement,li_measurement-driven_2019,choi_quantum_2020,bao_theory_2020,jian_measurement-induced_2020}, wherein measurements disentangle system qubits from each other and decrease the internal entanglement of the system. Here, in the mixed-state dynamics, measurements can play an additional role by curtailing decoherence. This occurs both by disentangling system qudits from the bath, allowing them to reentangle with each other, as well as by diminishing long-range entanglement structures with which the boundary dephasing channels could decohere the bulk. 
Remarkably, while measurements cannot protect the full volume-law entanglement from decoherence, the interplay between dephasing and measurements has revealed the ``critical" $L^{1/3}$ scaling of entanglement that was previously hidden as a subleading contribution in the pure-state dynamics \cite{li_entanglement_2021}.

\begin{figure}[t!]
    \centering
    \includegraphics[width=0.7\linewidth]{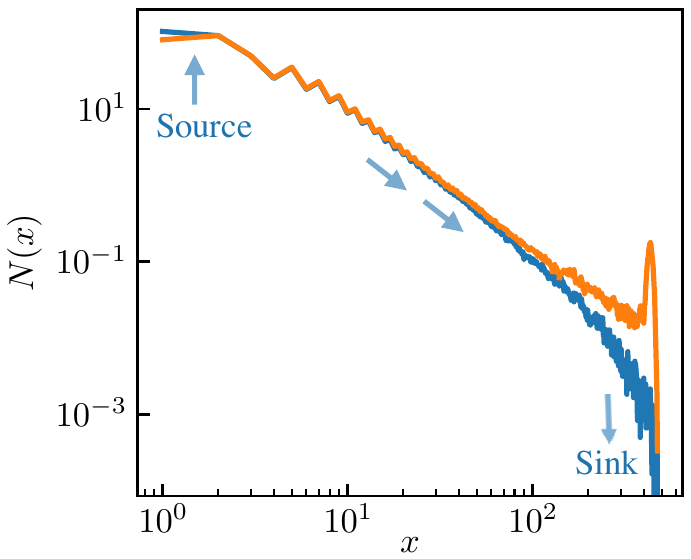}
    \caption{Late-time stabilizer length distribution in log-log scale. Here, we consider a circuit with decoherence at two random sites rather than two edges in each time step. Blue and red curve represent the results with and without dephasing baths at the boundary, respectively. The numerical simulation is performed in the circuits of size $L = 480$ and with measurement rate $p = 0.1$. The results are averaged over $200$ random circuit realizations.}
    \label{fig:3}
\end{figure}

In random Clifford circuits with strictly pure-state dynamics, the distribution of stabilizer lengths has previously offered insight on the bipartite entanglement entropy \cite{li_measurement-driven_2019}. It is therefore interesting to see how the stabilizer length distribution is modified in the presence of boundary dephasing channels (see Fig.~\ref{fig:3}). Although the length distribution does not directly determine the negativity in a mixed state, it can be used to compute the mutual information $I_{A:B} = S_A + S_B - S_{AB}$ \cite{nielsen_quantum_2010}, which shows qualitatively similar behavior to the negativity \cite{SOM} despite failing as a mixed-state entanglement measure. On one hand, we see that dephasing channels act as a stabilizer ``sink'' by destroying the buildup at lengths $x=L/2$, which was previously responsible for the volume-law contribution to the mutual information with no dephasing \cite{li_measurement-driven_2019}. On the other hand, measurements can act as a ``source,'' both by creating new short stabilizers, and by preventing stabilizers from becoming so long that they reach the system boundaries and dephase. This is evident in the power-law ramp, which is robust to dephasing and would be absent without measurements, and is responsible for the power-law scaling of $I_{A:B}$. The resulting steady-state dynamics of the stabilizer length distribution is reminiscent of energy transfer under turbulent cascade, and it is tempting to develop an effective classical model for the stabilizer dynamics to capture the power-law length distribution.

It is also interesting to consider how the negativity is affected by replacing the Markovian quantum channels with explicit bath qudits. For bath sizes smaller than the system, it is expected by Page's theorem that the negativity within the system can retain volume-law scaling in the absence of measurements. However, since the continuous monitoring of the system reduces the effective number of qudits participating in the system-bath entanglement dynamics, the negativity can undergo a first order Page-like transition within the volume-law entropy phase of the monitored circuit. The details of this Page-like negativity transition will be left for future work~\cite{Weinstein_Pagelike}.

In our analysis, it was crucial that decoherence occurred only at the boundary.
%
Instead, bulk decoherence will manifest as a symmetry-breaking field in the statistical mechanics model reducing $(S_{nk+1} \times S_{nk+1})\rtimes \mathbb{Z}_2$ down to a residual $S_{nk+1} \times \mathbb{Z}_2$ symmetry~\cite{bao2021symmetry}.
The decoherence pins the spins to the state $\mathbb{I}$, which is symmetric under the residual symmetry, resulting in a maximally mixed state in the circuit.
To establish a large-scale entanglement negativity in the presence of bulk decoherence, one needs to spontaneously break the residual $\mathbb{Z}_2$ Hermiticity symmetry.
%
%
One possibility is to introduce additional nonunitary elements, such as active feedback.
%
%
A designed feedback process using the knowledge of measurement results might possibly create preference of $\mathbb{C}$ and $\bar{\mathbb{C}}$ over $\mathbb{I}$, leading to a residual $\mathbb{Z}_2$ symmetry-breaking state with large-scale entanglement.

More broadly, we expect our results to be relevant beyond random circuit models to more realistic Hamiltonian dynamics. Given the significant recent interest in open-system quantum dynamics, it is interesting to consider whether the unique interplay between measurements and decoherence exhibited here can lead to new phases of nonequilibrium dynamics in settings accessible to modern experiments.

\begin{acknowledgements}
\textit{Acknowledgements.---} We thank Soonwon Choi, Zala Lenar\v{c}i\v{c}, and Yaodong Li for useful discussions. This work was supported in part by the NSF QLCI program through grant no. OMA-2016245. Z.W. is supported by the Berkeley Connect fellowship.
\end{acknowledgements}

\bibliographystyle{apsrev4-2}
\bibliography{refs}

\begin{thebibliography}{64}%
\makeatletter
\providecommand \@ifxundefined [1]{%
 \@ifx{#1\undefined}
}%
\providecommand \@ifnum [1]{%
 \ifnum #1\expandafter \@firstoftwo
 \else \expandafter \@secondoftwo
 \fi
}%
\providecommand \@ifx [1]{%
 \ifx #1\expandafter \@firstoftwo
 \else \expandafter \@secondoftwo
 \fi
}%
\providecommand \natexlab [1]{#1}%
\providecommand \enquote  [1]{``#1''}%
\providecommand \bibnamefont  [1]{#1}%
\providecommand \bibfnamefont [1]{#1}%
\providecommand \citenamefont [1]{#1}%
\providecommand \href@noop [0]{\@secondoftwo}%
\providecommand \href [0]{\begingroup \@sanitize@url \@href}%
\providecommand \@href[1]{\@@startlink{#1}\@@href}%
\providecommand \@@href[1]{\endgroup#1\@@endlink}%
\providecommand \@sanitize@url [0]{\catcode `\\12\catcode `\$12\catcode
  `\&12\catcode `\#12\catcode `\^12\catcode `\_12\catcode `\%12\relax}%
\providecommand \@@startlink[1]{}%
\providecommand \@@endlink[0]{}%
\providecommand \url  [0]{\begingroup\@sanitize@url \@url }%
\providecommand \@url [1]{\endgroup\@href {#1}{\urlprefix }}%
\providecommand \urlprefix  [0]{URL }%
\providecommand \Eprint [0]{\href }%
\providecommand \doibase [0]{https://doi.org/}%
\providecommand \selectlanguage [0]{\@gobble}%
\providecommand \bibinfo  [0]{\@secondoftwo}%
\providecommand \bibfield  [0]{\@secondoftwo}%
\providecommand \translation [1]{[#1]}%
\providecommand \BibitemOpen [0]{}%
\providecommand \bibitemStop [0]{}%
\providecommand \bibitemNoStop [0]{.\EOS\space}%
\providecommand \EOS [0]{\spacefactor3000\relax}%
\providecommand \BibitemShut  [1]{\csname bibitem#1\endcsname}%
\let\auto@bib@innerbib\@empty
\bibitem [{\citenamefont {Kim}\ and\ \citenamefont
  {Huse}(2013)}]{kim2013ballistic}%
  \BibitemOpen
  \bibfield  {author} {\bibinfo {author} {\bibfnamefont {H.}~\bibnamefont
  {Kim}}\ and\ \bibinfo {author} {\bibfnamefont {D.~A.}\ \bibnamefont {Huse}},\
  }\href@noop {} {\bibfield  {journal} {\bibinfo  {journal} {Physical review
  letters}\ }\textbf {\bibinfo {volume} {111}},\ \bibinfo {pages} {127205}
  (\bibinfo {year} {2013})}\BibitemShut {NoStop}%
\bibitem [{\citenamefont {Kaufman}\ \emph {et~al.}(2016)\citenamefont
  {Kaufman}, \citenamefont {Tai}, \citenamefont {Lukin}, \citenamefont
  {Rispoli}, \citenamefont {Schittko}, \citenamefont {Preiss},\ and\
  \citenamefont {Greiner}}]{kaufman2016quantum}%
  \BibitemOpen
  \bibfield  {author} {\bibinfo {author} {\bibfnamefont {A.~M.}\ \bibnamefont
  {Kaufman}}, \bibinfo {author} {\bibfnamefont {M.~E.}\ \bibnamefont {Tai}},
  \bibinfo {author} {\bibfnamefont {A.}~\bibnamefont {Lukin}}, \bibinfo
  {author} {\bibfnamefont {M.}~\bibnamefont {Rispoli}}, \bibinfo {author}
  {\bibfnamefont {R.}~\bibnamefont {Schittko}}, \bibinfo {author}
  {\bibfnamefont {P.~M.}\ \bibnamefont {Preiss}},\ and\ \bibinfo {author}
  {\bibfnamefont {M.}~\bibnamefont {Greiner}},\ }\href@noop {} {\bibfield
  {journal} {\bibinfo  {journal} {Science}\ }\textbf {\bibinfo {volume}
  {353}},\ \bibinfo {pages} {794} (\bibinfo {year} {2016})}\BibitemShut
  {NoStop}%
\bibitem [{\citenamefont {Nahum}\ \emph {et~al.}(2017)\citenamefont {Nahum},
  \citenamefont {Ruhman}, \citenamefont {Vijay},\ and\ \citenamefont
  {Haah}}]{nahum_quantum_2017}%
  \BibitemOpen
  \bibfield  {author} {\bibinfo {author} {\bibfnamefont {A.}~\bibnamefont
  {Nahum}}, \bibinfo {author} {\bibfnamefont {J.}~\bibnamefont {Ruhman}},
  \bibinfo {author} {\bibfnamefont {S.}~\bibnamefont {Vijay}},\ and\ \bibinfo
  {author} {\bibfnamefont {J.}~\bibnamefont {Haah}},\ }\href
  {https://doi.org/10.1103/PhysRevX.7.031016} {\bibfield  {journal} {\bibinfo
  {journal} {Physical Review X}\ }\textbf {\bibinfo {volume} {7}},\ \bibinfo
  {pages} {031016} (\bibinfo {year} {2017})}\BibitemShut {NoStop}%
\bibitem [{\citenamefont {Li}\ \emph {et~al.}(2018)\citenamefont {Li},
  \citenamefont {Chen},\ and\ \citenamefont {Fisher}}]{li_quantum_2018}%
  \BibitemOpen
  \bibfield  {author} {\bibinfo {author} {\bibfnamefont {Y.}~\bibnamefont
  {Li}}, \bibinfo {author} {\bibfnamefont {X.}~\bibnamefont {Chen}},\ and\
  \bibinfo {author} {\bibfnamefont {M.~P.~A.}\ \bibnamefont {Fisher}},\ }\href
  {https://doi.org/10.1103/PhysRevB.98.205136} {\bibfield  {journal} {\bibinfo
  {journal} {Physical Review B}\ }\textbf {\bibinfo {volume} {98}},\ \bibinfo
  {pages} {205136} (\bibinfo {year} {2018})}\BibitemShut {NoStop}%
\bibitem [{\citenamefont {Skinner}\ \emph {et~al.}(2019)\citenamefont
  {Skinner}, \citenamefont {Ruhman},\ and\ \citenamefont
  {Nahum}}]{skinner2019measurement}%
  \BibitemOpen
  \bibfield  {author} {\bibinfo {author} {\bibfnamefont {B.}~\bibnamefont
  {Skinner}}, \bibinfo {author} {\bibfnamefont {J.}~\bibnamefont {Ruhman}},\
  and\ \bibinfo {author} {\bibfnamefont {A.}~\bibnamefont {Nahum}},\
  }\href@noop {} {\bibfield  {journal} {\bibinfo  {journal} {Physical Review
  X}\ }\textbf {\bibinfo {volume} {9}},\ \bibinfo {pages} {031009} (\bibinfo
  {year} {2019})}\BibitemShut {NoStop}%
\bibitem [{\citenamefont {Arute}\ \emph {et~al.}(2019)\citenamefont {Arute},
  \citenamefont {Arya}, \citenamefont {Babbush}, \citenamefont {Bacon},
  \citenamefont {Bardin}, \citenamefont {Barends}, \citenamefont {Biswas},
  \citenamefont {Boixo}, \citenamefont {Brandao}, \citenamefont {Buell} \emph
  {et~al.}}]{arute2019quantum}%
  \BibitemOpen
  \bibfield  {author} {\bibinfo {author} {\bibfnamefont {F.}~\bibnamefont
  {Arute}}, \bibinfo {author} {\bibfnamefont {K.}~\bibnamefont {Arya}},
  \bibinfo {author} {\bibfnamefont {R.}~\bibnamefont {Babbush}}, \bibinfo
  {author} {\bibfnamefont {D.}~\bibnamefont {Bacon}}, \bibinfo {author}
  {\bibfnamefont {J.~C.}\ \bibnamefont {Bardin}}, \bibinfo {author}
  {\bibfnamefont {R.}~\bibnamefont {Barends}}, \bibinfo {author} {\bibfnamefont
  {R.}~\bibnamefont {Biswas}}, \bibinfo {author} {\bibfnamefont
  {S.}~\bibnamefont {Boixo}}, \bibinfo {author} {\bibfnamefont {F.~G.}\
  \bibnamefont {Brandao}}, \bibinfo {author} {\bibfnamefont {D.~A.}\
  \bibnamefont {Buell}}, \emph {et~al.},\ }\href@noop {} {\bibfield  {journal}
  {\bibinfo  {journal} {Nature}\ }\textbf {\bibinfo {volume} {574}},\ \bibinfo
  {pages} {505} (\bibinfo {year} {2019})}\BibitemShut {NoStop}%
\bibitem [{\citenamefont {Choi}\ \emph {et~al.}(2021)\citenamefont {Choi},
  \citenamefont {Shaw}, \citenamefont {Madjarov}, \citenamefont {Xie},
  \citenamefont {Covey}, \citenamefont {Cotler}, \citenamefont {Mark},
  \citenamefont {Huang}, \citenamefont {Kale}, \citenamefont {Pichler} \emph
  {et~al.}}]{choi2021emergent}%
  \BibitemOpen
  \bibfield  {author} {\bibinfo {author} {\bibfnamefont {J.}~\bibnamefont
  {Choi}}, \bibinfo {author} {\bibfnamefont {A.~L.}\ \bibnamefont {Shaw}},
  \bibinfo {author} {\bibfnamefont {I.~S.}\ \bibnamefont {Madjarov}}, \bibinfo
  {author} {\bibfnamefont {X.}~\bibnamefont {Xie}}, \bibinfo {author}
  {\bibfnamefont {J.~P.}\ \bibnamefont {Covey}}, \bibinfo {author}
  {\bibfnamefont {J.~S.}\ \bibnamefont {Cotler}}, \bibinfo {author}
  {\bibfnamefont {D.~K.}\ \bibnamefont {Mark}}, \bibinfo {author}
  {\bibfnamefont {H.-Y.}\ \bibnamefont {Huang}}, \bibinfo {author}
  {\bibfnamefont {A.}~\bibnamefont {Kale}}, \bibinfo {author} {\bibfnamefont
  {H.}~\bibnamefont {Pichler}}, \emph {et~al.},\ }\href@noop {} {\bibfield
  {journal} {\bibinfo  {journal} {arXiv preprint arXiv:2103.03535}\ } (\bibinfo
  {year} {2021})}\BibitemShut {NoStop}%
\bibitem [{\citenamefont {Noel}\ \emph {et~al.}(2021)\citenamefont {Noel},
  \citenamefont {Niroula}, \citenamefont {Risinger}, \citenamefont {Egan},
  \citenamefont {Biswas}, \citenamefont {Cetina}, \citenamefont {Gorshkov},
  \citenamefont {Gullans}, \citenamefont {Huse},\ and\ \citenamefont
  {Monroe}}]{noel2021observation}%
  \BibitemOpen
  \bibfield  {author} {\bibinfo {author} {\bibfnamefont {C.}~\bibnamefont
  {Noel}}, \bibinfo {author} {\bibfnamefont {P.}~\bibnamefont {Niroula}},
  \bibinfo {author} {\bibfnamefont {A.}~\bibnamefont {Risinger}}, \bibinfo
  {author} {\bibfnamefont {L.}~\bibnamefont {Egan}}, \bibinfo {author}
  {\bibfnamefont {D.}~\bibnamefont {Biswas}}, \bibinfo {author} {\bibfnamefont
  {M.}~\bibnamefont {Cetina}}, \bibinfo {author} {\bibfnamefont {A.~V.}\
  \bibnamefont {Gorshkov}}, \bibinfo {author} {\bibfnamefont {M.}~\bibnamefont
  {Gullans}}, \bibinfo {author} {\bibfnamefont {D.~A.}\ \bibnamefont {Huse}},\
  and\ \bibinfo {author} {\bibfnamefont {C.}~\bibnamefont {Monroe}},\
  }\href@noop {} {\bibfield  {journal} {\bibinfo  {journal} {arXiv preprint
  arXiv:2106.05881}\ } (\bibinfo {year} {2021})}\BibitemShut {NoStop}%
\bibitem [{\citenamefont {Ippoliti}\ \emph {et~al.}(2020)\citenamefont
  {Ippoliti}, \citenamefont {Gullans}, \citenamefont {Gopalakrishnan},
  \citenamefont {Huse},\ and\ \citenamefont
  {Khemani}}]{ippoliti_entanglement_2020}%
  \BibitemOpen
  \bibfield  {author} {\bibinfo {author} {\bibfnamefont {M.}~\bibnamefont
  {Ippoliti}}, \bibinfo {author} {\bibfnamefont {M.~J.}\ \bibnamefont
  {Gullans}}, \bibinfo {author} {\bibfnamefont {S.}~\bibnamefont
  {Gopalakrishnan}}, \bibinfo {author} {\bibfnamefont {D.~A.}\ \bibnamefont
  {Huse}},\ and\ \bibinfo {author} {\bibfnamefont {V.}~\bibnamefont
  {Khemani}},\ }\href {http://arxiv.org/abs/2004.09560} {\bibfield  {journal}
  {\bibinfo  {journal} {arXiv:2004.09560 [cond-mat, physics:quant-ph]}\ }
  (\bibinfo {year} {2020})},\ \bibinfo {note} {arXiv: 2004.09560}\BibitemShut
  {NoStop}%
\bibitem [{\citenamefont {Li}\ \emph {et~al.}(2019)\citenamefont {Li},
  \citenamefont {Chen},\ and\ \citenamefont
  {Fisher}}]{li_measurement-driven_2019}%
  \BibitemOpen
  \bibfield  {author} {\bibinfo {author} {\bibfnamefont {Y.}~\bibnamefont
  {Li}}, \bibinfo {author} {\bibfnamefont {X.}~\bibnamefont {Chen}},\ and\
  \bibinfo {author} {\bibfnamefont {M.~P.~A.}\ \bibnamefont {Fisher}},\ }\href
  {https://doi.org/10.1103/PhysRevB.100.134306} {\bibfield  {journal} {\bibinfo
   {journal} {Physical Review B}\ }\textbf {\bibinfo {volume} {100}},\ \bibinfo
  {pages} {134306} (\bibinfo {year} {2019})}\BibitemShut {NoStop}%
\bibitem [{\citenamefont {Choi}\ \emph {et~al.}(2020)\citenamefont {Choi},
  \citenamefont {Bao}, \citenamefont {Qi},\ and\ \citenamefont
  {Altman}}]{choi_quantum_2020}%
  \BibitemOpen
  \bibfield  {author} {\bibinfo {author} {\bibfnamefont {S.}~\bibnamefont
  {Choi}}, \bibinfo {author} {\bibfnamefont {Y.}~\bibnamefont {Bao}}, \bibinfo
  {author} {\bibfnamefont {X.-L.}\ \bibnamefont {Qi}},\ and\ \bibinfo {author}
  {\bibfnamefont {E.}~\bibnamefont {Altman}},\ }\href
  {https://doi.org/10.1103/PhysRevLett.125.030505} {\bibfield  {journal}
  {\bibinfo  {journal} {Physical Review Letters}\ }\textbf {\bibinfo {volume}
  {125}},\ \bibinfo {pages} {030505} (\bibinfo {year} {2020})}\BibitemShut
  {NoStop}%
\bibitem [{\citenamefont {Gullans}\ and\ \citenamefont
  {Huse}(2020)}]{gullans2020dynamical}%
  \BibitemOpen
  \bibfield  {author} {\bibinfo {author} {\bibfnamefont {M.~J.}\ \bibnamefont
  {Gullans}}\ and\ \bibinfo {author} {\bibfnamefont {D.~A.}\ \bibnamefont
  {Huse}},\ }\href@noop {} {\bibfield  {journal} {\bibinfo  {journal} {Physical
  Review X}\ }\textbf {\bibinfo {volume} {10}},\ \bibinfo {pages} {041020}
  (\bibinfo {year} {2020})}\BibitemShut {NoStop}%
\bibitem [{\citenamefont {Ippoliti}\ and\ \citenamefont
  {Khemani}(2020)}]{ippoliti_postselection-free_2020}%
  \BibitemOpen
  \bibfield  {author} {\bibinfo {author} {\bibfnamefont {M.}~\bibnamefont
  {Ippoliti}}\ and\ \bibinfo {author} {\bibfnamefont {V.}~\bibnamefont
  {Khemani}},\ }\href {http://arxiv.org/abs/2010.15840} {\bibfield  {journal}
  {\bibinfo  {journal} {arXiv:2010.15840 [cond-mat, physics:quant-ph]}\ }
  (\bibinfo {year} {2020})},\ \bibinfo {note} {arXiv: 2010.15840}\BibitemShut
  {NoStop}%
\bibitem [{\citenamefont {Ippoliti}\ \emph {et~al.}(2021)\citenamefont
  {Ippoliti}, \citenamefont {Rakovszky},\ and\ \citenamefont
  {Khemani}}]{ippoliti_fractal_2021}%
  \BibitemOpen
  \bibfield  {author} {\bibinfo {author} {\bibfnamefont {M.}~\bibnamefont
  {Ippoliti}}, \bibinfo {author} {\bibfnamefont {T.}~\bibnamefont
  {Rakovszky}},\ and\ \bibinfo {author} {\bibfnamefont {V.}~\bibnamefont
  {Khemani}},\ }\href {http://arxiv.org/abs/2103.06873} {\bibfield  {journal}
  {\bibinfo  {journal} {arXiv:2103.06873 [cond-mat, physics:hep-th,
  physics:quant-ph]}\ } (\bibinfo {year} {2021})},\ \bibinfo {note} {arXiv:
  2103.06873}\BibitemShut {NoStop}%
\bibitem [{\citenamefont {Lu}\ and\ \citenamefont
  {Grover}(2021)}]{lu_spacetime_2021}%
  \BibitemOpen
  \bibfield  {author} {\bibinfo {author} {\bibfnamefont {T.-C.}\ \bibnamefont
  {Lu}}\ and\ \bibinfo {author} {\bibfnamefont {T.}~\bibnamefont {Grover}},\
  }\href {https://doi.org/10.1103/PRXQuantum.2.040319} {\bibfield  {journal}
  {\bibinfo  {journal} {PRX Quantum}\ }\textbf {\bibinfo {volume} {2}},\
  \bibinfo {pages} {040319} (\bibinfo {year} {2021})},\ \bibinfo {note} {arXiv:
  2103.06356 version: 3}\BibitemShut {NoStop}%
\bibitem [{\citenamefont {Bao}\ \emph {et~al.}(2020)\citenamefont {Bao},
  \citenamefont {Choi},\ and\ \citenamefont {Altman}}]{bao_theory_2020}%
  \BibitemOpen
  \bibfield  {author} {\bibinfo {author} {\bibfnamefont {Y.}~\bibnamefont
  {Bao}}, \bibinfo {author} {\bibfnamefont {S.}~\bibnamefont {Choi}},\ and\
  \bibinfo {author} {\bibfnamefont {E.}~\bibnamefont {Altman}},\ }\href
  {https://doi.org/10.1103/PhysRevB.101.104301} {\bibfield  {journal} {\bibinfo
   {journal} {Physical Review B}\ }\textbf {\bibinfo {volume} {101}},\ \bibinfo
  {pages} {104301} (\bibinfo {year} {2020})}\BibitemShut {NoStop}%
\bibitem [{\citenamefont {Vidal}\ and\ \citenamefont
  {Werner}(2002)}]{vidal_computable_2002}%
  \BibitemOpen
  \bibfield  {author} {\bibinfo {author} {\bibfnamefont {G.}~\bibnamefont
  {Vidal}}\ and\ \bibinfo {author} {\bibfnamefont {R.~F.}\ \bibnamefont
  {Werner}},\ }\href {https://doi.org/10.1103/PhysRevA.65.032314} {\bibfield
  {journal} {\bibinfo  {journal} {Physical Review A}\ }\textbf {\bibinfo
  {volume} {65}},\ \bibinfo {pages} {032314} (\bibinfo {year}
  {2002})}\BibitemShut {NoStop}%
\bibitem [{\citenamefont {Plenio}(2005)}]{plenio_logarithmic_2005}%
  \BibitemOpen
  \bibfield  {author} {\bibinfo {author} {\bibfnamefont {M.~B.}\ \bibnamefont
  {Plenio}},\ }\href {https://doi.org/10.1103/PhysRevLett.95.090503} {\bibfield
   {journal} {\bibinfo  {journal} {Physical Review Letters}\ }\textbf {\bibinfo
  {volume} {95}},\ \bibinfo {pages} {090503} (\bibinfo {year}
  {2005})}\BibitemShut {NoStop}%
\bibitem [{\citenamefont {Calabrese}\ \emph {et~al.}(2012)\citenamefont
  {Calabrese}, \citenamefont {Cardy},\ and\ \citenamefont
  {Tonni}}]{calabrese_entanglement_2012}%
  \BibitemOpen
  \bibfield  {author} {\bibinfo {author} {\bibfnamefont {P.}~\bibnamefont
  {Calabrese}}, \bibinfo {author} {\bibfnamefont {J.}~\bibnamefont {Cardy}},\
  and\ \bibinfo {author} {\bibfnamefont {E.}~\bibnamefont {Tonni}},\ }\href
  {https://doi.org/10.1103/PhysRevLett.109.130502} {\bibfield  {journal}
  {\bibinfo  {journal} {Physical Review Letters}\ }\textbf {\bibinfo {volume}
  {109}},\ \bibinfo {pages} {130502} (\bibinfo {year} {2012})}\BibitemShut
  {NoStop}%
\bibitem [{\citenamefont {Calabrese}\ \emph {et~al.}(2013)\citenamefont
  {Calabrese}, \citenamefont {Cardy},\ and\ \citenamefont
  {Tonni}}]{calabrese_entanglement_2013}%
  \BibitemOpen
  \bibfield  {author} {\bibinfo {author} {\bibfnamefont {P.}~\bibnamefont
  {Calabrese}}, \bibinfo {author} {\bibfnamefont {J.}~\bibnamefont {Cardy}},\
  and\ \bibinfo {author} {\bibfnamefont {E.}~\bibnamefont {Tonni}},\ }\href
  {https://doi.org/10.1088/1742-5468/2013/02/P02008} {\bibfield  {journal}
  {\bibinfo  {journal} {Journal of Statistical Mechanics: Theory and
  Experiment}\ }\textbf {\bibinfo {volume} {2013}},\ \bibinfo {pages} {P02008}
  (\bibinfo {year} {2013})}\BibitemShut {NoStop}%
\bibitem [{\citenamefont {Lu}\ and\ \citenamefont
  {Grover}(2019)}]{lu_singularity_2019}%
  \BibitemOpen
  \bibfield  {author} {\bibinfo {author} {\bibfnamefont {T.-C.}\ \bibnamefont
  {Lu}}\ and\ \bibinfo {author} {\bibfnamefont {T.}~\bibnamefont {Grover}},\
  }\href {https://doi.org/10.1103/PhysRevB.99.075157} {\bibfield  {journal}
  {\bibinfo  {journal} {Physical Review B}\ }\textbf {\bibinfo {volume} {99}},\
  \bibinfo {pages} {075157} (\bibinfo {year} {2019})}\BibitemShut {NoStop}%
\bibitem [{\citenamefont {Lu}\ \emph {et~al.}(2020)\citenamefont {Lu},
  \citenamefont {Hsieh},\ and\ \citenamefont {Grover}}]{lu_detecting_2020}%
  \BibitemOpen
  \bibfield  {author} {\bibinfo {author} {\bibfnamefont {T.-C.}\ \bibnamefont
  {Lu}}, \bibinfo {author} {\bibfnamefont {T.~H.}\ \bibnamefont {Hsieh}},\ and\
  \bibinfo {author} {\bibfnamefont {T.}~\bibnamefont {Grover}},\ }\href
  {https://doi.org/10.1103/PhysRevLett.125.116801} {\bibfield  {journal}
  {\bibinfo  {journal} {Physical Review Letters}\ }\textbf {\bibinfo {volume}
  {125}},\ \bibinfo {pages} {116801} (\bibinfo {year} {2020})}\BibitemShut
  {NoStop}%
\bibitem [{\citenamefont {Wu}\ \emph {et~al.}(2020)\citenamefont {Wu},
  \citenamefont {Lu}, \citenamefont {Chung}, \citenamefont {Kao},\ and\
  \citenamefont {Grover}}]{wu_entanglement_2020}%
  \BibitemOpen
  \bibfield  {author} {\bibinfo {author} {\bibfnamefont {K.-H.}\ \bibnamefont
  {Wu}}, \bibinfo {author} {\bibfnamefont {T.-C.}\ \bibnamefont {Lu}}, \bibinfo
  {author} {\bibfnamefont {C.-M.}\ \bibnamefont {Chung}}, \bibinfo {author}
  {\bibfnamefont {Y.-J.}\ \bibnamefont {Kao}},\ and\ \bibinfo {author}
  {\bibfnamefont {T.}~\bibnamefont {Grover}},\ }\href
  {https://doi.org/10.1103/PhysRevLett.125.140603} {\bibfield  {journal}
  {\bibinfo  {journal} {Physical Review Letters}\ }\textbf {\bibinfo {volume}
  {125}},\ \bibinfo {pages} {140603} (\bibinfo {year} {2020})}\BibitemShut
  {NoStop}%
\bibitem [{\citenamefont {Lu}\ and\ \citenamefont
  {Grover}(2020)}]{lu_entanglement_2020}%
  \BibitemOpen
  \bibfield  {author} {\bibinfo {author} {\bibfnamefont {T.-C.}\ \bibnamefont
  {Lu}}\ and\ \bibinfo {author} {\bibfnamefont {T.}~\bibnamefont {Grover}},\
  }\href {https://doi.org/10.1103/PhysRevB.102.235110} {\bibfield  {journal}
  {\bibinfo  {journal} {Physical Review B}\ }\textbf {\bibinfo {volume}
  {102}},\ \bibinfo {pages} {235110} (\bibinfo {year} {2020})}\BibitemShut
  {NoStop}%
\bibitem [{\citenamefont {Sang}\ \emph {et~al.}(2021)\citenamefont {Sang},
  \citenamefont {Li}, \citenamefont {Zhou}, \citenamefont {Chen}, \citenamefont
  {Hsieh},\ and\ \citenamefont {Fisher}}]{sang_entanglement_2020}%
  \BibitemOpen
  \bibfield  {author} {\bibinfo {author} {\bibfnamefont {S.}~\bibnamefont
  {Sang}}, \bibinfo {author} {\bibfnamefont {Y.}~\bibnamefont {Li}}, \bibinfo
  {author} {\bibfnamefont {T.}~\bibnamefont {Zhou}}, \bibinfo {author}
  {\bibfnamefont {X.}~\bibnamefont {Chen}}, \bibinfo {author} {\bibfnamefont
  {T.~H.}\ \bibnamefont {Hsieh}},\ and\ \bibinfo {author} {\bibfnamefont
  {M.~P.~A.}\ \bibnamefont {Fisher}},\ }\href
  {https://doi.org/10.1103/PRXQuantum.2.030313} {\bibfield  {journal} {\bibinfo
   {journal} {PRX Quantum}\ }\textbf {\bibinfo {volume} {2}},\ \bibinfo {pages}
  {030313} (\bibinfo {year} {2021})}\BibitemShut {NoStop}%
\bibitem [{\citenamefont {Shapourian}\ \emph {et~al.}(2021)\citenamefont
  {Shapourian}, \citenamefont {Liu}, \citenamefont {Kudler-Flam},\ and\
  \citenamefont {Vishwanath}}]{shapourian_entanglement_2020}%
  \BibitemOpen
  \bibfield  {author} {\bibinfo {author} {\bibfnamefont {H.}~\bibnamefont
  {Shapourian}}, \bibinfo {author} {\bibfnamefont {S.}~\bibnamefont {Liu}},
  \bibinfo {author} {\bibfnamefont {J.}~\bibnamefont {Kudler-Flam}},\ and\
  \bibinfo {author} {\bibfnamefont {A.}~\bibnamefont {Vishwanath}},\ }\href
  {https://doi.org/10.1103/PRXQuantum.2.030347} {\bibfield  {journal} {\bibinfo
   {journal} {PRX Quantum}\ }\textbf {\bibinfo {volume} {2}},\ \bibinfo {pages}
  {030347} (\bibinfo {year} {2021})}\BibitemShut {NoStop}%
\bibitem [{\citenamefont {Hayden}\ \emph {et~al.}(2016)\citenamefont {Hayden},
  \citenamefont {Nezami}, \citenamefont {Qi}, \citenamefont {Thomas},
  \citenamefont {Walter},\ and\ \citenamefont
  {Yang}}]{hayden_holographic_2016}%
  \BibitemOpen
  \bibfield  {author} {\bibinfo {author} {\bibfnamefont {P.}~\bibnamefont
  {Hayden}}, \bibinfo {author} {\bibfnamefont {S.}~\bibnamefont {Nezami}},
  \bibinfo {author} {\bibfnamefont {X.-L.}\ \bibnamefont {Qi}}, \bibinfo
  {author} {\bibfnamefont {N.}~\bibnamefont {Thomas}}, \bibinfo {author}
  {\bibfnamefont {M.}~\bibnamefont {Walter}},\ and\ \bibinfo {author}
  {\bibfnamefont {Z.}~\bibnamefont {Yang}},\ }\href
  {https://doi.org/10.1007/JHEP11(2016)009} {\bibfield  {journal} {\bibinfo
  {journal} {Journal of High Energy Physics}\ }\textbf {\bibinfo {volume}
  {2016}},\ \bibinfo {pages} {9} (\bibinfo {year} {2016})}\BibitemShut
  {NoStop}%
\bibitem [{\citenamefont {Zhou}\ and\ \citenamefont
  {Nahum}(2019)}]{zhou_emergent_2019}%
  \BibitemOpen
  \bibfield  {author} {\bibinfo {author} {\bibfnamefont {T.}~\bibnamefont
  {Zhou}}\ and\ \bibinfo {author} {\bibfnamefont {A.}~\bibnamefont {Nahum}},\
  }\href {https://doi.org/10.1103/PhysRevB.99.174205} {\bibfield  {journal}
  {\bibinfo  {journal} {Physical Review B}\ }\textbf {\bibinfo {volume} {99}},\
  \bibinfo {pages} {174205} (\bibinfo {year} {2019})}\BibitemShut {NoStop}%
\bibitem [{\citenamefont {Jian}\ \emph {et~al.}(2020)\citenamefont {Jian},
  \citenamefont {You}, \citenamefont {Vasseur},\ and\ \citenamefont
  {Ludwig}}]{jian_measurement-induced_2020}%
  \BibitemOpen
  \bibfield  {author} {\bibinfo {author} {\bibfnamefont {C.-M.}\ \bibnamefont
  {Jian}}, \bibinfo {author} {\bibfnamefont {Y.-Z.}\ \bibnamefont {You}},
  \bibinfo {author} {\bibfnamefont {R.}~\bibnamefont {Vasseur}},\ and\ \bibinfo
  {author} {\bibfnamefont {A.~W.~W.}\ \bibnamefont {Ludwig}},\ }\href
  {https://doi.org/10.1103/PhysRevB.101.104302} {\bibfield  {journal} {\bibinfo
   {journal} {Physical Review B}\ }\textbf {\bibinfo {volume} {101}},\ \bibinfo
  {pages} {104302} (\bibinfo {year} {2020})}\BibitemShut {NoStop}%
\bibitem [{\citenamefont {Dong}\ \emph {et~al.}(2021)\citenamefont {Dong},
  \citenamefont {Qi},\ and\ \citenamefont {Walter}}]{dong_holographic_2021}%
  \BibitemOpen
  \bibfield  {author} {\bibinfo {author} {\bibfnamefont {X.}~\bibnamefont
  {Dong}}, \bibinfo {author} {\bibfnamefont {X.-L.}\ \bibnamefont {Qi}},\ and\
  \bibinfo {author} {\bibfnamefont {M.}~\bibnamefont {Walter}},\ }\href
  {https://doi.org/10.1007/JHEP06(2021)024} {\bibfield  {journal} {\bibinfo
  {journal} {Journal of High Energy Physics}\ }\textbf {\bibinfo {volume}
  {2021}},\ \bibinfo {pages} {24} (\bibinfo {year} {2021})}\BibitemShut
  {NoStop}%
\bibitem [{\citenamefont {Li}\ \emph {et~al.}(2021)\citenamefont {Li},
  \citenamefont {Vijay},\ and\ \citenamefont {Fisher}}]{li_entanglement_2021}%
  \BibitemOpen
  \bibfield  {author} {\bibinfo {author} {\bibfnamefont {Y.}~\bibnamefont
  {Li}}, \bibinfo {author} {\bibfnamefont {S.}~\bibnamefont {Vijay}},\ and\
  \bibinfo {author} {\bibfnamefont {M.~P.~A.}\ \bibnamefont {Fisher}},\ }\href
  {http://arxiv.org/abs/2105.13352} {\bibfield  {journal} {\bibinfo  {journal}
  {arXiv:2105.13352 [cond-mat, physics:quant-ph]}\ } (\bibinfo {year}
  {2021})}\BibitemShut {NoStop}%
\bibitem [{\citenamefont {Kardar}(1985)}]{kardar_roughening_1985}%
  \BibitemOpen
  \bibfield  {author} {\bibinfo {author} {\bibfnamefont {M.}~\bibnamefont
  {Kardar}},\ }\href {https://doi.org/10.1103/PhysRevLett.55.2923} {\bibfield
  {journal} {\bibinfo  {journal} {Physical Review Letters}\ }\textbf {\bibinfo
  {volume} {55}},\ \bibinfo {pages} {2923} (\bibinfo {year}
  {1985})}\BibitemShut {NoStop}%
\bibitem [{\citenamefont {Huse}\ \emph {et~al.}(1985)\citenamefont {Huse},
  \citenamefont {Henley},\ and\ \citenamefont {Fisher}}]{huse_huse_1985}%
  \BibitemOpen
  \bibfield  {author} {\bibinfo {author} {\bibfnamefont {D.~A.}\ \bibnamefont
  {Huse}}, \bibinfo {author} {\bibfnamefont {C.~L.}\ \bibnamefont {Henley}},\
  and\ \bibinfo {author} {\bibfnamefont {D.~S.}\ \bibnamefont {Fisher}},\
  }\href {https://doi.org/10.1103/PhysRevLett.55.2924} {\bibfield  {journal}
  {\bibinfo  {journal} {Physical Review Letters}\ }\textbf {\bibinfo {volume}
  {55}},\ \bibinfo {pages} {2924} (\bibinfo {year} {1985})}\BibitemShut
  {NoStop}%
\bibitem [{\citenamefont {Kardar}\ \emph {et~al.}(1986)\citenamefont {Kardar},
  \citenamefont {Parisi},\ and\ \citenamefont {Zhang}}]{kardar_dynamic_1986}%
  \BibitemOpen
  \bibfield  {author} {\bibinfo {author} {\bibfnamefont {M.}~\bibnamefont
  {Kardar}}, \bibinfo {author} {\bibfnamefont {G.}~\bibnamefont {Parisi}},\
  and\ \bibinfo {author} {\bibfnamefont {Y.-C.}\ \bibnamefont {Zhang}},\ }\href
  {https://doi.org/10.1103/PhysRevLett.56.889} {\bibfield  {journal} {\bibinfo
  {journal} {Physical Review Letters}\ }\textbf {\bibinfo {volume} {56}},\
  \bibinfo {pages} {889} (\bibinfo {year} {1986})}\BibitemShut {NoStop}%
\bibitem [{\citenamefont {Kardar}(2007)}]{kardar_statistical_2007}%
  \BibitemOpen
  \bibfield  {author} {\bibinfo {author} {\bibfnamefont {M.}~\bibnamefont
  {Kardar}},\ }\href@noop {} {\emph {\bibinfo {title} {Statistical physics of
  fields}}}\ (\bibinfo  {publisher} {Cambridge University Press},\ \bibinfo
  {address} {Cambridge; New York},\ \bibinfo {year} {2007})\ \bibinfo {note}
  {oCLC: ocn123113789}\BibitemShut {NoStop}%
\bibitem [{\citenamefont {Agrawal}\ \emph {et~al.}(2021)\citenamefont
  {Agrawal}, \citenamefont {Zabalo}, \citenamefont {Chen}, \citenamefont
  {Wilson}, \citenamefont {Potter}, \citenamefont {Pixley}, \citenamefont
  {Gopalakrishnan},\ and\ \citenamefont {Vasseur}}]{agrawal_entanglement_2021}%
  \BibitemOpen
  \bibfield  {author} {\bibinfo {author} {\bibfnamefont {U.}~\bibnamefont
  {Agrawal}}, \bibinfo {author} {\bibfnamefont {A.}~\bibnamefont {Zabalo}},
  \bibinfo {author} {\bibfnamefont {K.}~\bibnamefont {Chen}}, \bibinfo {author}
  {\bibfnamefont {J.~H.}\ \bibnamefont {Wilson}}, \bibinfo {author}
  {\bibfnamefont {A.~C.}\ \bibnamefont {Potter}}, \bibinfo {author}
  {\bibfnamefont {J.~H.}\ \bibnamefont {Pixley}}, \bibinfo {author}
  {\bibfnamefont {S.}~\bibnamefont {Gopalakrishnan}},\ and\ \bibinfo {author}
  {\bibfnamefont {R.}~\bibnamefont {Vasseur}},\ }\href
  {http://arxiv.org/abs/2107.10279} {\bibfield  {journal} {\bibinfo  {journal}
  {arXiv:2107.10279 [cond-mat, physics:quant-ph]}\ } (\bibinfo {year}
  {2021})}\BibitemShut {NoStop}%
\bibitem [{\citenamefont {Nielsen}\ and\ \citenamefont
  {Chuang}(2010)}]{nielsen_quantum_2010}%
  \BibitemOpen
  \bibfield  {author} {\bibinfo {author} {\bibfnamefont {M.~A.}\ \bibnamefont
  {Nielsen}}\ and\ \bibinfo {author} {\bibfnamefont {I.~L.}\ \bibnamefont
  {Chuang}},\ }\href@noop {} {\emph {\bibinfo {title} {Quantum computation and
  quantum information}}},\ \bibinfo {edition} {10th}\ ed.\ (\bibinfo
  {publisher} {Cambridge University Press},\ \bibinfo {address} {Cambridge; New
  York},\ \bibinfo {year} {2010})\BibitemShut {NoStop}%
\bibitem [{\citenamefont {Lidar}(2020)}]{lidar_lecture_2020}%
  \BibitemOpen
  \bibfield  {author} {\bibinfo {author} {\bibfnamefont {D.~A.}\ \bibnamefont
  {Lidar}},\ }\href {http://arxiv.org/abs/1902.00967} {\bibfield  {journal}
  {\bibinfo  {journal} {arXiv:1902.00967 [quant-ph]}\ } (\bibinfo {year}
  {2020})}\BibitemShut {NoStop}%
\bibitem [{\citenamefont {Bennett}\ \emph {et~al.}(1996)\citenamefont
  {Bennett}, \citenamefont {DiVincenzo}, \citenamefont {Smolin},\ and\
  \citenamefont {Wootters}}]{bennett_mixed-state_1996}%
  \BibitemOpen
  \bibfield  {author} {\bibinfo {author} {\bibfnamefont {C.~H.}\ \bibnamefont
  {Bennett}}, \bibinfo {author} {\bibfnamefont {D.~P.}\ \bibnamefont
  {DiVincenzo}}, \bibinfo {author} {\bibfnamefont {J.~A.}\ \bibnamefont
  {Smolin}},\ and\ \bibinfo {author} {\bibfnamefont {W.~K.}\ \bibnamefont
  {Wootters}},\ }\href {https://doi.org/10.1103/PhysRevA.54.3824} {\bibfield
  {journal} {\bibinfo  {journal} {Physical Review A}\ }\textbf {\bibinfo
  {volume} {54}},\ \bibinfo {pages} {3824} (\bibinfo {year}
  {1996})}\BibitemShut {NoStop}%
\bibitem [{\citenamefont {Horodecki}\ \emph {et~al.}(1998)\citenamefont
  {Horodecki}, \citenamefont {Horodecki},\ and\ \citenamefont
  {Horodecki}}]{horodecki_mixed-state_1998}%
  \BibitemOpen
  \bibfield  {author} {\bibinfo {author} {\bibfnamefont {M.}~\bibnamefont
  {Horodecki}}, \bibinfo {author} {\bibfnamefont {P.}~\bibnamefont
  {Horodecki}},\ and\ \bibinfo {author} {\bibfnamefont {R.}~\bibnamefont
  {Horodecki}},\ }\href {https://doi.org/10.1103/PhysRevLett.80.5239}
  {\bibfield  {journal} {\bibinfo  {journal} {Physical Review Letters}\
  }\textbf {\bibinfo {volume} {80}},\ \bibinfo {pages} {5239} (\bibinfo {year}
  {1998})}\BibitemShut {NoStop}%
\bibitem [{\citenamefont {Peres}(1996)}]{peres_separability_1996}%
  \BibitemOpen
  \bibfield  {author} {\bibinfo {author} {\bibfnamefont {A.}~\bibnamefont
  {Peres}},\ }\href {https://doi.org/10.1103/PhysRevLett.77.1413} {\bibfield
  {journal} {\bibinfo  {journal} {Physical Review Letters}\ }\textbf {\bibinfo
  {volume} {77}},\ \bibinfo {pages} {1413} (\bibinfo {year}
  {1996})}\BibitemShut {NoStop}%
\bibitem [{\citenamefont {Horodecki}\ \emph {et~al.}(1996)\citenamefont
  {Horodecki}, \citenamefont {Horodecki},\ and\ \citenamefont
  {Horodecki}}]{horodecki_separability_1996}%
  \BibitemOpen
  \bibfield  {author} {\bibinfo {author} {\bibfnamefont {M.}~\bibnamefont
  {Horodecki}}, \bibinfo {author} {\bibfnamefont {P.}~\bibnamefont
  {Horodecki}},\ and\ \bibinfo {author} {\bibfnamefont {R.}~\bibnamefont
  {Horodecki}},\ }\href {https://doi.org/10.1016/S0375-9601(96)00706-2}
  {\bibfield  {journal} {\bibinfo  {journal} {Physics Letters A}\ }\textbf
  {\bibinfo {volume} {223}},\ \bibinfo {pages} {1} (\bibinfo {year}
  {1996})}\BibitemShut {NoStop}%
\bibitem [{\citenamefont {Vidal}\ and\ \citenamefont
  {Cirac}(2001)}]{vidal_irreversibility_2001}%
  \BibitemOpen
  \bibfield  {author} {\bibinfo {author} {\bibfnamefont {G.}~\bibnamefont
  {Vidal}}\ and\ \bibinfo {author} {\bibfnamefont {J.~I.}\ \bibnamefont
  {Cirac}},\ }\href {https://doi.org/10.1103/PhysRevLett.86.5803} {\bibfield
  {journal} {\bibinfo  {journal} {Physical Review Letters}\ }\textbf {\bibinfo
  {volume} {86}},\ \bibinfo {pages} {5803} (\bibinfo {year}
  {2001})}\BibitemShut {NoStop}%
\bibitem [{\citenamefont {Horodecki}\ \emph {et~al.}(2020)\citenamefont
  {Horodecki}, \citenamefont {Rudnicki},\ and\ \citenamefont
  {Zyczkowski}}]{horodecki_five_2020}%
  \BibitemOpen
  \bibfield  {author} {\bibinfo {author} {\bibfnamefont {P.}~\bibnamefont
  {Horodecki}}, \bibinfo {author} {\bibfnamefont {L.}~\bibnamefont
  {Rudnicki}},\ and\ \bibinfo {author} {\bibfnamefont {K.}~\bibnamefont
  {Zyczkowski}},\ }\href {http://arxiv.org/abs/2002.03233} {\bibfield
  {journal} {\bibinfo  {journal} {arXiv:2002.03233 [quant-ph]}\ } (\bibinfo
  {year} {2020})},\ \bibinfo {note} {arXiv: 2002.03233}\BibitemShut {NoStop}%
\bibitem [{\citenamefont {Gottesman}(1998)}]{gottesman_heisenberg_1998}%
  \BibitemOpen
  \bibfield  {author} {\bibinfo {author} {\bibfnamefont {D.}~\bibnamefont
  {Gottesman}},\ }\href {http://arxiv.org/abs/quant-ph/9807006} {\bibfield
  {journal} {\bibinfo  {journal} {arXiv:quant-ph/9807006}\ } (\bibinfo {year}
  {1998})},\ \bibinfo {note} {arXiv: quant-ph/9807006}\BibitemShut {NoStop}%
\bibitem [{\citenamefont {Aaronson}\ and\ \citenamefont
  {Gottesman}(2004)}]{aaronson_improved_2004}%
  \BibitemOpen
  \bibfield  {author} {\bibinfo {author} {\bibfnamefont {S.}~\bibnamefont
  {Aaronson}}\ and\ \bibinfo {author} {\bibfnamefont {D.}~\bibnamefont
  {Gottesman}},\ }\href {https://doi.org/10.1103/PhysRevA.70.052328} {\bibfield
   {journal} {\bibinfo  {journal} {Physical Review A}\ }\textbf {\bibinfo
  {volume} {70}},\ \bibinfo {pages} {052328} (\bibinfo {year}
  {2004})}\BibitemShut {NoStop}%
\bibitem [{\citenamefont {Hamma}\ \emph
  {et~al.}(2005{\natexlab{a}})\citenamefont {Hamma}, \citenamefont
  {Ionicioiu},\ and\ \citenamefont {Zanardi}}]{hamma2005ground}%
  \BibitemOpen
  \bibfield  {author} {\bibinfo {author} {\bibfnamefont {A.}~\bibnamefont
  {Hamma}}, \bibinfo {author} {\bibfnamefont {R.}~\bibnamefont {Ionicioiu}},\
  and\ \bibinfo {author} {\bibfnamefont {P.}~\bibnamefont {Zanardi}},\
  }\href@noop {} {\bibfield  {journal} {\bibinfo  {journal} {Physics Letters
  A}\ }\textbf {\bibinfo {volume} {337}},\ \bibinfo {pages} {22} (\bibinfo
  {year} {2005}{\natexlab{a}})}\BibitemShut {NoStop}%
\bibitem [{\citenamefont {Hamma}\ \emph
  {et~al.}(2005{\natexlab{b}})\citenamefont {Hamma}, \citenamefont
  {Ionicioiu},\ and\ \citenamefont {Zanardi}}]{hamma2005bipartite}%
  \BibitemOpen
  \bibfield  {author} {\bibinfo {author} {\bibfnamefont {A.}~\bibnamefont
  {Hamma}}, \bibinfo {author} {\bibfnamefont {R.}~\bibnamefont {Ionicioiu}},\
  and\ \bibinfo {author} {\bibfnamefont {P.}~\bibnamefont {Zanardi}},\
  }\href@noop {} {\bibfield  {journal} {\bibinfo  {journal} {Physical Review
  A}\ }\textbf {\bibinfo {volume} {71}},\ \bibinfo {pages} {022315} (\bibinfo
  {year} {2005}{\natexlab{b}})}\BibitemShut {NoStop}%
\bibitem [{SOM()}]{SOM}%
  \BibitemOpen
  \href@noop {} {}\bibinfo {note} {See supplementary online material for
  details.}\BibitemShut {Stop}%
\bibitem [{\citenamefont {Webb}(2016)}]{webb2015clifford}%
  \BibitemOpen
  \bibfield  {author} {\bibinfo {author} {\bibfnamefont {Z.}~\bibnamefont
  {Webb}},\ }\href {http://arxiv.org/abs/1510.02769} {\bibfield  {journal}
  {\bibinfo  {journal} {arXiv:1510.02769 [quant-ph]}\ } (\bibinfo {year}
  {2016})}\BibitemShut {NoStop}%
\bibitem [{\citenamefont {Page}(1993)}]{page_average_1993}%
  \BibitemOpen
  \bibfield  {author} {\bibinfo {author} {\bibfnamefont {D.~N.}\ \bibnamefont
  {Page}},\ }\href {https://doi.org/10.1103/PhysRevLett.71.1291} {\bibfield
  {journal} {\bibinfo  {journal} {Physical Review Letters}\ }\textbf {\bibinfo
  {volume} {71}},\ \bibinfo {pages} {1291} (\bibinfo {year}
  {1993})}\BibitemShut {NoStop}%
\bibitem [{\citenamefont {Bhosale}\ \emph {et~al.}(2012)\citenamefont
  {Bhosale}, \citenamefont {Tomsovic},\ and\ \citenamefont
  {Lakshminarayan}}]{bhosale_entanglement_2012}%
  \BibitemOpen
  \bibfield  {author} {\bibinfo {author} {\bibfnamefont {U.~T.}\ \bibnamefont
  {Bhosale}}, \bibinfo {author} {\bibfnamefont {S.}~\bibnamefont {Tomsovic}},\
  and\ \bibinfo {author} {\bibfnamefont {A.}~\bibnamefont {Lakshminarayan}},\
  }\href {https://doi.org/10.1103/PhysRevA.85.062331} {\bibfield  {journal}
  {\bibinfo  {journal} {Physical Review A}\ }\textbf {\bibinfo {volume} {85}},\
  \bibinfo {pages} {062331} (\bibinfo {year} {2012})}\BibitemShut {NoStop}%
\bibitem [{\citenamefont {Zabalo}\ \emph {et~al.}(2020)\citenamefont {Zabalo},
  \citenamefont {Gullans}, \citenamefont {Wilson}, \citenamefont
  {Gopalakrishnan}, \citenamefont {Huse},\ and\ \citenamefont
  {Pixley}}]{zabalo2020critical}%
  \BibitemOpen
  \bibfield  {author} {\bibinfo {author} {\bibfnamefont {A.}~\bibnamefont
  {Zabalo}}, \bibinfo {author} {\bibfnamefont {M.~J.}\ \bibnamefont {Gullans}},
  \bibinfo {author} {\bibfnamefont {J.~H.}\ \bibnamefont {Wilson}}, \bibinfo
  {author} {\bibfnamefont {S.}~\bibnamefont {Gopalakrishnan}}, \bibinfo
  {author} {\bibfnamefont {D.~A.}\ \bibnamefont {Huse}},\ and\ \bibinfo
  {author} {\bibfnamefont {J.~H.}\ \bibnamefont {Pixley}},\ }\href@noop {}
  {\bibfield  {journal} {\bibinfo  {journal} {Physical Review B}\ }\textbf
  {\bibinfo {volume} {101}},\ \bibinfo {pages} {060301(R)} (\bibinfo {year}
  {2020})}\BibitemShut {NoStop}%
\bibitem [{\citenamefont {Nishimori}(2001)}]{nishimori_statistical_2001}%
  \BibitemOpen
  \bibfield  {author} {\bibinfo {author} {\bibfnamefont {H.}~\bibnamefont
  {Nishimori}},\ }\href@noop {} {\emph {\bibinfo {title} {Statistical physics
  of spin glasses and information processing: an introduction}}},\ \bibinfo
  {series} {International series of monographs on physics}\ No.\ \bibinfo
  {number} {111}\ (\bibinfo  {publisher} {Oxford University Press},\ \bibinfo
  {address} {Oxford; New York},\ \bibinfo {year} {2001})\ \bibinfo {note}
  {oCLC: ocm47063323}\BibitemShut {NoStop}%
\bibitem [{\citenamefont {Collins}(2003)}]{collins_moments_2003}%
  \BibitemOpen
  \bibfield  {author} {\bibinfo {author} {\bibfnamefont {B.}~\bibnamefont
  {Collins}},\ }\href {https://doi.org/10.1155/S107379280320917X} {\bibfield
  {journal} {\bibinfo  {journal} {International Mathematics Research Notices}\
  }\textbf {\bibinfo {volume} {2003}},\ \bibinfo {pages} {953} (\bibinfo {year}
  {2003})},\ \Eprint
  {https://arxiv.org/abs/https://academic.oup.com/imrn/article-pdf/2003/17/953/1881428/2003-17-953.pdf}
  {https://academic.oup.com/imrn/article-pdf/2003/17/953/1881428/2003-17-953.pdf}
  \BibitemShut {NoStop}%
\bibitem [{\citenamefont {Collins}\ and\ \citenamefont
  {{\'S}niady}(2006)}]{collins2006integration}%
  \BibitemOpen
  \bibfield  {author} {\bibinfo {author} {\bibfnamefont {B.}~\bibnamefont
  {Collins}}\ and\ \bibinfo {author} {\bibfnamefont {P.}~\bibnamefont
  {{\'S}niady}},\ }\href@noop {} {\bibfield  {journal} {\bibinfo  {journal}
  {Communications in Mathematical Physics}\ }\textbf {\bibinfo {volume}
  {264}},\ \bibinfo {pages} {773} (\bibinfo {year} {2006})}\BibitemShut
  {NoStop}%
\bibitem [{\citenamefont {Nahum}\ \emph {et~al.}(2018)\citenamefont {Nahum},
  \citenamefont {Vijay},\ and\ \citenamefont {Haah}}]{nahum_operator_2018}%
  \BibitemOpen
  \bibfield  {author} {\bibinfo {author} {\bibfnamefont {A.}~\bibnamefont
  {Nahum}}, \bibinfo {author} {\bibfnamefont {S.}~\bibnamefont {Vijay}},\ and\
  \bibinfo {author} {\bibfnamefont {J.}~\bibnamefont {Haah}},\ }\href
  {https://doi.org/10.1103/PhysRevX.8.021014} {\bibfield  {journal} {\bibinfo
  {journal} {Physical Review X}\ }\textbf {\bibinfo {volume} {8}},\ \bibinfo
  {pages} {021014} (\bibinfo {year} {2018})}\BibitemShut {NoStop}%
\bibitem [{\citenamefont {You}\ and\ \citenamefont
  {Gu}(2018)}]{you2018entanglement}%
  \BibitemOpen
  \bibfield  {author} {\bibinfo {author} {\bibfnamefont {Y.-Z.}\ \bibnamefont
  {You}}\ and\ \bibinfo {author} {\bibfnamefont {Y.}~\bibnamefont {Gu}},\
  }\href@noop {} {\bibfield  {journal} {\bibinfo  {journal} {Physical Review
  B}\ }\textbf {\bibinfo {volume} {98}},\ \bibinfo {pages} {014309} (\bibinfo
  {year} {2018})}\BibitemShut {NoStop}%
\bibitem [{Note1()}]{Note1}%
  \BibitemOpen
  \bibinfo {note} {In the zero temperature, late time limit, we may assume that
  the polymer is \protect \textit {directed} in the $x$ direction -- that is,
  it has no loops or overhangs \cite {kardar_statistical_2007}.}\BibitemShut
  {Stop}%
\bibitem [{\citenamefont {Huse}\ and\ \citenamefont
  {Henley}(1985)}]{huse_pinning_1985}%
  \BibitemOpen
  \bibfield  {author} {\bibinfo {author} {\bibfnamefont {D.~A.}\ \bibnamefont
  {Huse}}\ and\ \bibinfo {author} {\bibfnamefont {C.~L.}\ \bibnamefont
  {Henley}},\ }\href {https://doi.org/10.1103/PhysRevLett.54.2708} {\bibfield
  {journal} {\bibinfo  {journal} {Physical Review Letters}\ }\textbf {\bibinfo
  {volume} {54}},\ \bibinfo {pages} {2708} (\bibinfo {year}
  {1985})}\BibitemShut {NoStop}%
\bibitem [{\citenamefont {Gueudre}\ and\ \citenamefont
  {Doussal}(2012)}]{gueudre_directed_2012}%
  \BibitemOpen
  \bibfield  {author} {\bibinfo {author} {\bibfnamefont {T.}~\bibnamefont
  {Gueudre}}\ and\ \bibinfo {author} {\bibfnamefont {P.~L.}\ \bibnamefont
  {Doussal}},\ }\href {https://doi.org/10.1209/0295-5075/100/26006} {\bibfield
  {journal} {\bibinfo  {journal} {EPL (Europhysics Letters)}\ }\textbf
  {\bibinfo {volume} {100}},\ \bibinfo {pages} {26006} (\bibinfo {year}
  {2012})}\BibitemShut {NoStop}%
\bibitem [{\citenamefont {Barraquand}\ \emph {et~al.}(2020)\citenamefont
  {Barraquand}, \citenamefont {Krajenbrink},\ and\ \citenamefont
  {Le~Doussal}}]{barraquand_half-space_2020}%
  \BibitemOpen
  \bibfield  {author} {\bibinfo {author} {\bibfnamefont {G.}~\bibnamefont
  {Barraquand}}, \bibinfo {author} {\bibfnamefont {A.}~\bibnamefont
  {Krajenbrink}},\ and\ \bibinfo {author} {\bibfnamefont {P.}~\bibnamefont
  {Le~Doussal}},\ }\href {https://doi.org/10.1007/s10955-020-02622-z}
  {\bibfield  {journal} {\bibinfo  {journal} {Journal of Statistical Physics}\
  }\textbf {\bibinfo {volume} {181}},\ \bibinfo {pages} {1149} (\bibinfo {year}
  {2020})}\BibitemShut {NoStop}%
\bibitem [{Wei()}]{Weinstein_Pagelike}%
  \BibitemOpen
  \href@noop {} {}\bibinfo {note} {Z. Weinstein, Y. Bao, Z. Lenar\v{c}i\v{c},
  S. Choi, and E. Altman, \textit{In preparation}.}\BibitemShut {Stop}%
\bibitem [{\citenamefont {Bao}\ \emph {et~al.}(2021)\citenamefont {Bao},
  \citenamefont {Choi},\ and\ \citenamefont {Altman}}]{bao2021symmetry}%
  \BibitemOpen
  \bibfield  {author} {\bibinfo {author} {\bibfnamefont {Y.}~\bibnamefont
  {Bao}}, \bibinfo {author} {\bibfnamefont {S.}~\bibnamefont {Choi}},\ and\
  \bibinfo {author} {\bibfnamefont {E.}~\bibnamefont {Altman}},\ }\href@noop {}
  {\bibfield  {journal} {\bibinfo  {journal} {Annals of Physics}\ }\textbf
  {\bibinfo {volume} {435}},\ \bibinfo {pages} {168618} (\bibinfo {year}
  {2021})}\BibitemShut {NoStop}%
\end{thebibliography}%


\begin{thebibliography}{48}%
\makeatletter
\providecommand \@ifxundefined [1]{%
 \@ifx{#1\undefined}
}%
\providecommand \@ifnum [1]{%
 \ifnum #1\expandafter \@firstoftwo
 \else \expandafter \@secondoftwo
 \fi
}%
\providecommand \@ifx [1]{%
 \ifx #1\expandafter \@firstoftwo
 \else \expandafter \@secondoftwo
 \fi
}%
\providecommand \natexlab [1]{#1}%
\providecommand \enquote  [1]{``#1''}%
\providecommand \bibnamefont  [1]{#1}%
\providecommand \bibfnamefont [1]{#1}%
\providecommand \citenamefont [1]{#1}%
\providecommand \href@noop [0]{\@secondoftwo}%
\providecommand \href [0]{\begingroup \@sanitize@url \@href}%
\providecommand \@href[1]{\@@startlink{#1}\@@href}%
\providecommand \@@href[1]{\endgroup#1\@@endlink}%
\providecommand \@sanitize@url [0]{\catcode `\\12\catcode `\$12\catcode
  `\&12\catcode `\#12\catcode `\^12\catcode `\_12\catcode `\%12\relax}%
\providecommand \@@startlink[1]{}%
\providecommand \@@endlink[0]{}%
\providecommand \url  [0]{\begingroup\@sanitize@url \@url }%
\providecommand \@url [1]{\endgroup\@href {#1}{\urlprefix }}%
\providecommand \urlprefix  [0]{URL }%
\providecommand \Eprint [0]{\href }%
\providecommand \doibase [0]{https://doi.org/}%
\providecommand \selectlanguage [0]{\@gobble}%
\providecommand \bibinfo  [0]{\@secondoftwo}%
\providecommand \bibfield  [0]{\@secondoftwo}%
\providecommand \translation [1]{[#1]}%
\providecommand \BibitemOpen [0]{}%
\providecommand \bibitemStop [0]{}%
\providecommand \bibitemNoStop [0]{.\EOS\space}%
\providecommand \EOS [0]{\spacefactor3000\relax}%
\providecommand \BibitemShut  [1]{\csname bibitem#1\endcsname}%
\let\auto@bib@innerbib\@empty
\bibitem [{\citenamefont {Vidal}\ and\ \citenamefont
  {Werner}(2002)}]{vidal_computable_2002}%
  \BibitemOpen
  \bibfield  {author} {\bibinfo {author} {\bibfnamefont {G.}~\bibnamefont
  {Vidal}}\ and\ \bibinfo {author} {\bibfnamefont {R.~F.}\ \bibnamefont
  {Werner}},\ }\href {https://doi.org/10.1103/PhysRevA.65.032314} {\bibfield
  {journal} {\bibinfo  {journal} {Physical Review A}\ }\textbf {\bibinfo
  {volume} {65}},\ \bibinfo {pages} {032314} (\bibinfo {year}
  {2002})}\BibitemShut {NoStop}%
\bibitem [{\citenamefont {Plenio}(2005)}]{plenio_logarithmic_2005}%
  \BibitemOpen
  \bibfield  {author} {\bibinfo {author} {\bibfnamefont {M.~B.}\ \bibnamefont
  {Plenio}},\ }\href {https://doi.org/10.1103/PhysRevLett.95.090503} {\bibfield
   {journal} {\bibinfo  {journal} {Physical Review Letters}\ }\textbf {\bibinfo
  {volume} {95}},\ \bibinfo {pages} {090503} (\bibinfo {year}
  {2005})}\BibitemShut {NoStop}%
\bibitem [{\citenamefont {Calabrese}\ \emph {et~al.}(2012)\citenamefont
  {Calabrese}, \citenamefont {Cardy},\ and\ \citenamefont
  {Tonni}}]{calabrese_entanglement_2012}%
  \BibitemOpen
  \bibfield  {author} {\bibinfo {author} {\bibfnamefont {P.}~\bibnamefont
  {Calabrese}}, \bibinfo {author} {\bibfnamefont {J.}~\bibnamefont {Cardy}},\
  and\ \bibinfo {author} {\bibfnamefont {E.}~\bibnamefont {Tonni}},\ }\href
  {https://doi.org/10.1103/PhysRevLett.109.130502} {\bibfield  {journal}
  {\bibinfo  {journal} {Physical Review Letters}\ }\textbf {\bibinfo {volume}
  {109}},\ \bibinfo {pages} {130502} (\bibinfo {year} {2012})}\BibitemShut
  {NoStop}%
\bibitem [{\citenamefont {Calabrese}\ \emph {et~al.}(2013)\citenamefont
  {Calabrese}, \citenamefont {Cardy},\ and\ \citenamefont
  {Tonni}}]{calabrese_entanglement_2013}%
  \BibitemOpen
  \bibfield  {author} {\bibinfo {author} {\bibfnamefont {P.}~\bibnamefont
  {Calabrese}}, \bibinfo {author} {\bibfnamefont {J.}~\bibnamefont {Cardy}},\
  and\ \bibinfo {author} {\bibfnamefont {E.}~\bibnamefont {Tonni}},\ }\href
  {https://doi.org/10.1088/1742-5468/2013/02/P02008} {\bibfield  {journal}
  {\bibinfo  {journal} {Journal of Statistical Mechanics: Theory and
  Experiment}\ }\textbf {\bibinfo {volume} {2013}},\ \bibinfo {pages} {P02008}
  (\bibinfo {year} {2013})}\BibitemShut {NoStop}%
\bibitem [{\citenamefont {Lu}\ and\ \citenamefont
  {Grover}(2019)}]{lu_singularity_2019}%
  \BibitemOpen
  \bibfield  {author} {\bibinfo {author} {\bibfnamefont {T.-C.}\ \bibnamefont
  {Lu}}\ and\ \bibinfo {author} {\bibfnamefont {T.}~\bibnamefont {Grover}},\
  }\href {https://doi.org/10.1103/PhysRevB.99.075157} {\bibfield  {journal}
  {\bibinfo  {journal} {Physical Review B}\ }\textbf {\bibinfo {volume} {99}},\
  \bibinfo {pages} {075157} (\bibinfo {year} {2019})}\BibitemShut {NoStop}%
\bibitem [{\citenamefont {Lu}\ \emph {et~al.}(2020)\citenamefont {Lu},
  \citenamefont {Hsieh},\ and\ \citenamefont {Grover}}]{lu_detecting_2020}%
  \BibitemOpen
  \bibfield  {author} {\bibinfo {author} {\bibfnamefont {T.-C.}\ \bibnamefont
  {Lu}}, \bibinfo {author} {\bibfnamefont {T.~H.}\ \bibnamefont {Hsieh}},\ and\
  \bibinfo {author} {\bibfnamefont {T.}~\bibnamefont {Grover}},\ }\href
  {https://doi.org/10.1103/PhysRevLett.125.116801} {\bibfield  {journal}
  {\bibinfo  {journal} {Physical Review Letters}\ }\textbf {\bibinfo {volume}
  {125}},\ \bibinfo {pages} {116801} (\bibinfo {year} {2020})}\BibitemShut
  {NoStop}%
\bibitem [{\citenamefont {Wu}\ \emph {et~al.}(2020)\citenamefont {Wu},
  \citenamefont {Lu}, \citenamefont {Chung}, \citenamefont {Kao},\ and\
  \citenamefont {Grover}}]{wu_entanglement_2020}%
  \BibitemOpen
  \bibfield  {author} {\bibinfo {author} {\bibfnamefont {K.-H.}\ \bibnamefont
  {Wu}}, \bibinfo {author} {\bibfnamefont {T.-C.}\ \bibnamefont {Lu}}, \bibinfo
  {author} {\bibfnamefont {C.-M.}\ \bibnamefont {Chung}}, \bibinfo {author}
  {\bibfnamefont {Y.-J.}\ \bibnamefont {Kao}},\ and\ \bibinfo {author}
  {\bibfnamefont {T.}~\bibnamefont {Grover}},\ }\href
  {https://doi.org/10.1103/PhysRevLett.125.140603} {\bibfield  {journal}
  {\bibinfo  {journal} {Physical Review Letters}\ }\textbf {\bibinfo {volume}
  {125}},\ \bibinfo {pages} {140603} (\bibinfo {year} {2020})}\BibitemShut
  {NoStop}%
\bibitem [{\citenamefont {Lu}\ and\ \citenamefont
  {Grover}(2020)}]{lu_entanglement_2020}%
  \BibitemOpen
  \bibfield  {author} {\bibinfo {author} {\bibfnamefont {T.-C.}\ \bibnamefont
  {Lu}}\ and\ \bibinfo {author} {\bibfnamefont {T.}~\bibnamefont {Grover}},\
  }\href {https://doi.org/10.1103/PhysRevB.102.235110} {\bibfield  {journal}
  {\bibinfo  {journal} {Physical Review B}\ }\textbf {\bibinfo {volume}
  {102}},\ \bibinfo {pages} {235110} (\bibinfo {year} {2020})}\BibitemShut
  {NoStop}%
\bibitem [{\citenamefont {Sang}\ \emph {et~al.}(2021)\citenamefont {Sang},
  \citenamefont {Li}, \citenamefont {Zhou}, \citenamefont {Chen}, \citenamefont
  {Hsieh},\ and\ \citenamefont {Fisher}}]{sang_entanglement_2020}%
  \BibitemOpen
  \bibfield  {author} {\bibinfo {author} {\bibfnamefont {S.}~\bibnamefont
  {Sang}}, \bibinfo {author} {\bibfnamefont {Y.}~\bibnamefont {Li}}, \bibinfo
  {author} {\bibfnamefont {T.}~\bibnamefont {Zhou}}, \bibinfo {author}
  {\bibfnamefont {X.}~\bibnamefont {Chen}}, \bibinfo {author} {\bibfnamefont
  {T.~H.}\ \bibnamefont {Hsieh}},\ and\ \bibinfo {author} {\bibfnamefont
  {M.~P.}\ \bibnamefont {Fisher}},\ }\href
  {https://doi.org/10.1103/PRXQuantum.2.030313} {\bibfield  {journal} {\bibinfo
   {journal} {PRX Quantum}\ }\textbf {\bibinfo {volume} {2}},\ \bibinfo {pages}
  {030313} (\bibinfo {year} {2021})}\BibitemShut {NoStop}%
\bibitem [{\citenamefont {Shapourian}\ \emph {et~al.}(2021)\citenamefont
  {Shapourian}, \citenamefont {Liu}, \citenamefont {Kudler-Flam},\ and\
  \citenamefont {Vishwanath}}]{shapourian_entanglement_2020}%
  \BibitemOpen
  \bibfield  {author} {\bibinfo {author} {\bibfnamefont {H.}~\bibnamefont
  {Shapourian}}, \bibinfo {author} {\bibfnamefont {S.}~\bibnamefont {Liu}},
  \bibinfo {author} {\bibfnamefont {J.}~\bibnamefont {Kudler-Flam}},\ and\
  \bibinfo {author} {\bibfnamefont {A.}~\bibnamefont {Vishwanath}},\ }\href
  {https://doi.org/10.1103/PRXQuantum.2.030347} {\bibfield  {journal} {\bibinfo
   {journal} {PRX Quantum}\ }\textbf {\bibinfo {volume} {2}},\ \bibinfo {pages}
  {030347} (\bibinfo {year} {2021})}\BibitemShut {NoStop}%
\bibitem [{\citenamefont {Nielsen}\ and\ \citenamefont
  {Chuang}(2010)}]{nielsen_quantum_2010}%
  \BibitemOpen
  \bibfield  {author} {\bibinfo {author} {\bibfnamefont {M.~A.}\ \bibnamefont
  {Nielsen}}\ and\ \bibinfo {author} {\bibfnamefont {I.~L.}\ \bibnamefont
  {Chuang}},\ }\href@noop {} {\emph {\bibinfo {title} {Quantum computation and
  quantum information}}},\ \bibinfo {edition} {10th}\ ed.\ (\bibinfo
  {publisher} {Cambridge University Press},\ \bibinfo {address} {Cambridge; New
  York},\ \bibinfo {year} {2010})\BibitemShut {NoStop}%
\bibitem [{\citenamefont {Bennett}\ \emph {et~al.}(1996)\citenamefont
  {Bennett}, \citenamefont {DiVincenzo}, \citenamefont {Smolin},\ and\
  \citenamefont {Wootters}}]{bennett_mixed-state_1996}%
  \BibitemOpen
  \bibfield  {author} {\bibinfo {author} {\bibfnamefont {C.~H.}\ \bibnamefont
  {Bennett}}, \bibinfo {author} {\bibfnamefont {D.~P.}\ \bibnamefont
  {DiVincenzo}}, \bibinfo {author} {\bibfnamefont {J.~A.}\ \bibnamefont
  {Smolin}},\ and\ \bibinfo {author} {\bibfnamefont {W.~K.}\ \bibnamefont
  {Wootters}},\ }\href {https://doi.org/10.1103/PhysRevA.54.3824} {\bibfield
  {journal} {\bibinfo  {journal} {Physical Review A}\ }\textbf {\bibinfo
  {volume} {54}},\ \bibinfo {pages} {3824} (\bibinfo {year}
  {1996})}\BibitemShut {NoStop}%
\bibitem [{\citenamefont {Horodecki}\ \emph {et~al.}(1998)\citenamefont
  {Horodecki}, \citenamefont {Horodecki},\ and\ \citenamefont
  {Horodecki}}]{horodecki_mixed-state_1998}%
  \BibitemOpen
  \bibfield  {author} {\bibinfo {author} {\bibfnamefont {M.}~\bibnamefont
  {Horodecki}}, \bibinfo {author} {\bibfnamefont {P.}~\bibnamefont
  {Horodecki}},\ and\ \bibinfo {author} {\bibfnamefont {R.}~\bibnamefont
  {Horodecki}},\ }\href {https://doi.org/10.1103/PhysRevLett.80.5239}
  {\bibfield  {journal} {\bibinfo  {journal} {Physical Review Letters}\
  }\textbf {\bibinfo {volume} {80}},\ \bibinfo {pages} {5239} (\bibinfo {year}
  {1998})}\BibitemShut {NoStop}%
\bibitem [{\citenamefont {Peres}(1996)}]{peres_separability_1996}%
  \BibitemOpen
  \bibfield  {author} {\bibinfo {author} {\bibfnamefont {A.}~\bibnamefont
  {Peres}},\ }\href {https://doi.org/10.1103/PhysRevLett.77.1413} {\bibfield
  {journal} {\bibinfo  {journal} {Physical Review Letters}\ }\textbf {\bibinfo
  {volume} {77}},\ \bibinfo {pages} {1413} (\bibinfo {year}
  {1996})}\BibitemShut {NoStop}%
\bibitem [{\citenamefont {Horodecki}\ \emph {et~al.}(1996)\citenamefont
  {Horodecki}, \citenamefont {Horodecki},\ and\ \citenamefont
  {Horodecki}}]{horodecki_separability_1996}%
  \BibitemOpen
  \bibfield  {author} {\bibinfo {author} {\bibfnamefont {M.}~\bibnamefont
  {Horodecki}}, \bibinfo {author} {\bibfnamefont {P.}~\bibnamefont
  {Horodecki}},\ and\ \bibinfo {author} {\bibfnamefont {R.}~\bibnamefont
  {Horodecki}},\ }\href {https://doi.org/10.1016/S0375-9601(96)00706-2}
  {\bibfield  {journal} {\bibinfo  {journal} {Physics Letters A}\ }\textbf
  {\bibinfo {volume} {223}},\ \bibinfo {pages} {1} (\bibinfo {year}
  {1996})}\BibitemShut {NoStop}%
\bibitem [{\citenamefont {Vidal}\ and\ \citenamefont
  {Cirac}(2001)}]{vidal_irreversibility_2001}%
  \BibitemOpen
  \bibfield  {author} {\bibinfo {author} {\bibfnamefont {G.}~\bibnamefont
  {Vidal}}\ and\ \bibinfo {author} {\bibfnamefont {J.~I.}\ \bibnamefont
  {Cirac}},\ }\href {https://doi.org/10.1103/PhysRevLett.86.5803} {\bibfield
  {journal} {\bibinfo  {journal} {Physical Review Letters}\ }\textbf {\bibinfo
  {volume} {86}},\ \bibinfo {pages} {5803} (\bibinfo {year}
  {2001})}\BibitemShut {NoStop}%
\bibitem [{\citenamefont {Horodecki}\ \emph {et~al.}(2020)\citenamefont
  {Horodecki}, \citenamefont {Rudnicki},\ and\ \citenamefont
  {Zyczkowski}}]{horodecki_five_2020}%
  \BibitemOpen
  \bibfield  {author} {\bibinfo {author} {\bibfnamefont {P.}~\bibnamefont
  {Horodecki}}, \bibinfo {author} {\bibfnamefont {L.}~\bibnamefont
  {Rudnicki}},\ and\ \bibinfo {author} {\bibfnamefont {K.}~\bibnamefont
  {Zyczkowski}},\ }\href {http://arxiv.org/abs/2002.03233} {\bibfield
  {journal} {\bibinfo  {journal} {arXiv:2002.03233 [quant-ph]}\ } (\bibinfo
  {year} {2020})},\ \bibinfo {note} {arXiv: 2002.03233}\BibitemShut {NoStop}%
\bibitem [{\citenamefont {Page}(1993)}]{page_average_1993}%
  \BibitemOpen
  \bibfield  {author} {\bibinfo {author} {\bibfnamefont {D.~N.}\ \bibnamefont
  {Page}},\ }\href {https://doi.org/10.1103/PhysRevLett.71.1291} {\bibfield
  {journal} {\bibinfo  {journal} {Physical Review Letters}\ }\textbf {\bibinfo
  {volume} {71}},\ \bibinfo {pages} {1291} (\bibinfo {year}
  {1993})}\BibitemShut {NoStop}%
\bibitem [{\citenamefont {Bhosale}\ \emph {et~al.}(2012)\citenamefont
  {Bhosale}, \citenamefont {Tomsovic},\ and\ \citenamefont
  {Lakshminarayan}}]{bhosale_entanglement_2012}%
  \BibitemOpen
  \bibfield  {author} {\bibinfo {author} {\bibfnamefont {U.~T.}\ \bibnamefont
  {Bhosale}}, \bibinfo {author} {\bibfnamefont {S.}~\bibnamefont {Tomsovic}},\
  and\ \bibinfo {author} {\bibfnamefont {A.}~\bibnamefont {Lakshminarayan}},\
  }\href {https://doi.org/10.1103/PhysRevA.85.062331} {\bibfield  {journal}
  {\bibinfo  {journal} {Physical Review A}\ }\textbf {\bibinfo {volume} {85}},\
  \bibinfo {pages} {062331} (\bibinfo {year} {2012})}\BibitemShut {NoStop}%
\bibitem [{\citenamefont {Gottesman}(1998)}]{gottesman_heisenberg_1998}%
  \BibitemOpen
  \bibfield  {author} {\bibinfo {author} {\bibfnamefont {D.}~\bibnamefont
  {Gottesman}},\ }\href {http://arxiv.org/abs/quant-ph/9807006} {\bibfield
  {journal} {\bibinfo  {journal} {arXiv:quant-ph/9807006}\ } (\bibinfo {year}
  {1998})},\ \bibinfo {note} {arXiv: quant-ph/9807006}\BibitemShut {NoStop}%
\bibitem [{\citenamefont {Aaronson}\ and\ \citenamefont
  {Gottesman}(2004)}]{aaronson_improved_2004}%
  \BibitemOpen
  \bibfield  {author} {\bibinfo {author} {\bibfnamefont {S.}~\bibnamefont
  {Aaronson}}\ and\ \bibinfo {author} {\bibfnamefont {D.}~\bibnamefont
  {Gottesman}},\ }\href {https://doi.org/10.1103/PhysRevA.70.052328} {\bibfield
   {journal} {\bibinfo  {journal} {Physical Review A}\ }\textbf {\bibinfo
  {volume} {70}},\ \bibinfo {pages} {052328} (\bibinfo {year}
  {2004})}\BibitemShut {NoStop}%
\bibitem [{\citenamefont {Nahum}\ \emph {et~al.}(2017)\citenamefont {Nahum},
  \citenamefont {Ruhman}, \citenamefont {Vijay},\ and\ \citenamefont
  {Haah}}]{nahum_quantum_2017}%
  \BibitemOpen
  \bibfield  {author} {\bibinfo {author} {\bibfnamefont {A.}~\bibnamefont
  {Nahum}}, \bibinfo {author} {\bibfnamefont {J.}~\bibnamefont {Ruhman}},
  \bibinfo {author} {\bibfnamefont {S.}~\bibnamefont {Vijay}},\ and\ \bibinfo
  {author} {\bibfnamefont {J.}~\bibnamefont {Haah}},\ }\href
  {https://doi.org/10.1103/PhysRevX.7.031016} {\bibfield  {journal} {\bibinfo
  {journal} {Physical Review X}\ }\textbf {\bibinfo {volume} {7}},\ \bibinfo
  {pages} {031016} (\bibinfo {year} {2017})}\BibitemShut {NoStop}%
\bibitem [{\citenamefont {Li}\ \emph {et~al.}(2019)\citenamefont {Li},
  \citenamefont {Chen},\ and\ \citenamefont
  {Fisher}}]{li_measurement-driven_2019}%
  \BibitemOpen
  \bibfield  {author} {\bibinfo {author} {\bibfnamefont {Y.}~\bibnamefont
  {Li}}, \bibinfo {author} {\bibfnamefont {X.}~\bibnamefont {Chen}},\ and\
  \bibinfo {author} {\bibfnamefont {M.~P.~A.}\ \bibnamefont {Fisher}},\ }\href
  {https://doi.org/10.1103/PhysRevB.100.134306} {\bibfield  {journal} {\bibinfo
   {journal} {Physical Review B}\ }\textbf {\bibinfo {volume} {100}},\ \bibinfo
  {pages} {134306} (\bibinfo {year} {2019})}\BibitemShut {NoStop}%
\bibitem [{\citenamefont {Hamma}\ \emph
  {et~al.}(2005{\natexlab{a}})\citenamefont {Hamma}, \citenamefont
  {Ionicioiu},\ and\ \citenamefont {Zanardi}}]{hamma2005bipartite}%
  \BibitemOpen
  \bibfield  {author} {\bibinfo {author} {\bibfnamefont {A.}~\bibnamefont
  {Hamma}}, \bibinfo {author} {\bibfnamefont {R.}~\bibnamefont {Ionicioiu}},\
  and\ \bibinfo {author} {\bibfnamefont {P.}~\bibnamefont {Zanardi}},\
  }\href@noop {} {\bibfield  {journal} {\bibinfo  {journal} {Physical Review
  A}\ }\textbf {\bibinfo {volume} {71}},\ \bibinfo {pages} {022315} (\bibinfo
  {year} {2005}{\natexlab{a}})}\BibitemShut {NoStop}%
\bibitem [{\citenamefont {Hamma}\ \emph
  {et~al.}(2005{\natexlab{b}})\citenamefont {Hamma}, \citenamefont
  {Ionicioiu},\ and\ \citenamefont {Zanardi}}]{hamma2005ground}%
  \BibitemOpen
  \bibfield  {author} {\bibinfo {author} {\bibfnamefont {A.}~\bibnamefont
  {Hamma}}, \bibinfo {author} {\bibfnamefont {R.}~\bibnamefont {Ionicioiu}},\
  and\ \bibinfo {author} {\bibfnamefont {P.}~\bibnamefont {Zanardi}},\
  }\href@noop {} {\bibfield  {journal} {\bibinfo  {journal} {Physics Letters
  A}\ }\textbf {\bibinfo {volume} {337}},\ \bibinfo {pages} {22} (\bibinfo
  {year} {2005}{\natexlab{b}})}\BibitemShut {NoStop}%
\bibitem [{\citenamefont {Skinner}\ \emph {et~al.}(2019)\citenamefont
  {Skinner}, \citenamefont {Ruhman},\ and\ \citenamefont
  {Nahum}}]{skinner2019measurement}%
  \BibitemOpen
  \bibfield  {author} {\bibinfo {author} {\bibfnamefont {B.}~\bibnamefont
  {Skinner}}, \bibinfo {author} {\bibfnamefont {J.}~\bibnamefont {Ruhman}},\
  and\ \bibinfo {author} {\bibfnamefont {A.}~\bibnamefont {Nahum}},\
  }\href@noop {} {\bibfield  {journal} {\bibinfo  {journal} {Physical Review
  X}\ }\textbf {\bibinfo {volume} {9}},\ \bibinfo {pages} {031009} (\bibinfo
  {year} {2019})}\BibitemShut {NoStop}%
\bibitem [{\citenamefont {Choi}\ \emph {et~al.}(2020)\citenamefont {Choi},
  \citenamefont {Bao}, \citenamefont {Qi},\ and\ \citenamefont
  {Altman}}]{choi_quantum_2020}%
  \BibitemOpen
  \bibfield  {author} {\bibinfo {author} {\bibfnamefont {S.}~\bibnamefont
  {Choi}}, \bibinfo {author} {\bibfnamefont {Y.}~\bibnamefont {Bao}}, \bibinfo
  {author} {\bibfnamefont {X.-L.}\ \bibnamefont {Qi}},\ and\ \bibinfo {author}
  {\bibfnamefont {E.}~\bibnamefont {Altman}},\ }\href
  {https://doi.org/10.1103/PhysRevLett.125.030505} {\bibfield  {journal}
  {\bibinfo  {journal} {Physical Review Letters}\ }\textbf {\bibinfo {volume}
  {125}},\ \bibinfo {pages} {030505} (\bibinfo {year} {2020})}\BibitemShut
  {NoStop}%
\bibitem [{\citenamefont {Zabalo}\ \emph {et~al.}(2020)\citenamefont {Zabalo},
  \citenamefont {Gullans}, \citenamefont {Wilson}, \citenamefont
  {Gopalakrishnan}, \citenamefont {Huse},\ and\ \citenamefont
  {Pixley}}]{zabalo2020critical}%
  \BibitemOpen
  \bibfield  {author} {\bibinfo {author} {\bibfnamefont {A.}~\bibnamefont
  {Zabalo}}, \bibinfo {author} {\bibfnamefont {M.~J.}\ \bibnamefont {Gullans}},
  \bibinfo {author} {\bibfnamefont {J.~H.}\ \bibnamefont {Wilson}}, \bibinfo
  {author} {\bibfnamefont {S.}~\bibnamefont {Gopalakrishnan}}, \bibinfo
  {author} {\bibfnamefont {D.~A.}\ \bibnamefont {Huse}},\ and\ \bibinfo
  {author} {\bibfnamefont {J.}~\bibnamefont {Pixley}},\ }\href@noop {}
  {\bibfield  {journal} {\bibinfo  {journal} {Physical Review B}\ }\textbf
  {\bibinfo {volume} {101}},\ \bibinfo {pages} {060301} (\bibinfo {year}
  {2020})}\BibitemShut {NoStop}%
\bibitem [{\citenamefont {Li}\ \emph {et~al.}(2018)\citenamefont {Li},
  \citenamefont {Chen},\ and\ \citenamefont {Fisher}}]{li_quantum_2018}%
  \BibitemOpen
  \bibfield  {author} {\bibinfo {author} {\bibfnamefont {Y.}~\bibnamefont
  {Li}}, \bibinfo {author} {\bibfnamefont {X.}~\bibnamefont {Chen}},\ and\
  \bibinfo {author} {\bibfnamefont {M.~P.~A.}\ \bibnamefont {Fisher}},\ }\href
  {https://doi.org/10.1103/PhysRevB.98.205136} {\bibfield  {journal} {\bibinfo
  {journal} {Physical Review B}\ }\textbf {\bibinfo {volume} {98}},\ \bibinfo
  {pages} {205136} (\bibinfo {year} {2018})}\BibitemShut {NoStop}%
\bibitem [{\citenamefont {Zhou}\ and\ \citenamefont
  {Nahum}(2019)}]{zhou_emergent_2019}%
  \BibitemOpen
  \bibfield  {author} {\bibinfo {author} {\bibfnamefont {T.}~\bibnamefont
  {Zhou}}\ and\ \bibinfo {author} {\bibfnamefont {A.}~\bibnamefont {Nahum}},\
  }\href {https://doi.org/10.1103/PhysRevB.99.174205} {\bibfield  {journal}
  {\bibinfo  {journal} {Physical Review B}\ }\textbf {\bibinfo {volume} {99}},\
  \bibinfo {pages} {174205} (\bibinfo {year} {2019})}\BibitemShut {NoStop}%
\bibitem [{\citenamefont {Bao}\ \emph {et~al.}(2020)\citenamefont {Bao},
  \citenamefont {Choi},\ and\ \citenamefont {Altman}}]{bao_theory_2020}%
  \BibitemOpen
  \bibfield  {author} {\bibinfo {author} {\bibfnamefont {Y.}~\bibnamefont
  {Bao}}, \bibinfo {author} {\bibfnamefont {S.}~\bibnamefont {Choi}},\ and\
  \bibinfo {author} {\bibfnamefont {E.}~\bibnamefont {Altman}},\ }\href
  {https://doi.org/10.1103/PhysRevB.101.104301} {\bibfield  {journal} {\bibinfo
   {journal} {Physical Review B}\ }\textbf {\bibinfo {volume} {101}},\ \bibinfo
  {pages} {104301} (\bibinfo {year} {2020})}\BibitemShut {NoStop}%
\bibitem [{\citenamefont {Jian}\ \emph {et~al.}(2020)\citenamefont {Jian},
  \citenamefont {You}, \citenamefont {Vasseur},\ and\ \citenamefont
  {Ludwig}}]{jian_measurement-induced_2020}%
  \BibitemOpen
  \bibfield  {author} {\bibinfo {author} {\bibfnamefont {C.-M.}\ \bibnamefont
  {Jian}}, \bibinfo {author} {\bibfnamefont {Y.-Z.}\ \bibnamefont {You}},
  \bibinfo {author} {\bibfnamefont {R.}~\bibnamefont {Vasseur}},\ and\ \bibinfo
  {author} {\bibfnamefont {A.~W.~W.}\ \bibnamefont {Ludwig}},\ }\href
  {https://doi.org/10.1103/PhysRevB.101.104302} {\bibfield  {journal} {\bibinfo
   {journal} {Physical Review B}\ }\textbf {\bibinfo {volume} {101}},\ \bibinfo
  {pages} {104302} (\bibinfo {year} {2020})}\BibitemShut {NoStop}%
\bibitem [{\citenamefont {Li}\ \emph {et~al.}(2021)\citenamefont {Li},
  \citenamefont {Vijay},\ and\ \citenamefont {Fisher}}]{li_entanglement_2021}%
  \BibitemOpen
  \bibfield  {author} {\bibinfo {author} {\bibfnamefont {Y.}~\bibnamefont
  {Li}}, \bibinfo {author} {\bibfnamefont {S.}~\bibnamefont {Vijay}},\ and\
  \bibinfo {author} {\bibfnamefont {M.~P.~A.}\ \bibnamefont {Fisher}},\ }\href
  {http://arxiv.org/abs/2105.13352} {\bibfield  {journal} {\bibinfo  {journal}
  {arXiv:2105.13352 [cond-mat, physics:quant-ph]}\ } (\bibinfo {year}
  {2021})}\BibitemShut {NoStop}%
\bibitem [{\citenamefont {Agrawal}\ \emph {et~al.}(2021)\citenamefont
  {Agrawal}, \citenamefont {Zabalo}, \citenamefont {Chen}, \citenamefont
  {Wilson}, \citenamefont {Potter}, \citenamefont {Pixley}, \citenamefont
  {Gopalakrishnan},\ and\ \citenamefont {Vasseur}}]{agrawal_entanglement_2021}%
  \BibitemOpen
  \bibfield  {author} {\bibinfo {author} {\bibfnamefont {U.}~\bibnamefont
  {Agrawal}}, \bibinfo {author} {\bibfnamefont {A.}~\bibnamefont {Zabalo}},
  \bibinfo {author} {\bibfnamefont {K.}~\bibnamefont {Chen}}, \bibinfo {author}
  {\bibfnamefont {J.~H.}\ \bibnamefont {Wilson}}, \bibinfo {author}
  {\bibfnamefont {A.~C.}\ \bibnamefont {Potter}}, \bibinfo {author}
  {\bibfnamefont {J.~H.}\ \bibnamefont {Pixley}}, \bibinfo {author}
  {\bibfnamefont {S.}~\bibnamefont {Gopalakrishnan}},\ and\ \bibinfo {author}
  {\bibfnamefont {R.}~\bibnamefont {Vasseur}},\ }\href
  {http://arxiv.org/abs/2107.10279} {\bibfield  {journal} {\bibinfo  {journal}
  {arXiv:2107.10279 [cond-mat, physics:quant-ph]}\ } (\bibinfo {year}
  {2021})}\BibitemShut {NoStop}%
\bibitem [{\citenamefont {Kardar}(2007)}]{kardar_statistical_2007}%
  \BibitemOpen
  \bibfield  {author} {\bibinfo {author} {\bibfnamefont {M.}~\bibnamefont
  {Kardar}},\ }\href@noop {} {\emph {\bibinfo {title} {Statistical physics of
  fields}}}\ (\bibinfo  {publisher} {Cambridge University Press},\ \bibinfo
  {address} {Cambridge; New York},\ \bibinfo {year} {2007})\ \bibinfo {note}
  {oCLC: ocn123113789}\BibitemShut {NoStop}%
\bibitem [{\citenamefont {Nishimori}(2001)}]{nishimori_statistical_2001}%
  \BibitemOpen
  \bibfield  {author} {\bibinfo {author} {\bibfnamefont {H.}~\bibnamefont
  {Nishimori}},\ }\href@noop {} {\emph {\bibinfo {title} {Statistical physics
  of spin glasses and information processing: an introduction}}},\ \bibinfo
  {series} {International series of monographs on physics}\ No.\ \bibinfo
  {number} {111}\ (\bibinfo  {publisher} {Oxford University Press},\ \bibinfo
  {address} {Oxford; New York},\ \bibinfo {year} {2001})\ \bibinfo {note}
  {oCLC: ocm47063323}\BibitemShut {NoStop}%
\bibitem [{\citenamefont {You}\ and\ \citenamefont
  {Gu}(2018)}]{you2018entanglement}%
  \BibitemOpen
  \bibfield  {author} {\bibinfo {author} {\bibfnamefont {Y.-Z.}\ \bibnamefont
  {You}}\ and\ \bibinfo {author} {\bibfnamefont {Y.}~\bibnamefont {Gu}},\
  }\href@noop {} {\bibfield  {journal} {\bibinfo  {journal} {Physical Review
  B}\ }\textbf {\bibinfo {volume} {98}},\ \bibinfo {pages} {014309} (\bibinfo
  {year} {2018})}\BibitemShut {NoStop}%
\bibitem [{\citenamefont {Collins}(2003)}]{collins_moments_2003}%
  \BibitemOpen
  \bibfield  {author} {\bibinfo {author} {\bibfnamefont {B.}~\bibnamefont
  {Collins}},\ }\href {https://doi.org/10.1155/S107379280320917X} {\bibfield
  {journal} {\bibinfo  {journal} {International Mathematics Research Notices}\
  }\textbf {\bibinfo {volume} {2003}},\ \bibinfo {pages} {953} (\bibinfo {year}
  {2003})},\ \Eprint
  {https://arxiv.org/abs/https://academic.oup.com/imrn/article-pdf/2003/17/953/1881428/2003-17-953.pdf}
  {https://academic.oup.com/imrn/article-pdf/2003/17/953/1881428/2003-17-953.pdf}
  \BibitemShut {NoStop}%
\bibitem [{\citenamefont {Collins}\ and\ \citenamefont
  {{\'S}niady}(2006)}]{collins2006integration}%
  \BibitemOpen
  \bibfield  {author} {\bibinfo {author} {\bibfnamefont {B.}~\bibnamefont
  {Collins}}\ and\ \bibinfo {author} {\bibfnamefont {P.}~\bibnamefont
  {{\'S}niady}},\ }\href@noop {} {\bibfield  {journal} {\bibinfo  {journal}
  {Communications in Mathematical Physics}\ }\textbf {\bibinfo {volume}
  {264}},\ \bibinfo {pages} {773} (\bibinfo {year} {2006})}\BibitemShut
  {NoStop}%
\bibitem [{\citenamefont {Nahum}\ \emph {et~al.}(2018)\citenamefont {Nahum},
  \citenamefont {Vijay},\ and\ \citenamefont {Haah}}]{nahum_operator_2018}%
  \BibitemOpen
  \bibfield  {author} {\bibinfo {author} {\bibfnamefont {A.}~\bibnamefont
  {Nahum}}, \bibinfo {author} {\bibfnamefont {S.}~\bibnamefont {Vijay}},\ and\
  \bibinfo {author} {\bibfnamefont {J.}~\bibnamefont {Haah}},\ }\href
  {https://doi.org/10.1103/PhysRevX.8.021014} {\bibfield  {journal} {\bibinfo
  {journal} {Physical Review X}\ }\textbf {\bibinfo {volume} {8}},\ \bibinfo
  {pages} {021014} (\bibinfo {year} {2018})}\BibitemShut {NoStop}%
\bibitem [{\citenamefont {Bao}\ \emph {et~al.}(2021)\citenamefont {Bao},
  \citenamefont {Choi},\ and\ \citenamefont {Altman}}]{bao2021symmetry}%
  \BibitemOpen
  \bibfield  {author} {\bibinfo {author} {\bibfnamefont {Y.}~\bibnamefont
  {Bao}}, \bibinfo {author} {\bibfnamefont {S.}~\bibnamefont {Choi}},\ and\
  \bibinfo {author} {\bibfnamefont {E.}~\bibnamefont {Altman}},\ }\href@noop {}
  {\bibfield  {journal} {\bibinfo  {journal} {Annals of Physics}\ }\textbf
  {\bibinfo {volume} {435}},\ \bibinfo {pages} {168618} (\bibinfo {year}
  {2021})}\BibitemShut {NoStop}%
\bibitem [{\citenamefont {Dong}\ \emph {et~al.}(2021)\citenamefont {Dong},
  \citenamefont {Qi},\ and\ \citenamefont {Walter}}]{dong_holographic_2021}%
  \BibitemOpen
  \bibfield  {author} {\bibinfo {author} {\bibfnamefont {X.}~\bibnamefont
  {Dong}}, \bibinfo {author} {\bibfnamefont {X.-L.}\ \bibnamefont {Qi}},\ and\
  \bibinfo {author} {\bibfnamefont {M.}~\bibnamefont {Walter}},\ }\href
  {https://doi.org/10.1007/JHEP06(2021)024} {\bibfield  {journal} {\bibinfo
  {journal} {Journal of High Energy Physics}\ }\textbf {\bibinfo {volume}
  {2021}},\ \bibinfo {pages} {24} (\bibinfo {year} {2021})}\BibitemShut
  {NoStop}%
\bibitem [{\citenamefont {Kardar}(1985)}]{kardar_roughening_1985}%
  \BibitemOpen
  \bibfield  {author} {\bibinfo {author} {\bibfnamefont {M.}~\bibnamefont
  {Kardar}},\ }\href {https://doi.org/10.1103/PhysRevLett.55.2923} {\bibfield
  {journal} {\bibinfo  {journal} {Physical Review Letters}\ }\textbf {\bibinfo
  {volume} {55}},\ \bibinfo {pages} {2923} (\bibinfo {year}
  {1985})}\BibitemShut {NoStop}%
\bibitem [{\citenamefont {Kardar}\ \emph {et~al.}(1986)\citenamefont {Kardar},
  \citenamefont {Parisi},\ and\ \citenamefont {Zhang}}]{kardar_dynamic_1986}%
  \BibitemOpen
  \bibfield  {author} {\bibinfo {author} {\bibfnamefont {M.}~\bibnamefont
  {Kardar}}, \bibinfo {author} {\bibfnamefont {G.}~\bibnamefont {Parisi}},\
  and\ \bibinfo {author} {\bibfnamefont {Y.-C.}\ \bibnamefont {Zhang}},\ }\href
  {https://doi.org/10.1103/PhysRevLett.56.889} {\bibfield  {journal} {\bibinfo
  {journal} {Physical Review Letters}\ }\textbf {\bibinfo {volume} {56}},\
  \bibinfo {pages} {889} (\bibinfo {year} {1986})}\BibitemShut {NoStop}%
\bibitem [{\citenamefont {Huse}\ \emph {et~al.}(1985)\citenamefont {Huse},
  \citenamefont {Henley},\ and\ \citenamefont {Fisher}}]{huse_huse_1985}%
  \BibitemOpen
  \bibfield  {author} {\bibinfo {author} {\bibfnamefont {D.~A.}\ \bibnamefont
  {Huse}}, \bibinfo {author} {\bibfnamefont {C.~L.}\ \bibnamefont {Henley}},\
  and\ \bibinfo {author} {\bibfnamefont {D.~S.}\ \bibnamefont {Fisher}},\
  }\href {https://doi.org/10.1103/PhysRevLett.55.2924} {\bibfield  {journal}
  {\bibinfo  {journal} {Physical Review Letters}\ }\textbf {\bibinfo {volume}
  {55}},\ \bibinfo {pages} {2924} (\bibinfo {year} {1985})}\BibitemShut
  {NoStop}%
\bibitem [{\citenamefont {Huse}\ and\ \citenamefont
  {Henley}(1985)}]{huse_pinning_1985}%
  \BibitemOpen
  \bibfield  {author} {\bibinfo {author} {\bibfnamefont {D.~A.}\ \bibnamefont
  {Huse}}\ and\ \bibinfo {author} {\bibfnamefont {C.~L.}\ \bibnamefont
  {Henley}},\ }\href {https://doi.org/10.1103/PhysRevLett.54.2708} {\bibfield
  {journal} {\bibinfo  {journal} {Physical Review Letters}\ }\textbf {\bibinfo
  {volume} {54}},\ \bibinfo {pages} {2708} (\bibinfo {year}
  {1985})}\BibitemShut {NoStop}%
\bibitem [{\citenamefont {Barraquand}\ \emph {et~al.}(2020)\citenamefont
  {Barraquand}, \citenamefont {Krajenbrink},\ and\ \citenamefont
  {Le~Doussal}}]{barraquand_half-space_2020}%
  \BibitemOpen
  \bibfield  {author} {\bibinfo {author} {\bibfnamefont {G.}~\bibnamefont
  {Barraquand}}, \bibinfo {author} {\bibfnamefont {A.}~\bibnamefont
  {Krajenbrink}},\ and\ \bibinfo {author} {\bibfnamefont {P.}~\bibnamefont
  {Le~Doussal}},\ }\href {https://doi.org/10.1007/s10955-020-02622-z}
  {\bibfield  {journal} {\bibinfo  {journal} {Journal of Statistical Physics}\
  }\textbf {\bibinfo {volume} {181}},\ \bibinfo {pages} {1149} (\bibinfo {year}
  {2020})}\BibitemShut {NoStop}%
\bibitem [{\citenamefont {Gueudre}\ and\ \citenamefont
  {Doussal}(2012)}]{gueudre_directed_2012}%
  \BibitemOpen
  \bibfield  {author} {\bibinfo {author} {\bibfnamefont {T.}~\bibnamefont
  {Gueudre}}\ and\ \bibinfo {author} {\bibfnamefont {P.~L.}\ \bibnamefont
  {Doussal}},\ }\href {https://doi.org/10.1209/0295-5075/100/26006} {\bibfield
  {journal} {\bibinfo  {journal} {EPL (Europhysics Letters)}\ }\textbf
  {\bibinfo {volume} {100}},\ \bibinfo {pages} {26006} (\bibinfo {year}
  {2012})}\BibitemShut {NoStop}%
\end{thebibliography}%

\end{document}


\title{Supplemental Material: Measurement-induced power law negativity in an open monitored quantum circuit}

\author{Zack Weinstein}
\affiliation{Department of Physics, University of California, Berkeley, CA 94720, USA}

\author{Yimu Bao}
\affiliation{Department of Physics, University of California, Berkeley, CA 94720, USA}

\author{Ehud Altman}
\affiliation{Department of Physics, University of California, Berkeley, CA 94720, USA}
\affiliation{Materials Sciences Division, Lawrence Berkeley National Laboratory, Berkeley, CA 94720, USA}

\date{\today}

\maketitle

\section{Logarithmic Entanglement Negativity}
In this section, we collect and review important features of the logarithmic entanglement negativity as a measure of mixed-state entanglement. Additional discussion can be found in Refs. \cite{vidal_computable_2002,plenio_logarithmic_2005,calabrese_entanglement_2012,calabrese_entanglement_2013,lu_singularity_2019,lu_detecting_2020,wu_entanglement_2020,lu_entanglement_2020,sang_entanglement_2020,shapourian_entanglement_2020}.

Consider a state $\rho$ defined on a bipartite system $A \cup B$. If $\rho$ is a pure state -- that is, $\rho = \dyad{\psi}$ is a rank-one projector -- then the entanglement between subsystems $A$ and $B$ in state $\rho$ is canonically defined by the bipartite entanglement entropy \cite{nielsen_quantum_2010},
\begin{equation}
\label{eq:EE}
	S_A[\rho] = - \tr (\rho_A \log \rho_A) ,
\end{equation}
which is simply the von Neumann entropy of the reduced density matrix $\rho_A = \tr_B \rho$. The bipartite entanglement entropy satisfies many important desiderata of pure state entanglement measures; most importantly, it is an entanglement monotone, meaning that $S_A[\rho]$ cannot increase under local operations or classical communication (LOCC) between $A$ and $B$. This is no longer the case when $\rho$ is a mixed state, in which case the bipartite entanglement entropy is no longer a meaningful metric of entanglement \cite{bennett_mixed-state_1996,horodecki_mixed-state_1998}. Indeed, a \textit{separable} state \cite{peres_separability_1996,horodecki_separability_1996} of the form
\begin{equation}
\label{eq:separable}
	\rho = \sum_{\alpha} p_{\alpha} \rho_{A,\alpha} \otimes \rho_{B,\alpha} ,
\end{equation}
which can contain classical but not quantum correlations between $A$ and $B$, can be prepared from an initial product state using only LOCC but nevertheless can have large subsystem entropy. A dramatic example of the above is the maximally mixed state $\rho \propto \mathds{1}$, which contains neither classical nor quantum correlations between $A$ and $B$ but nevertheless yields a maximal bipartite entanglement entropy. Intuitively, $S_A$ quantifies our lack of information of subsystem $A$ upon discarding subsystem $B$; in a pure state, any lack of knowledge of $A$ necessarily arises due to its entanglement with $B$. In contrast, a mixed state may contain other sources of ignorance such as classical correlations with $B$ or classical/quantum correlations with a further environment $E$.

While there exist several viable measures of mixed state entanglement, a recently popular and easily computable choice is the entanglement negativity \cite{vidal_computable_2002,plenio_logarithmic_2005}. We first define the partial transpose $\rho^{T_B}$ of $\rho$ in region $B$ in terms of a particular product basis $\ket{ab} \equiv \ket{a}_A \otimes \ket{b}_B$ as 
\begin{equation}
	\begin{split}
		\rho^{T_B} = \qty( \sum_{ab\bar{a}\bar{b}} \rho_{ab,\bar{a}\bar{b}} \dyad{ab}{\bar{a} \bar{b}} )^{T_B} = \sum_{ab\bar{a}\bar{b}} \rho_{ab,\bar{a}\bar{b}} \dyad{a\bar{b}}{\bar{a}b} = \sum_{ab\bar{a}\bar{b}} \rho_{a\bar{b},\bar{a}b} \dyad{ab}{\bar{a}\bar{b}} .
	\end{split}
\end{equation}
In other words, we simply exchange the subsystem $B$ ket and bra labels of the matrix element $\dyad{ab}{cd}$, and extend by linearity. Just as the ordinary transpose is well-defined under a change of basis on the whole Hilbert space, the partial transpose is well-defined under a change of basis which does not mix subsystems $A$ and $B$. Importantly, while $\rho$ by definition has only positive eigenvalues, $\rho^{T_B}$ can have negative eigenvalues. While a density matrix with a positive-definite partial transpose (PPT), such as the separable state (\ref{eq:separable}), is known to have no distillable entanglement \cite{horodecki_mixed-state_1998,vidal_irreversibility_2001}, it is an open question whether a state with a non-positive-definite partial transpose (NPT) necessarily has distillable entanglement \cite{horodecki_five_2020}.

Using the partial transpose, the entanglement negativity $\mathcal{N}_{A:B}$ and the logarithmic entanglement negativity $\mathcal{E}_{A:B}$ are defined as 
\begin{equation}
	\begin{split}
		\mathcal{N}_{A:B}[\rho] &= \frac{\norm{\rho^{T_B}}_1 - 1}{2} = \sum_{\lambda_{\alpha} < 0} \abs{\lambda_{\alpha}}, \\
		\mathcal{E}_{A:B}[\rho] &= \log \norm{\rho^{T_B}}_1 = \log \qty( 1 + 2 \mathcal{N}_{A:B}[\rho] ) ,
	\end{split}
\end{equation}
where $\norm{\rho^{T_B}}_1 = \tr \sqrt{ (\rho^{T_B})^2 }$ is the trace-norm\footnote{For a general operator $\mathcal{O}$, the trace norm $\norm{\mathcal{O}}_1 = \tr \sqrt{\mathcal{O}^{\dagger} \mathcal{O}}$ is the sum of the singular values of $\mathcal{O}$.}, most easily understood here as the sum of the absolute values of the eigenvalues $\lambda_{\alpha}$ of $\rho^{T_B}$. Since $\tr \rho^{T_B} = \tr \rho = 1$, it is easy to see that $\mathcal{N}_{A:B}$ is minus the sum of the negative eigenvalues of $\rho^{T_B}$. Thus, both $\mathcal{N}_{A:B}$ and $\mathcal{E}_{A:B}$ are zero for PPT states. When $\rho$ is a pure state, $\mathcal{E}_{A:B}$ is proportional to the one-half R\'enyi entropy, $S^{(1/2)}_A[\rho] = 2 \log \tr \rho_A^{1/2}$. Unlike the entanglement entropies and R\'enyi entropies, however, both $\mathcal{N}_{A:B}$ and $\mathcal{E}_{A:B}$ remain meaningful measures of entanglement in mixed states. In particular, both $\mathcal{N}_{A:B}$ and $\mathcal{E}_{A:B}$ are entanglement monotones \cite{vidal_computable_2002,plenio_logarithmic_2005}, and $\mathcal{E}_{A:B}$ is an upper-bound to the distillable entanglement between $A$ and $B$.

Just as the R\'enyi entropies can be used to calculate the von Neumann entanglement entropy, it is useful to define R\'enyi negativities to calculate the  logarithmic negativity. The $n$th R\'enyi negativity is defined as
\begin{equation}
\label{eq:RenyiNegativity}
	\mathcal{E}^{(n)}_{A:B}[\rho] = b_n \log \qty{ \frac{\tr[ (\rho^{T_B})^n ]}{\tr \rho^n} }, \ b_n = \begin{dcases}
		\frac{1}{1-n}, & \! \! \text{odd } n \\
		\frac{1}{2-n}, & \! \!  \text{even } n 
	\end{dcases} .
\end{equation}
The normalization $b_n$ is chosen so that for pure states, the R\'enyi negativity agrees with the R\'enyi entropies $S^{(n)}_A$ and $S^{(n/2)}_A$ for odd and even $n$ respectively \cite{lu_detecting_2020}. Additionally, the factor $\tr \rho^n$ is introduced so that $\mathcal{E}^{(n)}_{A:B}$ remains zero for PPT states. By writing $\tr[ (\rho^{T_B})^{2n} ] = \sum_i \abs{\lambda_i}^{2n}$, the logarithmic entanglement negativity can be recovered from the R\'enyi negativities along the peculiar limit \cite{calabrese_entanglement_2012,calabrese_entanglement_2013}
\begin{equation}
	\mathcal{E}_{A:B}[\rho] = \underset{\text{even}}{\lim_{n \rightarrow 1}} \mathcal{E}^{(n)}_{A:B}[\rho] ,
\end{equation}
in which only the even R\'enyi negativities are used to construct the analytical continuation $n \rightarrow 1$. 

\begin{figure}[h]
	\centering
	\includegraphics{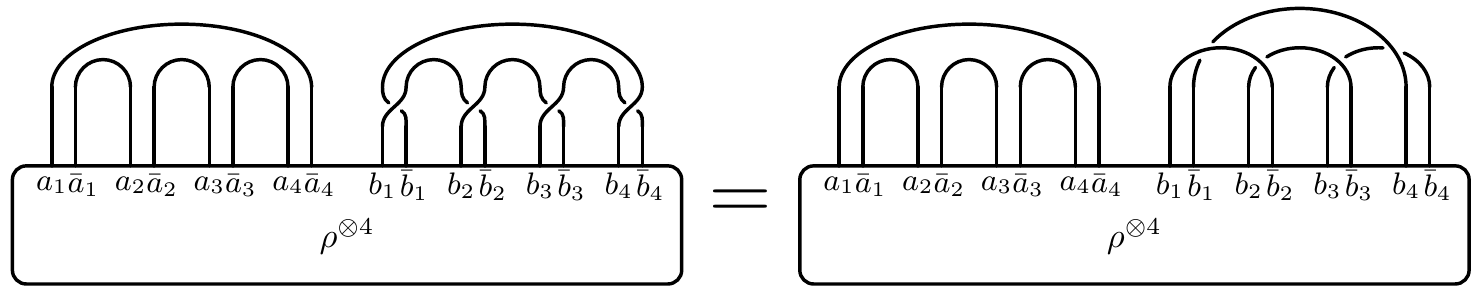}
	\caption{Tensor network representation of Eq. (\ref{eq:trRhoTbIden}) for $n=4$. Indices $a_{\ell}$ and $b_{\ell}$ represent ket indices of systems $A$ and $B$ respectively, while $\bar{a}_{\ell}$ and $\bar{b}_{\ell}$ represent bra indices; $\ell = 1,\ldots,n$ represents the $n$ copies of $\rho$: $\rho^{\otimes n} = \sum_{\qty{ a, \bar{a}, b, \bar{b} }} \rho_{a_1 b_1, \bar{a}_1 \bar{b}_1} \ldots \rho_{a_n b_n, \bar{a}_n \bar{b}_n} \ket{a_1 b_1, \ldots , a_n b_n} \bra{\bar{a}_1 \bar{b}_1, \ldots , \bar{a}_n \bar{b}_n}$.}
	\label{fig:renyiEN_TN}
\end{figure}

Similar to the R\'enyi entropies, one benefit of studying the R\'enyi negativities is that $\tr[ (\rho^{T_B})^n ]$ is linear in the $n$th moment of $\rho$, $\rho^{\otimes n}$. This makes $\mathcal{E}^{(n)}_{A:B}$ more directly amenable both to tensor network methods and to disorder averaging techniques in random circuit analyses. Specifically, we can write $\tr [(\rho^{T_B})^n]$ as
\begin{equation}
\label{eq:trRhoTbIden}
	\tr[ (\rho^{T_B})^n] = \Tr[ (C_A \otimes C_B) (\rho^{T_B})^{\otimes n} ] = \Tr[ (C_A \otimes \bar{C}_B) \rho^{\otimes n} ] ,
\end{equation}
where $\Tr$ denotes the trace over the $n$-fold replicated Hilbert space, $C_A$ is the cyclic permutation among the $n$ ket indices of subsystem $A$, and $\bar{C}_B$ is the \textit{anti}-cyclic permutation among the $n$ kets of subsystem $B$:
\begin{equation}
\label{eq:cyclic_anticyclic}
	C_A = \bigotimes_{i \in A} \qty[\sum_{a_1 \ldots a_n} \ket{a_2 a_3 \ldots a_n a_1} \bra{a_1 a_2 \ldots a_{n-1} a_n}_i], \quad \bar{C}_B = \bigotimes_{i \in B} \qty[\sum_{b_1 \ldots b_n} \ket{b_n b_1 \ldots b_{n-2} b_{n-1}} \bra{b_1 b_2 \ldots b_{n-1} b_n}_i ] .
\end{equation}
The most straightforward way to verify this is to simply express $\rho = \sum_{ab\bar{a} \bar{b}} \rho_{ab,\bar{a} \bar{b}} \ket{ab} \bra{\bar{a} \bar{b}}$ and compute both sides of (\ref{eq:trRhoTbIden}) in terms of the matrix elements $\rho_{ab,\bar{a} \bar{b}}$. It can also be visualized using tensor network methods as in Fig. \ref{fig:renyiEN_TN}.

Finally, we consider the behavior of entanglement negativity in a (Haar) random state. In a pure state, Page's theorem \cite{page_average_1993} states that the bipartite entanglement entropy in a random state on a bipartite system $A \cup B$ of $L = L_A + L_B$ qubits is equal to $\min \qty{ L_A, L_B }$ up to sub-extensive corrections -- that is, a random state will have near maximal entanglement between $A$ and $B$. Using the entanglement negativity, we may extend the intuition from Page's theorem to mixed states by considering random states on the \textit{tripartite} system $A \cup B \cup C$ of $L=L_A+L_B+L_C$ qubits. The extensive contribution to the negativity between $A$ and $B$ may be understood by the following heuristic argument \cite{lu_entanglement_2020}, which can be verified by more sophisticated means \cite{bhosale_entanglement_2012}. First, consider the case $L_A + L_B > L_C$, in which the ``system'' is greater in size than the ``bath''. By Page's theorem, we expect approximately $L_C$ Bell pairs to be shared between $AB$ and $C$, leaving $L_{AB} - L_C$ unpaired qubits in $AB$. Assuming $L_{A} = L_{B} = L_{AB}/2$ for simplicity, we expect the logarithmic negativity to be equal to the number of Bell pairs which can be formed between $A$ and $B$, $\mathcal{E}_{A:B} \sim \frac{1}{2}(L_{AB} - L_C)$, up to sub-extensive corrections. On the other hand, when $L_{AB} < L_C$ and the ``system'' is smaller than the ``bath'', Page's theorem predicts that $L_{AB}$ Bell pairs will be shared between $AB$ and $C$. This leaves no unpaired qubits in the system to pair with each other, and we expect no extensive contribution to the negativity. Altogether,
\begin{equation}
\label{eq:app_EN_random_state}
	\mathcal{E}_{A:B}[\rho] \sim \begin{cases}
		\frac{1}{2} (L_A + L_B - L_C), & L_A + L_B > L_C \\
		o(L), & L_A + L_B < L_C 
	\end{cases},
\end{equation}
again up to sub-extensive corrections.

\section{Mixed State Clifford Simulation}
In this section, we review the tools required for efficient classical simulation of random Clifford circuits \cite{gottesman_heisenberg_1998,aaronson_improved_2004}, as employed in the main text. While recent works have reviewed the technique in the context of pure-state dynamics \cite{nahum_quantum_2017,li_measurement-driven_2019}, fewer works thus far have utilized mixed-state dynamics. We therefore focus the most attention on details specific to mixed-state Clifford simulation.

We consider a system of $L$ $d=2$ qubits. Let $\mathcal{P}^L_+$ be the set of Pauli strings squaring to one -- that is, $g \in \mathcal{P}^L_+$ is an $L$-fold tensor product of $1$ and the Pauli matrices $X = \sigma^x$, $Y = \sigma^y$, and $Z = \sigma^z$, together with the phases $\pm 1$. A \textit{stabilizer group} $\mathcal{S}$ is a subset of $\mathcal{P}^L_+$ forming an Abelian group \cite{gottesman_heisenberg_1998,nielsen_quantum_2010}. Since the elements $g \in \mathcal{S}$ all commute, they can be simultaneously diagonalized; the simultaneous $+1$ eigenspace of the elements of $\mathcal{S}$ form the \textit{stabilizer space} $V_{\mathcal{S}}$. A state $\ket{\psi} \in V_{\mathcal{S}}$ is said to be stabilized by $\mathcal{S}$, and its elements $g \in \mathcal{S}$ are referred to as stabilizers.

As an Abelian group, $\mathcal{S}$ is generated by a smaller number of generators $\mathcal{G} = \qty{g_1, \ldots , g_k}$, such that an arbitrary stabilizer element can be written as
\begin{equation}
    g = g_1^{r_1} \ldots g_k^{r_k}, \quad \vec{r} = (r_1, \ldots , r_k) \in \qty{0,1}^{\otimes k} .
\end{equation}
Written this way, $\mathcal{S}$ can be understood as a vector space over the finite field $\mathbb{F}_2 = \qty{0,1}$ with (non-unique) basis elements $g_{\ell} \in \mathcal{G}$. It can be proven by elementary means \cite{nielsen_quantum_2010} that $V_{\mathcal{S}}$ is a $2^{L-k}$ dimensional subspace of the full Hilbert space, so that $k = L$ stabilizers are required to specify a unique stabilizer state $\ket{\psi}$ up to a phase. When $k < L$ there are a priori many possible mixed states one can form with support on $V_{\mathcal{S}}$; we will be particularly concerned with the uniquely defined stabilizer mixed state $\rho$ given by
\begin{equation}
    \rho = \frac{2^k}{2^L} \prod_{i = 1}^k \qty( \frac{1 + g_i}{2} ) = \frac{1}{2^L} \sum_{g \in \mathcal{S}} g ,
\end{equation}
which can be understood as an equal weight statistical mixture of all states in $V_{\mathcal{S}}$. Note that $\rho$ reduces to $\dyad{\psi}$ when $k=L$. More generally, the purity can be immediately read off as $\tr \rho^2 = 2^{k-L}$ for general $k$.

Any stabilizer state of the above form can be efficiently represented by $\mathcal{O}(kL)$ classical bits specifying the generating set $\mathcal{G}$, a big improvement over the a priori $2^{kL}$ complex numbers required for an arbitrary state \cite{gottesman_heisenberg_1998,aaronson_improved_2004}. So long as $\rho$ remains a stabilizer state under time evolution, the evolution of $\rho$ can be efficiently simulated on a classical computer by keeping track of the evolution of $\mathcal{G}$ alone. The set of unitary gates which map Pauli strings to Pauli strings under conjugation, and therefore map stabilizer states to stabilizer states, is known as the Clifford group:
\begin{equation}
    \text{Cl}(L) = \qty{ U \in \text{U}(L) : U g U^{\dagger} \in \mathcal{P}^L_+ \text{ for } g \in \mathcal{P}^L_+ } .
\end{equation}
Under time evolution by a Clifford gate $U \in \text{Cl}(L)$, the stabilizers $g$ (including their generators) evolve to $U g U^{\dagger}$. It turns out that the Clifford group can be completely generated by Hadamard, CNOT, and phase gates \cite{aaronson_improved_2004,nielsen_quantum_2010}.

Measurements of Pauli string observables, including single qubit measurements in the computational basis, also map stabilizer states to stabilizer states \cite{aaronson_improved_2004}. Suppose the Pauli string $h$ is measured on the stabilizer state $\rho$. Then, there are two possibilities: either $h$ or $-h$ is contained in $\mathcal{S}$ and the measurement outcome is deterministic, or $h$ is not contained in $\mathcal{S}$ and the measurement outcome is random and yields $\pm 1$ with equal probability. In the latter case, if $k=L$ and $\rho = \dyad{\psi}$ is a pure state, then $h$ must anticommute with at least one generator $g_{\ell} \in \mathcal{G}$. By performing a change of basis, we can always arrange that no other generators anticommute with $h$, in which case the new generating set following the measurement is obtained by replacing $g_{\ell}$ with $\pm h$ depending on the measurement outcome. On the other hand, if $k < L$ and $\rho$ is mixed, then it is possible for $h$ to commute with all of $\mathcal{S}$ without being an element of $\mathcal{S}$ itself; in this case, $\pm h$ is simply added to the list of generators following the measurement, and the purity of $\rho$ is increased.

The dephasing channel $\mathcal{D}_i[\rho] = \sum_{a = 0}^1 P^a_i \rho P^a_i$ on the $i$th qubit in the computational basis, as considered in the main text, also preserves stabilizer states. To determine its action on $\rho$, it is first useful (and always possible) to perform a change of basis on $\mathcal{S}$ so that $\mathcal{G}$ contains at most one generator $g_{\ell}$ which anticommutes with $Z_i$. Then, the projection operators $P^a_i$ commute with all other generators, and their action on $g_{\ell}$ is given by
\begin{equation}
    \sum_{a = 0}^1 P^a_i \qty( \frac{1 + g_{\ell}}{2} ) P^a_i = \sum_{\sigma = \pm 1} \qty( \frac{1 + \sigma Z_i}{2} ) \qty( \frac{1 + g_{\ell}}{2} ) \qty( \frac{1 + \sigma Z_i}{2} ) = \frac{1}{2} ,
\end{equation}
where we've used $\acomm{Z_i}{g_{\ell}} = 0$. In this way, each dephasing channel eliminates at most one generator from the generating set $\mathcal{G}$, decreasing the purity by a factor of two. On the other hand, if $\mathcal{S}$ contains no stabilizers which anticommute with $Z_i$, then $\rho$ is unaffected by dephasing.

The von Neumann entropy of a stabilizer state can be efficiently computed from the stabilizer group. Letting $\mathcal{G}_R$ be a generating set for the subgroup $\mathcal{S}_R$ of $\mathcal{S}$ containing elements with trivial support outside of the region $R$, it is straightforward to show \cite{hamma2005bipartite,hamma2005ground,nahum_quantum_2017,li_measurement-driven_2019} that the entanglement spectrum is flat, and therefore  the von Neumann entropy $S_R[\rho]$ and all R\'enyi entropies $S^{(n)}_R[\rho]$ are equal and given by
\begin{equation}
    S_R[\rho] = S_R^{(n)}[\rho] = L_R - \abs{\mathcal{G}_R} ,
\end{equation}
where $L_R$ is the number of qubits in region $R$, and $\abs{\mathcal{G}_R}$ is the number of generators of $\mathcal{S}_R$. 

The logarithmic negativity can also be computed efficiently from the stabilizer group \cite{sang_entanglement_2020}. Dividing the total system of $L$ qubits into subsystems $A$ and $B$, Let $g_{\ell} \eval_A$ be the generator $g_{\ell}$ restricted to subsystem $A$ -- that is, all nontrivial Pauli content of $g_{\ell}$ on subsystem $B$ is set to the identity. Upon restricting each generator to $A$, it is no longer guaranteed that the generators commute; defining the matrix $K$ with entries $K_{\ell \ell'}$ to equal $+1$ if $g_{\ell}$ and $g_{\ell'}$ commute and $-1$ if $g_{\ell}$ and $g_{\ell'}$ anticommute, the negativity $\mathcal{E}_{A:B}$ is given by
\begin{equation}
    \mathcal{E}_{A:B} = \frac{1}{2} \text{rank } K .
\end{equation}

\section{Additional Numerical Results}
In this section, we provide additional numerical results.

\subsection{Mutual Information}
\begin{figure}[h]
    \centering
    \includegraphics[width=0.5\linewidth]{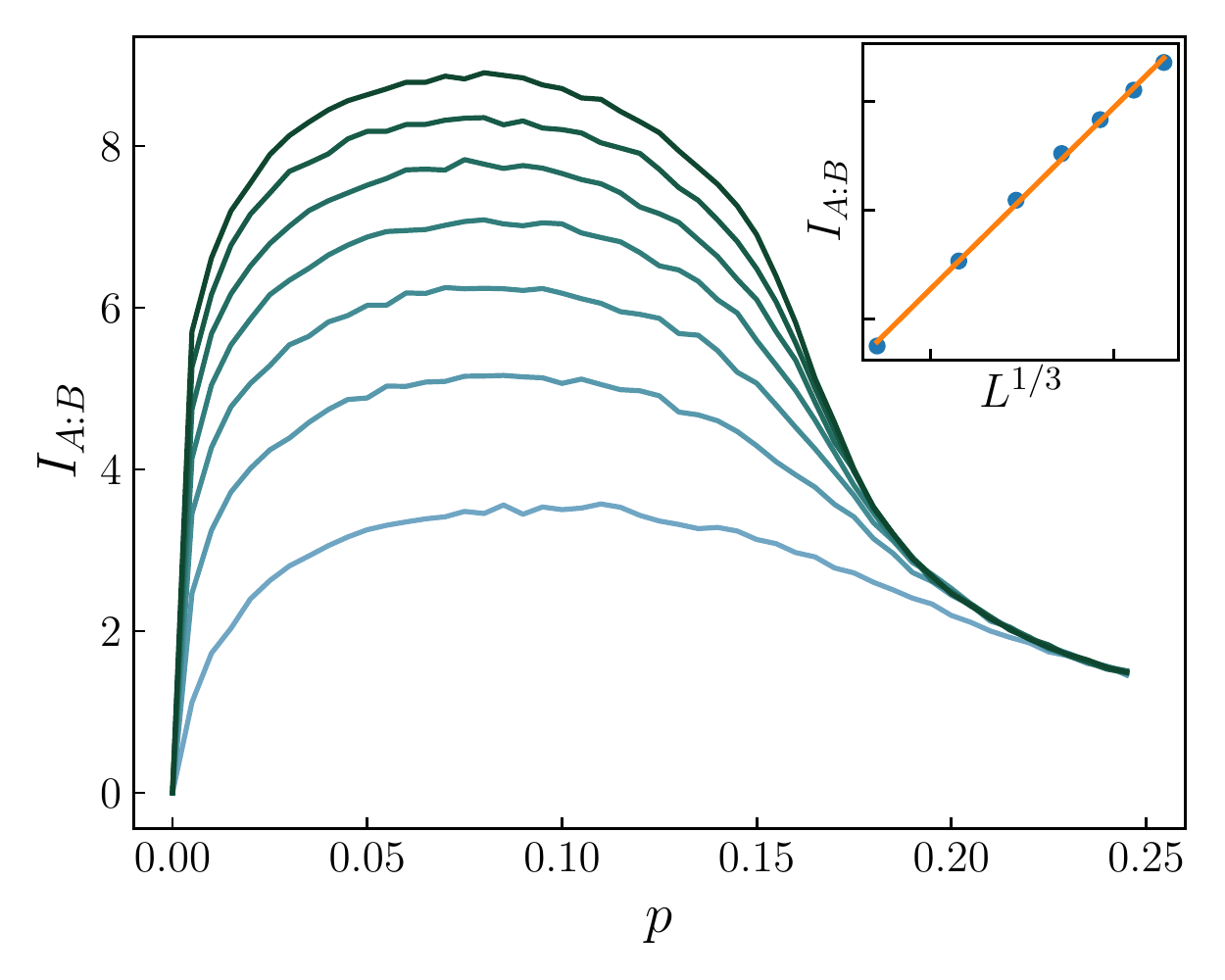}
    \caption{Late time mutual information between subsystems $A$ and $B$ as a function of measurement rate $p$, for various system sizes $L$ ranging from 40 (light blue) to 280 (dark green). Inset: mutual information as a function of $L^{1/3}$ (blue dots) at $p=0.1$, along with the fitting curve $y=b_1 L^{1/3} + b_2$ (orange line) with $b_1 \approx 1.670$ and $b_2 \approx -2.134$. The numerical results are averaged over 200 random circuit realizations.}
    \label{fig:MI_numerics}
\end{figure}

In the discussion of the main text, we mention that the mutual information $I_{A:B} = S_A + S_B - S_{AB}$ between the two halves of the system obeys qualitatively similar scaling to the logarithmic negativity. The mutual information as a function of measurement rate is shown in Fig. \ref{fig:MI_numerics}, which can be compared with Fig. 1 of the main text. We note that the mutual information is, roughly speaking, just over twice the size of the negativity, which is in line with the general bounds derived in Ref. \cite{sang_entanglement_2020}.

\subsection{Finite-Size Scaling}
\begin{figure}[h]
    \centering
    \includegraphics[width=0.7\linewidth]{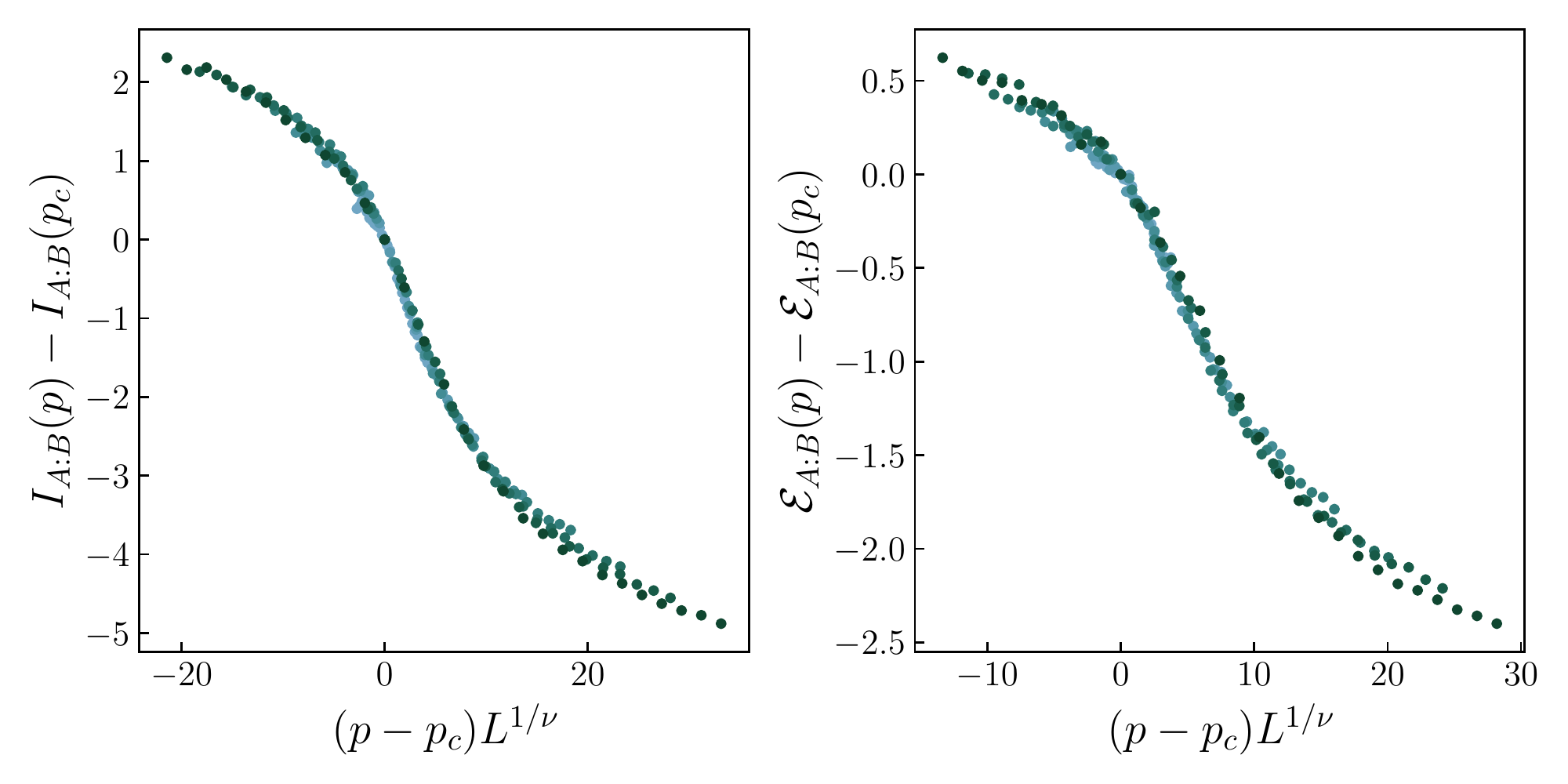}
    \caption{Finite-size scaling data collapses of the mutual information $I_{A:B}$ and the logarithmic negativity $\mathcal{E}_{A:B}$, for seven system sizes ranging from $L=40$ (light blue) to $L=280$ (dark green), for measurement rates $0.1 \leq p \leq 0.245$. The scaling collapse of the mutual information using the scaling formula $I_{A:B}(p) - I_{A:B}(p_c) = F((p-p_c) L^{1/\nu})$ yields the critical point $p_c \simeq 0.16$ and correlation length exponent $\nu \simeq 0.94$. A similar scaling collapse of the logarithmic negativity yields $p_c \simeq 0.15$ and $\nu \simeq 0.99$.}
    \label{fig:scaling_collapse}
\end{figure}

To determine the critical point $p_c$ and critical exponent $\nu$ of the measurement-induced transition, we perform a finite-size scaling analysis. Since the mutual information obeys similar scaling to the negativity but is larger in size, it is convenient to extract $p_c$ and $\nu$ from the mutual information rather than from the negativity. Using the scaling formula $I_{A:B}(p) - I_{A:B}(p_c) = F((p-p_c) L^{1/\nu})$ proposed in Ref. \cite{skinner2019measurement}, we perform a two-parameter fit following the method in Ref.~\cite{choi_quantum_2020} and find the best scaling collapse for $p_c \simeq 0.16$ and $\nu \simeq 0.94$. Scaling collapses for both the mutual information and negativity are shown in Fig. \ref{fig:scaling_collapse}.

While $p_c$ is consistent with the critical points found in previous works \cite{zabalo2020critical,li_quantum_2018,li_measurement-driven_2019}, $\nu$ appears quite different. This may be due to the numerical smallness of the mutual information, or it may indicate new universal features in the absence of a volume law contribution to the negativity or mutual information.

The logarithmic negativity yields slightly different values for $p_c$ and $\nu$ than the mutual information. From the negativity, we obtain $p_c \simeq 0.15$ and $\nu \simeq 0.99$. Since the negativity is smaller in size, we take $p_c$ and $\nu$ as extracted from the mutual information as our putative transition point and correlation length exponent.

\section{Effective statistical mechanics model}
In this section, we provide explicit details of our calculation mapping the negativity onto the free energy of directed polymers in a random environment. Much of the mapping follows identically to previous works \cite{zhou_emergent_2019,bao_theory_2020,jian_measurement-induced_2020,li_entanglement_2021,agrawal_entanglement_2021} mapping the pure state bipartite entanglement entropy to similar statistical mechanics models; the new features present in this work are the application of these techniques to the negativity, as well as the particular method relating the effective statistical mechanics model to directed polymers in a random environment. Specifically, we follow the approach of Ref. \cite{agrawal_entanglement_2021} and leave the spacetime locations of measurements as unaveraged quenched disorder. Then, in the limit of large qudit dimension $d \rightarrow \infty$, domain walls in the effective ferromagnetic spin model can be viewed as directed polymers and measurement locations can be viewed as an attractive random potential on the polymers. This approach shows naturally how Kardar-Parisi-Zhang (KPZ) fluctuations in the polymer free energy arise from the addition of measurements to the circuit.
%

\subsection{Setup}
We consider an open chain of $L$ $d$-dimensional qudits, each with a local Hilbert space $\mathcal{H}_1 = \text{span} \qty{ \ket{a} }_{a= 0}^{d-1}$. We take $L$ to be even for convenience, and we are primarily interested in the thermodynamic limit $L \rightarrow \infty$. Starting from an initial product state $\rho_0 = \dyad{0}^{\otimes L}$, we apply the quantum circuit shown in Fig. 1(a) of the main text: At each time step $t$, we first apply a layer of two-qudit Haar random unitary gates $U_{t,ij}$ on nearest neighbor qudits $i$ and $j = i+1$. Then, each $i$th qudit is measured in the computational basis with probability $p$, which collapses the system onto the state
\begin{equation}
	\rho \mapsto \frac{P^a_i \rho P^a_i}{\tr(P^a_i \rho)}
\end{equation}
with probability $\tr (P^a_i \rho)$ according to the Born rule, where $P^a_i = \dyad{a}_i$ projects the $i$th qudit onto the state $\ket{a}$. Finally, on each even time step the boundary qudits $i = 1$ and $i = L$ are subjected to the dephasing channel
\begin{equation}
\label{eq:dephasing}
	\mathcal{D}_i[\rho] = \sum_{a = 0}^{d-1} P^a_i \rho P^a_i ,
\end{equation}
which can be understood as a measurement in which we directly average the density matrix over all possible measurement outcomes. This is to be contrasted with the projective measurements occurring at every qudit, in which the average over measurement outcomes is performed only after computing an observable (namely, the negativity) which is nonlinear in the density matrix.

Each circuit realization is specified by the realization of random unitary gates $\mathcal{U} = \qty{U_{t,ij}}$, the spacetime locations $\vec{X}$ of measurements in the circuit, and the measurement trajectory $\vec{m}$ giving the outcomes of those measurements. Each unitary gate and measurement location may be sampled independently, while the probability of each measurement outcome is conditioned both on the unitary realization $\mathcal{U}$ and the measurement locations $\vec{X}$, as well as all prior measurements in the trajectory. After $T = \mathcal{O}(L)$ time steps, each circuit realization yields the (unnormalized) state
\begin{equation}
	\rho_{\vec{m}} = \rho_{\vec{m}}(\mathcal{U},\vec{X},T) = \mathcal{D} \qty{ P_T U_T \ldots \mathcal{D} \qty{ P_2 U_2 P_1 U_1 \rho_0 U_1^{\dagger} P_1 U_2^{\dagger} P_2 } \ldots U_T^{\dagger} P_T },
\end{equation}
where $U_t$ is the product of two-qudit unitary gates within the $t$th layer, $P_t$ projects onto the outcomes of measurements performed immediately after the $t$th layer of unitaries, and $\mathcal{D}$ is the dephasing channel (\ref{eq:dephasing}) applied to qudits $i=1$ and $i=L$. For a fixed configuration of unitary gates and measurement locations, each measurement trajectory $\vec{m}$ is achieved with probability
\begin{equation}
	p_{\vec{m}} = p_{\vec{m}}(\mathcal{U},\vec{X},T) = \tr \rho_{\vec{m}} .
\end{equation}
The properly normalized density matrix is thus $\rho_{\vec{m}} / p_{\vec{m}}$. It is simple to verify that $\sum_{\vec{m}} p_{\vec{m}} = 1$. Hereafter, we drop the explicit dependence on $\mathcal{U}$, $\vec{X}$, and $T$ in $\rho_{\vec{m}}$ and $p_{\vec{m}}$ and other related quantities.

\subsection{Replica Method}

We are interested in the entanglement between subsystems $A$ and $B$ of our qudit chain, taken to be the left and right halves of the chain respectively. Our central quantity of interest is the $n$th R\'enyi negativity (\ref{eq:RenyiNegativity}), from which we can obtain the logarithmic negativity by performing the limit $n \rightarrow 1$ along even $n$. For sake of comparison with previous works \cite{zhou_emergent_2019,bao_theory_2020,jian_measurement-induced_2020,li_entanglement_2021,agrawal_entanglement_2021}, we will also compute the R\'enyi entropy $S^{(n)}_A[\rho] = (1-n)^{-1} \log \tr \rho_A^n$ of subsystem $A$, despite its poor utility as a mixed state entanglement measure. The R\'enyi entropy can also be used to compute the mutual information, $I_{A:B} = S_A + S_B - S_{AB}$, which shares similar qualitative features to the negativity despite also failing as a mixed state entanglement measure.

The R\'enyi negativity and entropy for a \textit{fixed} set of measurement locations $\vec{X}$, averaged over unitary realizations $\mathcal{U}$ and measurement trajectories $\vec{m}$, are given by
\begin{equation}
\label{eq:overline_EN_EE}
	\overline{\mathcal{E}^{(n)}_{A:B}(\vec{X})} = \mathbb{E}_{\mathcal{U}} \sum_{\vec{m}} p_{\vec{m}} \frac{1}{2-n} \log \qty{ \frac{\tr[(\rho_{\vec{m}}^{T_B})^n]}{\tr \rho_{\vec{m}}^n} } , \quad \overline{S^{(n)}_A(\vec{X})} = \mathbb{E}_{\mathcal{U}} \sum_{\vec{m}} p_{\vec{m}} \frac{1}{1-n} \log \qty{ \frac{ \tr \rho_{A,\vec{m}}^n}{(\tr \rho_{\vec{m}})^n} } ,
\end{equation}
where $\overline{\cdot}$ denotes the averages over unitary realizations and measurement trajectories, but \textit{not} measurement locations. The fully averaged entropy and negativity are
\begin{equation}
	\fullavg{\mathcal{E}^{(n)}_{A:B}}  = \sum_{\vec{X}} (1-p)^{LT-\abs{\vec{X}}} p^{\abs{\vec{X}}} \overline{\mathcal{E}^{(n)}_{A:B}(\vec{X})}, \quad \fullavg{ S^{(n)}_A}  = \sum_{\vec{X}} (1-p)^{LT-\abs{\vec{X}}} p^{\abs{\vec{X}}} \overline{S^{(n)}_A(\vec{X})} ,
\end{equation}
where $\abs{\vec{X}}$ denotes the number of measurement locations. Following Ref. \cite{agrawal_entanglement_2021}, and in contrast to previous works \cite{bao_theory_2020,jian_measurement-induced_2020}, we will leave the locations of measurements as quenched disorder rather than performing their average. This has the benefit of resulting in simpler Boltzmann weights in the effective spin model. Previously, averaging over measurement locations led to the interpretation of the measurement rate $p$ as an effective temperature which drives the effective model through a finite temperature phase transition at a critical measurement rate \cite{bao_theory_2020,jian_measurement-induced_2020}, resulting in the measurement-induced phase transition in the entanglement entropy. In contrast, leaving the measurement locations as quenched disorder partially obscures the mechanism behind this transition. Nevertheless, this approach will be particularly well suited towards analyzing the ferromagnetic phase of the effective model, which is our primary concern in this work.

The quantities (\ref{eq:overline_EN_EE}) are structurally similar; the numerator and denominator inside the logarithm in both cases can be written using an $n$-fold replicated density matrix $\rho_{\vec{m}}^{\otimes n}$ in the form\footnote{Throughout, we use $\tr$ to denote the trace over the original Hilbert space, $\Tr$ to denote the trace over the $n$-fold replicated Hilbert space, and $\textbf{Tr}$ to denote the trace over the $r=nk+1$-fold replicated Hilbert space.}
\begin{equation}
	Z^{(n)}_{\Sigma} = \Tr \qty[ \Sigma \rho_{\vec{m}}^{\otimes n} ] ,
\end{equation}
which depends implicitly on the full circuit realization $(\mathcal{U},\vec{X}, \vec{m},T)$, and explicitly on a ``boundary operator'' $\Sigma = \Sigma_A \otimes \Sigma_B$ which affects a permutation among the $n$ ket indices at each qudit\footnote{Note the abuse of notation: we are using $\Sigma_A$ to stand for both the permutation itself, and for the operator which implements this permutation on the kets of subsystem $A$. The meaning should be clear from context.}:
\begin{equation}
	\Sigma_A = \bigotimes_{i \in A} \qty[ \sum_{a_1, \ldots , a_n = 0}^{d-1} \ket{a_{\Sigma_A(1)} \ldots a_{\Sigma_A(n)}} \bra{a_1 \ldots a_n}_i ] ,
\end{equation}
and similarly for $\Sigma_B$. The entropy and negativity can then be written
\begin{equation}
	\overline{\mathcal{E}^{(n)}_{A:B}(\vec{X})} = \frac{1}{2-n} \mathbb{E}_{\mathcal{U}} \sum_{\vec{m}} p_{\vec{m}} \log \qty{ \frac{Z^{(n)}_{\mathcal{E}}}{Z^{(n)}_{\mathcal{E}0}} }, \quad \overline{S^{(n)}_A(\vec{X})} = \frac{1}{1-n} \mathbb{E}_{\mathcal{U}} \sum_{\vec{m}} p_{\vec{m}} \log \qty{ \frac{Z^{(n)}_S}{Z^{(n)}_{S0}} } ,
\end{equation}
where $Z^{(n)}_{\mathcal{E}}$, $Z^{(n)}_{\mathcal{E}0}$, $Z^{(n)}_S$, and $Z^{(n)}_{S0}$ are defined using the four boundary operators
\begin{equation}
\label{eq:Boundary_Op}
	\begin{split}
	\Sigma_{\mathcal{E}} &= C_A \otimes \bar{C}_B, \quad \Sigma_S = C_A \otimes I_B, \\
	\Sigma_{\mathcal{E}0} &= C_A \otimes C_B, \quad \Sigma_{S0} = I_A \otimes I_B .
	\end{split}
\end{equation}
The permutation $C \equiv (1 \ 2 \ \ldots \ n)$ represents the cyclic permutation, while $\bar{C} \equiv C^{-1} = (1 \ n \ \ldots \ 2)$ represents the anti-cyclic permutation, as given in Eq. (\ref{eq:cyclic_anticyclic}). $I = \mathds{1}^{\otimes n}$ is the identity operator on the $n$-fold replicated Hilbert space, which corresponds to the identity permutation.

In order to perform the average over unitary realizations and measurement locations, we employ the replica trick \cite{kardar_statistical_2007,nishimori_statistical_2001}. For each of the four quantities $Z^{(n)}_{\Sigma}$,
\begin{equation}
	\mathbb{E}_{\mathcal{U}} \sum_{\vec{m}} p_{\vec{m}} \log Z^{(n)}_{\Sigma} = \lim_{k \rightarrow 0} \frac{1}{k} \log \qty{ \mathbb{E}_{\mathcal{U}} \sum_{\vec{m}} p_{\vec{m}} \qty(Z^{(n)}_{\Sigma})^k } \equiv \lim_{k \rightarrow 0} \frac{1}{k} \log Z^{(n,k)}_{\Sigma}(\vec{X}) ,
\end{equation}
where
\begin{equation}
\label{eq:Znk}
	Z^{(n,k)}_{\Sigma}(\vec{X}) = \mathbb{E}_{\mathcal{U}} \sum_{\vec{m}} p_{\vec{m}} \qty(Z^{(n)}_{\Sigma})^k = \mathbb{E}_{\mathcal{U}} \sum_{\vec{m}} \textbf{Tr} \qty[ (\Sigma^{\otimes k} \otimes \mathds{1}) \rho_{\vec{m}}^{\otimes nk+1} ] = \textbf{Tr} \qty{ (\Sigma^{\otimes k} \otimes \mathds{1}) \qty[ \mathbb{E}_{\mathcal{U}} \sum_{\vec{m}} \rho_{\vec{m}}^{\otimes nk+1} ] } .
\end{equation}
We therefore obtain formal expressions for the R\'enyi negativity $\mathcal{E}^{(n)}_{A:B}$ and the R\'enyi entropy $S^{(n)}_A$ as the $k\rightarrow 0$ limits of the replica negativity and entropy, $\mathcal{E}^{(n,k)}_{A:B}$ and $S^{(n,k)}_A$:
\begin{equation}
\label{eq:replicaSE}
	\mathcal{E}^{(n,k)}_{A:B}(\vec{X}) = \frac{1}{k(2-n)} \log \qty{ \frac{Z^{(n,k)}_{\mathcal{E}}(\vec{X})}{Z^{(n,k)}_{\mathcal{E}0}(\vec{X})} }, \quad S^{(n,k)}_A(\vec{X}) = \frac{1}{k(1-n)} \log \qty{ \frac{Z^{(n,k)}_S(\vec{X})}{Z^{(n,k)}_{S0}(\vec{X})} } .
\end{equation}
We will soon see that each $Z^{(n,k)}_{\Sigma}$ can be thought of as a partition function of spins $\sigma$ taking values in the permutation group $S_r$ for $r \equiv nk+1$. These partition functions describe identical spin models in the bulk, and only differ at the final time boundary conditions. In this case, each of $\mathcal{E}^{(n,k)}_{A:B}$ and $S^{(n,k)}_A$ can be thought of as the difference in free energies between two particular boundary conditions of a spin model:
\begin{equation}
	\mathcal{E}^{(n,k)}_{A:B}(\vec{X}) = \frac{1}{k(n-2)} \qty[ F^{(n,k)}_{\mathcal{E}}(\vec{X}) - F^{(n,k)}_{\mathcal{E}0}(\vec{X}) ], \quad S^{(n,k)}_A(\vec{X}) = \frac{1}{k(n-1)} \qty[ F^{(n,k)}_S(\vec{X}) - F^{(n,k)}_{S0}(\vec{X}) ] .
\end{equation}

\subsection{Averaging}
\begin{figure}
\centering
\includegraphics{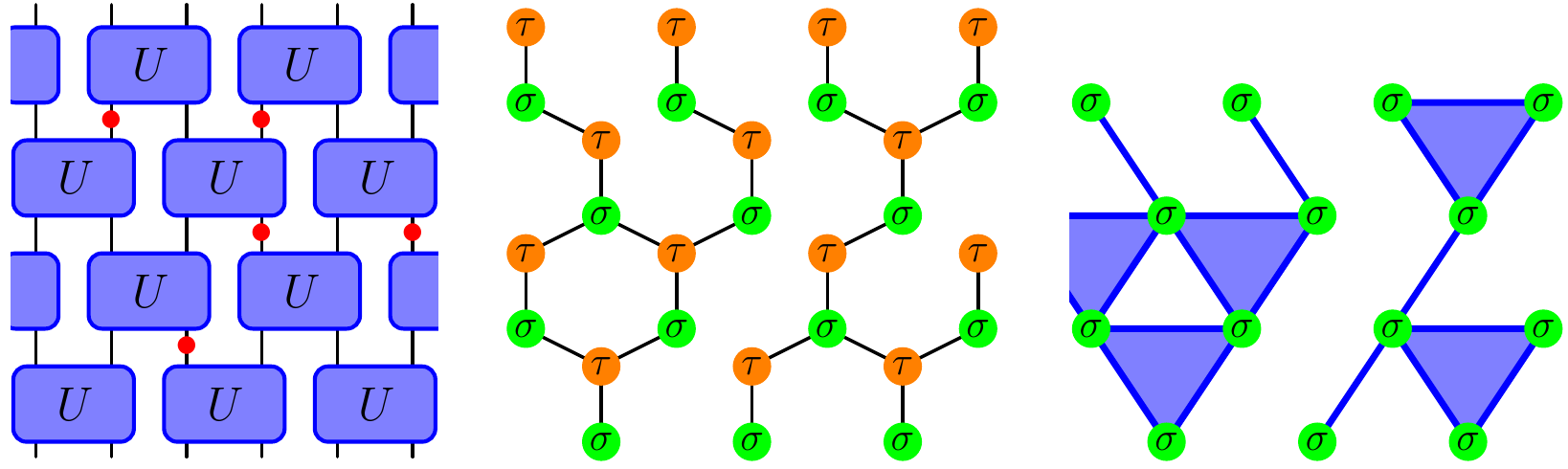}
	\caption{Outline of the mapping to the effective statistical mechanics model. Starting with the bulk circuit (left), performing the average over unitary gates in Eq. (\ref{eq:Znk}) yields two permutation-valued spins $\sigma$ and $\tau$ for each gate (center); vertical bonds are weighted by the Weingarten function in Eq. (\ref{eq:weingarten_moeb}), while diagonal bonds are weighted by (\ref{eq:diagonal_weight}) in the absence of a measurement. In the presence of a measurement, the diagonal bonds are instead missing. After integrating out the $\tau$ spins, we obtain a model with three-body weights on downward facing triangles (right) in the absence of measurements. The three-body weight is replaced by a two-body weight when one of the bonds is measured.}
	\label{fig:outline}
\end{figure}

As shown in Eq. (\ref{eq:Znk}), the quantities $Z^{(n,k)}_{\Sigma}(\vec{X})$ are linear in $\rho_{\vec{m}}^{\otimes r}$ with $r \equiv nk+1$, allowing the averages over random unitary gates and the sum over measurement outcomes to be performed. The result is the identification of each $Z^{(n,k)}_{\Sigma}$ as the partition function of an effective statistical mechanics model of permutation-valued spins. An outline of this mapping is shown in figure \ref{fig:outline}.

For notational purposes, it is convenient to first introduce a vectorized density matrix $\ket*{\rho^{(r)}_{\vec{m}}}$ with matrix elements
\begin{equation}
	\bra{\vec{a}_1 \bar{\vec{a}}_1, \ldots , \vec{a}_L \bar{\vec{a}}_L} \ket*{\rho^{(r)}_{\vec{m}}} = \bra{\vec{a}_1, \ldots , \vec{a}_L} \rho^{\otimes r}_{\vec{m}} \ket{\bar{\vec{a}}_1, \ldots , \bar{\vec{a}}_L} ,
\end{equation}
where $\vec{a}_i = (a_{1,i}, \ldots , a_{r,i} )$ and $\bar{\vec{a}}_i = (\bar{a}_{1,i}, \ldots , \bar{a}_{r,i} )$ respectively give the ket and bra indices for the $r$ copies of the $i$th qudit. Each of the traces of Eq. (\ref{eq:Znk}) can then be written in the form $\bra{\sigma_1 \ldots \sigma_L} \ket*{\rho^{(r)}_{\vec{m}}}$ \cite{you2018entanglement}, where $\ket{\sigma}$ for a permutation $\sigma \in S_r$ is a state on the $2r$-fold replicated Hilbert space $(\mathcal{H}_1 \otimes \mathcal{H}_1^*)^{\otimes r}$ of a single qudit, with matrix elements
\begin{equation}
	\bra{\vec{a} \bar{\vec{a}}} \ket{\sigma} = \bra{a_1 \bar{a}_1, \ldots ,a_r \bar{a}_r } \ket{\sigma} = \prod_{\ell = 1}^r \delta_{a_{\ell}, \bar{a}_{\sigma(\ell)}} .
\end{equation}
We will discuss the precise form of the boundary conditions below. The states $\ket{\sigma}$ will prove useful in defining the effective statistical mechanics model. In particular, the average over each unitary $U_{t,ij}$ yields \cite{collins_moments_2003,collins2006integration,nahum_quantum_2017,nahum_operator_2018,zhou_emergent_2019,bao_theory_2020,jian_measurement-induced_2020,agrawal_entanglement_2021}
\begin{equation}
	\mathbb{E}_{\mathcal{U}} (U_{t,ij} \otimes U_{t,ij}^*)^{\otimes r} = \sum_{\sigma, \tau \in S_r} \Wg(\sigma \tau^{-1}) \ket{\tau \tau} \bra{\sigma \sigma}_{ij} ,
\end{equation}
where $\Wg(\sigma)$ is the Weingarten function \cite{collins_moments_2003,collins2006integration}. It has a known (but complicated) closed form expression, and the following asymptotic expansion for large $d$ \cite{zhou_emergent_2019,collins2006integration}:
\begin{equation}
\label{eq:weingarten_moeb}
	\Wg(\sigma) = \frac{1}{d^{2r}} \qty[ \frac{\text{Moeb}(\sigma)}{d^{2\abs{\sigma}}} + \mathcal{O}(d^{-2\abs{\sigma}-4}) ] ,
\end{equation}
where $\abs{\sigma}$ is the number of transpositions required to construct $\sigma$ from the identity permutation $\mathbb{I}$, and $\text{Moeb}(\sigma)$ is the Moebius number of $\sigma$ \cite{collins2006integration}. In particular, $\Wg(\sigma)$ is maximized when $\sigma = \mathbb{I}$, and other values are suppressed to order $1/d^2$. Note, however, that the Weingarten function is not strictly positive.

The permutation valued ``spins'' $\sigma$ and $\tau$ will form the degrees of freedom for our statistical mechanics model. For a given configuration of spins, the remainder of the tensor network representing $\rho_{\vec{m}}^{\otimes r}$ is simple to evaluate: between two diagonally adjacent permutation spins, if no measurement is present, the tensor contraction between adjacent spins yields
\begin{equation}
\label{eq:diagonal_weight}
	w_d(\sigma, \tau) = \bra{\sigma} \ket{\tau} = d^{r - \abs{\sigma^{-1}\tau}} .
\end{equation}
On the other hand, if a measurement is present, we instead obtain
\begin{equation}
	w_d'(\sigma,\tau) = \bra{\sigma} (P^a \otimes P^a)^{\otimes r} \ket{\tau} = 1
\end{equation}
regardless of the measurement outcome $a$ or the values of the spins $\sigma$ and $\tau$. Since each Boltzmann weight no longer depends on the particular measurement outcome after performing the Haar average, the sum $\sum_{\vec{m}} = d^{\abs{\vec{X}}}$ cancels between the numerator and denominator in each expression (\ref{eq:replicaSE}) and can be ignored hereafter.

In the bulk of the circuit, we obtain a partition function of permutation-spins on the honeycomb lattice with nearest neighbor interactions. The interaction on vertical bonds is given by the Weingarten function, while the interaction on diagonal bonds is given either by $w_d$ or $w_d'$ depending on the absence or presence of a measurement. Due to the Weingarten function, these Boltzmann weights are not strictly positive. Some simplification arises when the $\tau$ spins are integrated out, yielding effective three-body weights of downward facing triangles in the absence of measurements of the form
\begin{equation}
\label{eq:triangle}
	W^{(0)}(\sigma_1, \sigma_2; \sigma_3) = \sum_{\tau \in S_r} \Wg(\sigma_3 \tau^{-1}) d^{2r - \abs{\sigma_1^{-1} \tau} - \abs{\sigma_2^{-1} \tau}} .
\end{equation}
These weights have been investigated in great detail in Ref. \cite{zhou_emergent_2019}. They can be calculated with ease for $r=2$ and with considerably more difficulty when $r=3$, and several exact results are known for certain weights for arbitrary $r$. They simplify considerably in the $d \rightarrow \infty$ limit, which we consider below.

In the presence of a measurement -- say, on the bond between $\sigma_1$ and $\sigma_3$ -- we instead obtain the weights (up to an irrelevant overall factor of $d^r$ on all Boltzmann weights)
\begin{equation}
\label{eq:triangle_1}
	W^{(1)}(\sigma_2, \sigma_3) = \sum_{\tau \in S_r} \Wg (\sigma_3 \tau^{-1}) d^{2r - \abs{\sigma_2^{-1} \tau}} .
\end{equation}
This is the same weight as before, with the coupling to $\sigma_1$ removed. We see that the effect of measurements is to decouple spins from the corresponding triangle weight.

In the case that both bonds on a given triangle are measured, we find a triangle weight $W^{(2)} = \sum_{\tau} \Wg(\tau)$ which is independent of all three spins. This constant prefactor can again be neglected, as it cancels between the numerator and denominator of (\ref{eq:replicaSE}). Hereafter, we will simply set $W^{(2)} = 1$.

Finally, we note the symmetry group of the bulk triangle weights \cite{bao_theory_2020,bao2021symmetry}. Since each of the triangle weights only depend on the conjugacy class of the combination $\sigma_i^{-1} \sigma_j$, they are invariant under transformations of the form $\sigma_i \mapsto \xi_1 \sigma_i \xi_2^{-1}$; these correspond to independent permutation of ket and bra indices. They are also invariant under inversion, $\sigma_i \mapsto \sigma_i^{-1}$, corresponding to swapping all ket and bra indices. Altogether, the symmetry group is $(S_r \times S_r) \rtimes \mathbb{Z}_2$.

With the unitary gates averaged, tensor contractions between adjacent spins performed, and $\tau$ spins integrated out, the quantities (\ref{eq:Znk}) are given by a partition function of spins of the form
\begin{equation}
\label{eq:partitionfn}
	Z^{(n,k)}_{\Sigma}(\vec{X}) = \sum_{\qty{\sigma}} W_{\Sigma} \qty(\qty{\sigma_T}) W_{\mathcal{D}}(\qty{ \sigma_L, \sigma_R }) \prod_{(\sigma_1, \sigma_2; \sigma_3) \in \triangledown} W^{(\vec{X})}(\sigma_1, \sigma_2; \sigma_3) .
\end{equation}
In the above, the product over $(\sigma_1, \sigma_2; \sigma_3) \in \triangledown$ indicates a product over all downward-facing triangles of three spins in the bulk, with $\sigma_3$ at the bottom vertex. $W^{(\vec{X})}$ is either $W^{(0)}$, $W^{(1)}$, or $W^{(2)}$ as required by the locations of the measurements. $W_{\Sigma}$ is an additional weighting on the top layer of spins $\qty{\sigma_T}$ at the final time due to the top boundary condition on the tensor network, while $W_{\mathcal{D}}$ is an additional weighting of spins $\qty{\sigma_L}$ and $\qty{\sigma_R}$ at the left and right boundaries of the system, due to the presence of the dephasing channels; we will explain the boundary condition weightings below.

\subsection{Boundary Conditions}
The boundary conditions at the bottom of the circuit are simple: since the contraction of the spins $\bra{\sigma}$ from the first layer of unitaries with any initial product state is unity independent of $\sigma$ or the product state, the bottom of the effective model has open boundary conditions. Spins in the very bottom row of the model are coupled only to the spins diagonally above them in a downward facing triangle via the triangle weights.

The left and right boundary conditions are influenced by the presence of dephasing channels, and are different than in previous works \cite{zhou_emergent_2019,bao_theory_2020,jian_measurement-induced_2020,agrawal_entanglement_2021}. As a super-operator on the $2r$-fold single qudit Hilbert space, $\mathcal{D}$ of Eq. (\ref{eq:dephasing}) has matrix elements with respect to the states $\ket{\sigma}$ of the form
\begin{equation}
	\bra{\sigma} \mathcal{D}^{\otimes r} \ket{\tau} = \sum_{\qty{a, \bar{a}}} \qty[ \prod_{\ell = 1}^r \delta_{a_{\ell}, \bar{a}_{\sigma(\ell)}} ] \qty[ \prod_{\ell = 1}^r \delta_{a_{\ell} \bar{a}_{\ell}} ] \qty[ \prod_{\ell = 1}^r \delta_{a_{\ell}, \bar{a}_{\tau(\ell)}} ] = \sum_{\qty{a}} \qty[ \prod_{\ell = 1}^r \delta_{a_{\ell}, a_{\sigma(\ell)}} ] \qty[ \prod_{\ell = 1}^r \delta_{a_{\ell}, a_{\tau(\ell)}} ] .
\end{equation}
The above is equal to $d^{\alpha}$, where $\alpha$ is the number of orbits of the subgroup of $S_r$ generated by $\sigma$ and $\tau$. For general $\sigma$ and $\tau$, there is no simple closed-form expression (to our knowledge) for the above expression in terms of $\sigma$ and $\tau$. However, there are a few simple observations that can be made: for example,
\begin{equation}
\label{eq:dephasing_weights}
	\bra{\sigma} \mathcal{D}^{\otimes r} \ket{\mathbb{I}} = \bra{\mathbb{I}} \mathcal{D}^{\otimes r} \ket{\sigma} = \bra{\sigma} \mathcal{D}^{\otimes r} \ket{\sigma} = d^{r - \abs{\sigma}}.
\end{equation}
Additionally, it is clear that $\bra{\sigma} \mathcal{D}^{\otimes r} \ket{\tau}$ for $\sigma \neq \tau$ is smaller than both $\bra{\sigma} \mathcal{D}^{\otimes r} \ket{\sigma}$ and $\bra{\tau} \mathcal{D}^{\otimes r} \ket{\tau}$ by at least a factor of $1/d$. When we later take the $d \rightarrow \infty$ limit, this will effectively lock the boundary spins to the identity permutation\footnote{In this work, we've used dephasing channels to contrast the effect of measurements averaged at the density matrix level with measurements averaged at the trajectory level. An alternative scheme is to use depolarizing channels, $\mathcal{F}[\rho] = \frac{1}{d} \mathds{1}$, which acts directly as a boundary of $\mathbb{I}$ permutations. The distinction between the two approaches is inconsequential in the large $d$ limit.} $\mathbb{I}$. More generally, we see that dephasing channels break the $(S_r \times S_r) \rtimes \mathbb{Z}_2$ symmetry into just $S_r \times \mathbb{Z}_2$, given by the two group actions $\sigma_i \mapsto \xi \sigma_i \xi^{-1}$ and $\sigma_i \mapsto \sigma_i^{-1}$. This is because the dephasing channels couple ket and bra within individual copies, so that the weights remain invariant under permutations of the $r$ copies of the qubit and under Hermitian conjugation, but not under independent permutations of kets and bras.

The top boundary conditions are different for each of the four partition functions $Z^{(n,k)}_{\Sigma}$ in (\ref{eq:replicaSE}). Since the trace of $\rho^{\otimes r}_{\vec{m}}$ can equivalently be described by the inner product between the vectorized density matrix $\ket*{\rho^{(r)}_{\vec{m}}}$ and the reference state $\bra{\mathbb{I}_1, \ldots , \mathbb{I}_L} \equiv \bra{\mathbb{I}_A, \mathbb{I}_B}$, each of the four partition functions (\ref{eq:Znk}) are given by an expression of the form
\begin{equation}
	Z^{(n,k)}_{\Sigma} = \bra{\mathbb{I}_A, \mathbb{I}_B} \qty[ \Sigma^{\otimes k} \otimes \mathds{1} ] \ket*{\overline{\rho^{(r)}_{\vec{m}}}} ,
\end{equation}
with $\Sigma$ given by one of the four boundary operators in (\ref{eq:Boundary_Op}) and $\ket*{\overline{\rho^{(r)}_{\vec{m}}}}$ denoting $\ket*{\rho^{(r)}_{\vec{m}}}$ averaged over unitary gates and measurement outcomes\footnote{Note the slight abuse of notation: $\ket{\rho^{(r)}_{\vec{m}}}$ contains within its $nk+1$st moment the probability factor $p_{\vec{m}}$ of the measurement outcome. Really, $\ket*{\overline{\rho^{(r)}_{\vec{m}}}}$ means $\mathbb{E}_{\mathcal{U}} \sum_{\vec{m}} \ket*{\rho^{(r)}_{\vec{m}}}$.}. Acting the permutation operator to the left, we see that the four different partition functions only differ in their contraction with a top layer of permutations $\bra{\sigma_1, \ldots , \sigma_L }$. We denote by $\mathbb{C} = C^{\otimes k} \otimes 1$ the cyclic permutation on $k$ groups of $n$ elements each; in cycle notation,
\begin{equation}
	\mathbb{C} = (1 \ \ldots \ n) (n \! +\! 1 \ \ldots \ 2n)\ldots ([n(k \! - \! 1)] \ \ldots \ nk ) (nk\! +\! 1) .
\end{equation}
We also denote by $\bar{\mathbb{{C}}}$ its inverse anti-cyclic permutation. With this notation, the four different boundary operators are given by
\begin{equation}
\label{eq:Z_BCs}
	\begin{split}
		Z^{(n,k)}_{\mathcal{E}} &= \bra{\mathbb{C}_A, \bar{\mathbb{C}}_B}\ket*{\overline{\rho^{(r)}_{\vec{m}}}}, \quad Z^{(n,k)}_S = \bra{\mathbb{C}_A, \mathbb{I}_B}\ket*{\overline{\rho^{(r)}_{\vec{m}}}}\\
		Z^{(n,k)}_{\mathcal{E}0} &= \bra{\mathbb{C}_A, \mathbb{C}_B}\ket*{\overline{\rho^{(r)}_{\vec{m}}}}, \quad Z^{(n,k)}_{S0} = \bra{\mathbb{I}_A, \mathbb{I}_B}\ket*{\overline{\rho^{(r)}_{\vec{m}}}} .
	\end{split}
\end{equation}
Note in particular that both $Z^{(n,k)}_{\mathcal{E}}$ and $Z^{(n,k)}_S$ enforce domain walls between two different permutations in regions $A$ and $B$ at the final timelike boundary, while $Z^{(n,k)}_{\mathcal{E}0}$ and $Z^{(n,k)}_{S0}$ are uniform at the top boundary.

\subsection{Large $d$ Limit}
The expression (\ref{eq:partitionfn}) takes the form of a partition function of spins with local three-body weights. However, for arbitrary $d$ we cannot properly regard $Z^{(n,k)}_{\Sigma}$ as a partition function of a statistical mechanics model due to negative weights contained in the Weingarten function. These technical difficulties are largely alleviated by working in the limit $d \rightarrow \infty$, wherein the statistical mechanics model becomes well-defined. This limit has previously been systematically studied in Ref. \cite{zhou_emergent_2019}, as well as in \cite{bao_theory_2020,jian_measurement-induced_2020,agrawal_entanglement_2021}.

Using the asymptotic form (\ref{eq:weingarten_moeb}) for the Weingarten function, one is tempted to suggest that the $d \rightarrow \infty$ limit of the three-body weight (\ref{eq:triangle}) in the absence of measurements is given by
\begin{equation}
\label{eq:large_d}
	W^{(0)}(\sigma_1, \sigma_2; \sigma_3) = \sum_{\tau \in S_r} \qty[ \frac{1}{d^{2r}} \delta_{\sigma_3,\tau} + \mathcal{O}(d^{-2r-2}) ] d^{2r - \abs{\sigma_1^{-1} \tau} - \abs{\sigma_2^{-1} \tau}} \stackrel{?}{\rightarrow} d^{-\abs{\sigma_1^{-1} \sigma_3} - \abs{\sigma_2^{-1} \sigma_3}} .
\end{equation}
This would suggest that the three-body Boltzmann weights factorize into two-body Boltzmann weights in the $d \rightarrow \infty$ limit, yielding a particularly simple effective model. We would find a spin model on the tilted square lattice with nearest-neighbor interactions of the form
\begin{equation}
\label{eq:DWenergy}
	\beta E(\sigma_i, \sigma_j) = \abs{\sigma_i^{-1} \sigma_j} \log d .
\end{equation}
The $d \rightarrow \infty$ limit is analogous to a zero temperature limit, and the only spin configurations contributing to the partition function will be those which minimize the energy by breaking as few bonds as possible. The effect of measurements is then simply to eliminate the measured bonds from the spin Hamiltonian.

Unfortunately, this limit cannot be entirely correct, and the righthand limiting expression does not consistently keep track of the $1/d$ expansion. To see that this is the case, note that the three-body weight satisfies an exact \textit{unitarity constraint} \cite{collins_moments_2003,collins2006integration,zhou_emergent_2019,bao_theory_2020}, valid to all orders in $d$, when $\sigma_1 = \sigma_2 = \sigma$:
\begin{equation}
\label{eq:unitarity}
	W^{(0)}(\sigma,\sigma; \sigma_3) = \sum_{\tau \in S_r} \Wg(\sigma_3^{-1} \tau) (d^2)^{r-\abs{\sigma^{-1} \tau}} = \delta_{\sigma,\sigma_3} ,
\end{equation}
which is clearly incompatible with the naive $d \rightarrow \infty$ limit of (\ref{eq:large_d}). If we consider triangle configurations in which two out of three spins agree, so that the third spin is separated by a domain wall, then the unitarity constraint implies that domain walls cannot travel horizontally through the downward facing triangle. Instead, domain walls necessarily must travel downwards from the top of the triangle and then turn left or right. In this case, we have triangle weights in the $d \rightarrow \infty$ limit of the form \cite{zhou_emergent_2019}
\begin{equation}
	W^{(0)}(\sigma,\sigma'; \sigma) = W^{(0)}(\sigma',\sigma;\sigma) \stackrel{d \rightarrow \infty}{=} d^{-\abs{\sigma^{-1}\sigma'}}, 
\end{equation}
which, together with (\ref{eq:unitarity}), give the correct triangle weights for a single domain wall passing through the downward facing triangle. If the domain wall does not split within a downward facing triangle into multiple domain walls, then the unitarity constraint can be regarded as a configurational constraint on the allowed shape of the domain walls. Since the $d \rightarrow \infty$ limit also requires that domain walls travel along the shortest possible trajectory, we will find that domain wall splittings will be very rare. In summary, we may use the effective domain wall energy costs (\ref{eq:DWenergy}) for domain walls traveling diagonally within the system without splitting. 

The weight $W^{(1)}$ due to a single measurement is simpler: it is indeed given by the naive  $d \rightarrow \infty$ limit:
\begin{equation}
	W^{(1)}(\sigma_2,\sigma_3) \stackrel{d\rightarrow \infty}{=} d^{-\abs{\sigma_2^{-1} \sigma_3}} .
\end{equation}
For each $\tau$ in the sum (\ref{eq:triangle_1}), the contribution to the sum at leading order in $d$ is proportional to $d^{-2\abs{\sigma_3 \tau^{-1}} - \abs{\sigma_2^{-1} \tau}}$, and it is straightforward to show that this expession is minimized for $\tau = \sigma_3$.

\subsection{Domain Wall Configurations}

\begin{figure}[t]
	\centering
	\includegraphics{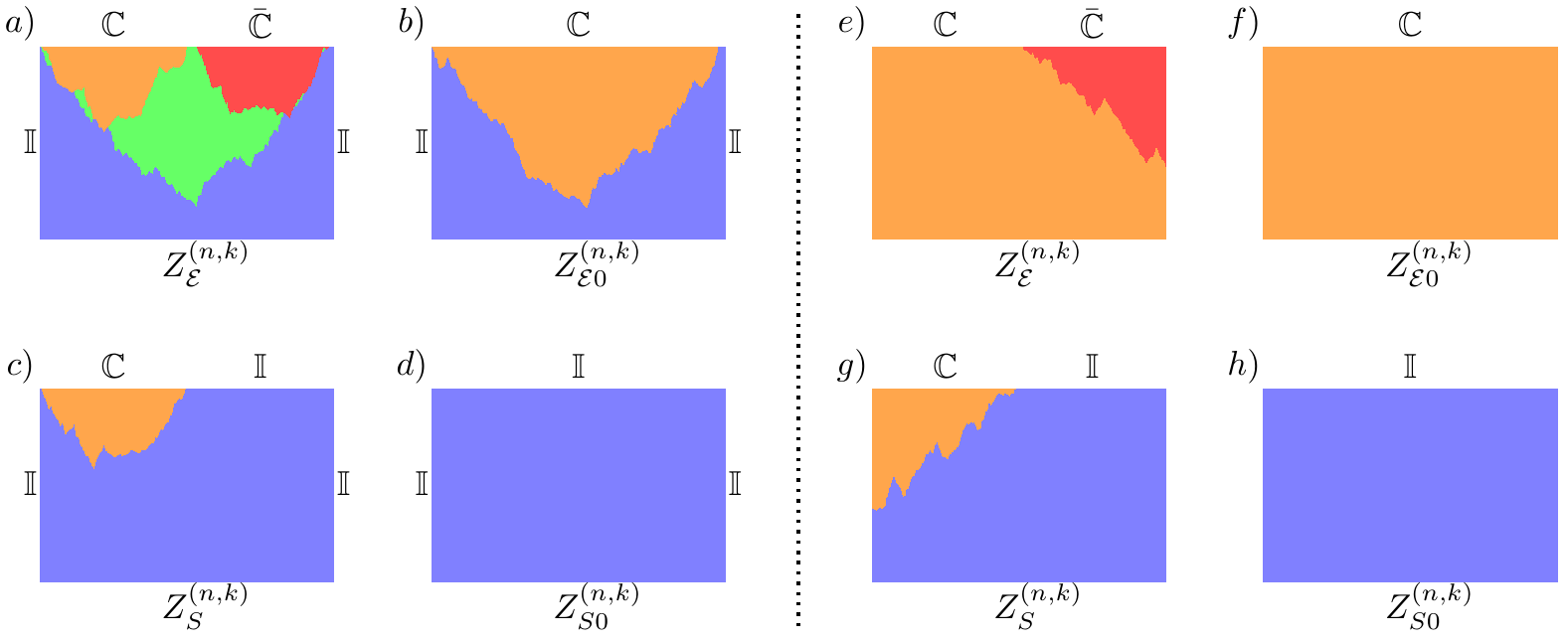}
	\caption{Left: domain wall configurations for the replica negativity (top) and entropy (bottom) in the presence of dephasing. Right: domain wall configurations for the replica negativity (top) and entropy (bottom) in the absence of dephasing, for sake of comparison with previous works.}
	\label{fig:DWs}
\end{figure}

With the relevant bulk Boltzmann weights identified in the $d \rightarrow \infty$ limit, and the boundary conditions properly accounted for, we can determine which spin configurations contribute to each partition function. In the $d \rightarrow \infty$ limit, only the single most energetically favorable configuration contributes. This is determined by the boundary conditions at the top and sides of the model. 

The top boundary conditions in each partition function are given in Eq. (\ref{eq:Z_BCs}): the replica negativity $\mathcal{E}^{(n,k)}_{A:B}$ is given by the free energy cost of a $\mathbb{C}$-$\bar{\mathbb{C}}$ domain wall at the top boundary relative to a uniform $\mathbb{C}$ boundary, while the replica entropy $S^{(n,k)}_A$ is given by the free energy cost of a $\mathbb{C}$-$\mathbb{I}$ domain wall at the top boundary relative to a uniform $\mathbb{I}$ boundary. Due to dephasing at the left and right boundaries, the weights (\ref{eq:dephasing_weights}) pin the left and right boundaries to $\mathbb{I}$ in the $d \rightarrow \infty$ limit. 

For sake of comparison with previous works, we start with the case in which dephasing is absent and the left and right boundaries are open. Examples of relevant domain wall configurations are depicted in the right half of Fig. \ref{fig:DWs}, subfigures $e$ through $h$. In both $Z^{(n,k)}_{\mathcal{E}}$ and $Z^{(n,k)}_S$, the structure of domain walls is simple: domains emerging at the middle of the top boundary run either to the left or right and terminate on the side boundary\footnote{In the absence of measurements, note that the domain walls could not only run to the left or right, but also split in two: in $Z^{(n,k)}_{\mathcal{E}}$, for instance, one could consider configurations in which the leftmost domain is $\mathbb{C}$ and the rightmost domain is $\bar{\mathbb{C}}$, while the middle domain is any $\mathbb{D}$ such that $\abs{\mathbb{C}^{-1} \mathbb{D}} + \abs{\mathbb{D}^{-1} \bar{C}} = \abs{\mathbb{C}^{-1} \bar{\mathbb{C}}}$. The multiplicity of minimal domain wall configurations with the same energy only gives an $\mathcal{O}(1)$ entropic contribution to the negativity of entropy, and does not affect the scaling with system size. Upon adding measurements, one side will yield a shorter path to the side boundary than the other with exponentially high probability, and this multiplicity is anyway expected to be destroyed.}. Since $\abs{\mathbb{C}^{-1} \bar{\mathbb{C}}} = \abs{\mathbb{C}^2} = k(n-2)$ and $\abs{\mathbb{C} \mathbb{I}} = \abs{\mathbb{C}} = k(n-1)$, the replica negativity and entropy in the absence of dephasing will be given by
\begin{equation}
\label{eq:no_dephasing}
	\mathcal{E}^{(n,k)}_{A:B}(\vec{X}) = \frac{1}{k(n-2)} [\beta E(\mathbb{C},\bar{\mathbb{C}})] \ell = \ell \log d, \quad S_A^{(n,k)}(\vec{X}) = \frac{1}{k(n-1)}[\beta E(\mathbb{C},\mathbb{I})] \ell = \ell \log d ,
\end{equation}
where $\ell \sim \mathcal{O}(L/2)$ is the length of the shortest possible domain wall from the center of the top boundary to either the left or right boundary, given the locations $\vec{X}$ of missing bonds and subject to the unitarity constraint. Since the dependence on $n$ and $k$ has canceled out, the replica limit can be trivially taken, yielding $\overline{\mathcal{E}_{A:B}(\vec{X})} = \overline{S_A(\vec{X})} = \ell \log d$. It is not surprising that the entropy and negativity agree with each other in the absence of dephasing; indeed, the negativity is equal to the 1/2 R\'enyi entropy $S^{(1/2)}_A$ in the case of pure states, and in the $d \rightarrow \infty$ limit the R\'enyi entropies $S^{(n)}_A$ are flat in $n$ as seen above. For the case of zero measurements $\ell = L/2$, yielding exactly the result expected from the Page curve \cite{page_average_1993,zhou_emergent_2019}. For nonzero measurement rates $\ell < L/2$, as the domain wall optimizes to pass through as many measurement locations as possible to minimize its energy.

We now turn to the effect of dephasing channels at the left and right boundaries. For the entropy, the change is not very significant: while the shape of the domain wall changes due to the left boundary condition, its length remains unchanged, leading to the same scaling as in (\ref{eq:no_dephasing}). This is depicted in subfigures $c$ and $d$ of Fig. \ref{fig:DWs}. 

On the other hand, the domain wall structure of the negativity is dramatically altered by the dephasing, as shown in Fig. \ref{fig:DWs}$a$ and $b$. In subfigure $a$, the domain walls separating $\mathbb{C}$ domains (orange) and $\bar{\mathbb{C}}$ domains (red) lie along minimal trajectories separating the top boundaries of subsystems $A$ and $B$ from the bulk, with lengths $\ell_A$ and $\ell_B$ respectively. The green domain represents any permutation $\mathbb{D}$ satisfying the criteria \cite{dong_holographic_2021}
\begin{equation}
\label{eq:D_criteria}
	\beta E(\mathbb{C}, \mathbb{D}) + \beta E(\mathbb{D}, \mathbb{I}) = \beta E(\bar{\mathbb{C}}, \mathbb{D}) + \beta E(\mathbb{D},\mathbb{I}) = \beta E(\mathbb{C},\mathbb{I}), \quad \beta E(\mathbb{C},\mathbb{D}) + \beta E(\mathbb{D},\bar{\mathbb{C}}) = \beta E(\mathbb{C}, \bar{\mathbb{C}}) ,
\end{equation}
and the domain wall boundary between the $\mathbb{D}$ and $\mathbb{I}$ domains lies along a minimal trajectory of length $\ell_{AB}$ separating the entire top boundary from the bulk. To see that such a $\mathbb{D}$ domain can exist, First consider the case in which figure \ref{fig:DWs}a were given by just three domains of $\mathbb{C}$, $\bar{\mathbb{C}}$, and $\mathbb{I}$ respectively. Then, using the bulk domain wall costs (\ref{eq:DWenergy}), one can separate the $\mathbb{C} \mathbb{I}$ domain wall into $\mathbb{C} \mathbb{D}$ and $\mathbb{D} \mathbb{I}$ domain walls with a thin $\mathbb{D}$ domain in between for no energy cost, and similarly for the $\bar{\mathbb{C}} \mathbb{I}$ domain wall. Then, the resulting $\mathbb{D}\mathbb{I}$ domain wall can reduce its energy by spanning the minimal length $\ell_{AB} \leq \ell_A + \ell_B$. Note that a similar observation has been made in Ref. \cite{dong_holographic_2021} in the context of random tensor networks.

Through simple algebraic manipulations, one can show from (\ref{eq:D_criteria}) that $\abs{\mathbb{C}^{-1} \mathbb{D}} = \abs{\bar{\mathbb{C}}^{-1} \mathbb{D}} = k \qty( \frac{n}{2} - 1 )$ and $\abs{\mathbb{D} \mathbb{I}} = k \frac{n}{2}$. Combining figures \ref{fig:DWs}a and \ref{fig:DWs}b, the replica negativity is given by
\begin{equation}
	\begin{split}
		\mathcal{E}^{(n,k)}_{A:B}(\vec{X}) &= \frac{1}{k(n-2)} \Big\{ [\beta E(\mathbb{C},\mathbb{D})] \ell_A + [\beta E(\bar{\mathbb{C}},\mathbb{D})]\ell_B + [\beta E(\mathbb{D},\mathbb{I})] \ell_{AB} - [\beta E(\mathbb{C},\mathbb{I})] \ell_{AB} \Big\} \\
		&= \frac{\log d}{k(n-2)} \qty{ k \qty( \frac{n}{2} - 1 ) (\ell_A + \ell_B) + k \frac{n}{2} \ell_{AB} - k(n-1) \ell_{AB} } \\
		&= \frac{\log d}{2} \Big\{ \ell_A + \ell_B - \ell_{AB} \Big\} ,
	\end{split}
\end{equation}
which is positive by definition. Once again, the $k$ and $n$ dependence has canceled out, allowing for the replica limit to be taken. Note in particular that when the measurement rate is zero, we have $\ell_A = \ell_B = \ell_{AB}/2 = L/2$, giving zero extensive negativity. This is in agreement both with the general arguments using Page's theorem leading to Eq. (\ref{eq:app_EN_random_state}), as well as the observed numerical results in the main text. Upon increasing the measurement rate from zero, we expect $\ell_A = \ell_B = \ell_{AB}/2 = \lambda(p) L/2 + o(L)$, where $\lambda(p)$ is a monotonically decreasing function of $p$. The above result then predicts that the $\mathcal{O}(L)$ contribution to the negativity vanishes at all nonzero measurement rates, once again in agreement with the results of the main text.

We also note that the mutual information $I_{A:B}$ can be obtained in a similar manner using the appropriate domain wall configurations for the R\'enyi entropies. Defining $I_{A:B}^{(n,k)} = S_A^{(n,k)} + S_B^{(n,k)} - S_{AB}^{(n,k)}$, the mutual information is given by
\begin{equation}
\label{eq:mutual_info}
    \begin{split}
        I^{(n,k)}_{A:B}(\vec{X}) &= \frac{1}{k(n-1)} \beta E(\mathbb{C},\mathbb{I}) \Big\{ \ell_A + \ell_B - \ell_{AB} \Big\} \\
        &= \log d \Big\{\ell_A + \ell_B - \ell_{AB} \Big\} ,
    \end{split}
\end{equation}
which is exactly twice as big as $\mathcal{E}^{(n,k)}_{A:B}$.

\subsection{KPZ Fluctuations}
In order to finally explain the $L^{1/3}$ power law negativity observed in the numerical results of the main text, we must address exactly how the domain wall lengths $\ell$ depend on the measurement rate. Towards this end, we relate the free energies of the domain walls in figure \ref{fig:DWs} to the free energies of a collection of directed polymers in a random environment (DPRE). This result will immediately suggest an $L^{1/3}$ subleading contribution to the scaling of each $\ell$ due to Kardar-Parisi-Zhang (KPZ) scaling of the free energy, yielding the desired result. Similar subleading $t^{1/3}$ KPZ fluctuations in time have previously been observed in the growth of entanglement in the pure unitary circuit, arising due to a slightly different mechanism \cite{zhou_emergent_2019}. Additionally, Ref. \cite{li_entanglement_2021} conjectured using both numerical and analytical evidence that the von Neumann entropy of a pure state monitored quantum circuit is given by the free energy of a directed polymer in a random environment, leading similarly to $L^{1/3}$ subleading scaling in the entanglement entropy. Our analytical arguments here provide additional support for this conjecture.

We start with $Z^{(n,k)}_{\mathcal{E}0}$ depicted in figure \ref{fig:DWs}b, since its domain wall structure is simpler than $Z^{(n,k)}_{\mathcal{E}}$ in figure \ref{fig:DWs}a. We will also neglect the bulk free energy of the domains, which are independent of the domain species and cancel in the numerator and denominator of Eq. (\ref{eq:replicaSE}), and focus on the additional free energy cost of the domain wall. In the limit $d \rightarrow \infty$, $Z^{(n,k)}_{\mathcal{E}0}$ is then given by the single Boltzmann weight of a domain wall between $\mathbb{C}$ and $\mathbb{I}$ domains of minimal length $\ell_{AB}$. However, according to the energy costs (\ref{eq:DWenergy}), we can alternatively think of this single domain wall of energy $k(n-1) \log d$ per length as $k (n-1)$ independent, noninteracting domain walls each of energy $\log d$ per length. Each of these domain walls can be thought of as an ``elementary'' domain wall \cite{zhou_emergent_2019,jian_measurement-induced_2020} between two spin domains differing by a single permutation, with $k(n-1)$ total permutations needed to reach an $\mathbb{I}$ domain from a $\mathbb{C}$ domain. For $d < \infty$ these elementary domain walls can in principle fluctuate away from each other slightly, but for $d \rightarrow \infty$ they must each follow the same minimal trajectory to pass through as many measurement locations as possible, assuming this minimal length trajectory is unique.

We can view each elementary domain wall as a single polymer. Since the polymer will never minimize its energy using looping or overhanging spatial configurations\footnote{It is possible to imagine extremely rare measurement configurations in which the polymer indeed minimizes its energy using an overhang. Such overhangs are expected to be lost upon coarse-graining.}, the polymer is \textit{directed}, and can be described by a spatial profile $y(x)$ restricted to the half-plane $y < 0$ with boundary conditions $y(0) = y(L) = 0$. Measurements can then be viewed as a random attractive potential on the polymer: it costs $\log d$ less energy for the polymer to pass through a measured bond than an unmeasured bond. 

In a coarse-grained picture, each set of spacetime measurement locations $\vec{X}$ gives rise to a disordered potential landscape $V(x,y)$ on the polymers with mean and variance
\begin{equation}
	\fullavg{ V(x,y) } = p \log d, \quad \fullavg{ V(x,y) V(x',y') } = p(1-p) (\log d)^2 \delta(x-x') \delta(y-y') .
\end{equation}
If we assume that coarse-graining occurs over a sufficiently large scale so as to include many measurement locations, then we can think of $V(x,y)$ as Gaussian distributed via the central limit theorem. If we make the final additional assumption that the unitary constraint can be largely neglected in the presence of nonzero measurement rates upon coarse-graining, then the polymer trajectory $y(x)$ for a given set of measurement locations can be obtained by minimizing the effective Hamiltonian
\begin{equation}
\label{eq:cont_polymer}
	\beta H[y(x)] = \log d \int \dd{x} \qty{ \sqrt{1 + \qty( \dv{y}{x} )^2} + V(x,y) } \simeq \log d \int \dd{x} \qty{ 1 + \frac{1}{2}(\partial_x y)^2 + V(x,y) } ,
\end{equation}
where we have Taylor expanded the square root and dropped irrelevant higher order derivative terms. It is well-known that the free energy of a directed polymer in a random Gaussian potential satisfies the KPZ equation \cite{kardar_roughening_1985,kardar_dynamic_1986,huse_huse_1985,huse_pinning_1985}, both at zero and finite temperature. The KPZ equation restricted to the half-plane has itself been studied extensively \cite{barraquand_half-space_2020,gueudre_directed_2012}. These previous analyses immediately suggest that the free energy $F^{(n,k)}_{\mathcal{E}0}$ of $n(k-1)$ directed polymers described by the effective Hamiltonian (\ref{eq:cont_polymer}), averaged over disorder potentials $V(x,y)$, scales as \cite{li_entanglement_2021}
\begin{equation}
	\fullavg{F^{(n,k)}_{\mathcal{E}0}(\vec{X})} = k(n-1)\qty( s_0 L + s_1 L^{1/3} ) ,
\end{equation}
where $s_0$ and $s_1$ are nonuniversal positive constants which depend on the measurement rate.

Similar analysis can be applied to $Z^{(n,k)}_{\mathcal{E}}$. Given the domain wall configurations depicted in Fig. \ref{fig:DWs}$a$ and described in the previous section, $Z^{(n,k)}_{\mathcal{E}}$ is described by $k\qty(\frac{n}{2} - 1)$ polymers with boundary conditions $y(0) = 0 = y(L/2) = 0$, $k \qty(\frac{n}{2} - 1)$ polymers with boundary conditions $y(L/2) = y(L) = 0$, and $k \frac{n}{2}$ polymers with boundary conditions $y(0) = y(L) = 0$. The free energy of this collection of polymers, averaged over disorder potentials, yields
\begin{equation}
	\fullavg{ F^{(n,k)}_{\mathcal{E}} } = 2k \qty(\frac{n}{2} - 1) \qty( s_0 \frac{L}{2} + s_1 \qty(\frac{L}{2})^{1/3} ) + k \frac{n}{2} \qty( s_0 L + s_1 L^{1/3} ) .
\end{equation}
The averaged replica negativity is then finally given by
\begin{equation}
	\fullavg{ \mathcal{E}^{(n,k)}_{A:B}(\vec{X}) } = \frac{1}{k(n-2)} \qty(\fullavg{ F^{(n,k)}_{\mathcal{E}} } - \fullavg{ F^{(n,k)}_{\mathcal{E}0} }) = \qty( \frac{1}{2^{1/3}} - \frac{1}{2} )s_1 L^{1/3} .
\end{equation}
Note in particular that the $\mathcal{O}(L)$ contribution has dropped out as expected, and that the result is independent of $k$ and $n$ so that the replica limit may be performed. Additionally, from (\ref{eq:mutual_info}) we see that the mutual information $\fullavg{I_{A:B}}$ satisfies the same $L^{1/3}$ scaling.

As a final comment, we mention some considerations for the case $d < \infty$. While the $d \rightarrow \infty$ limit has been used throughout for analytical convenience, both Clifford simulations and real physical systems have finite qudit dimensions. It is therefore important to understand to what extent the above results can be applied to finite systems. Ref. \cite{zhou_emergent_2019} has performed detailed perturbative expansions of the three-body weights (\ref{eq:triangle}), showing in particular that the elementary domain walls forming the noninteracting polymers of the above discussion attain interactions as $1/d$ is increased from zero. While the details of this interaction can be quite complex, it is worth noting that a collection of polymers with pointwise attractive interactions are also equivalent to the KPZ equation: applying the replica trick to a partition function with Hamiltonian of the form (\ref{eq:cont_polymer}), one obtains a model of several polymers with pointwise attractive interactions. Indeed, even without a direct mapping to the KPZ equation or a model with known KPZ dynamics, it is difficult to escape the KPZ universality class, as it contains no relevant perturbations. Thus, we expect any $1/d$ perturbative corrections to the results detailed above will not change the qualitative physics, and will only alter the nonunversal coefficients $s_{0/1}$.

\bibliographystyle{apsrev4-2}
\bibliography{supp_refs}